\documentclass[11pt]{article}
\usepackage[T1]{fontenc}
\usepackage[letterpaper,margin=1.15in]{geometry}
\usepackage{microtype}
\usepackage{amsmath,amssymb,amsthm}
\usepackage{newtxtext,newtxmath}
\usepackage{booktabs,array,tabularx,graphicx}
\usepackage{placeins}
\graphicspath{{figures/}}
\usepackage[round,authoryear]{natbib}
\usepackage[colorlinks=true,linkcolor=black,citecolor=black,urlcolor=black]{hyperref}
\newtheoremstyle{journal}{6pt}{6pt}{\normalfont}{}{\bfseries}{.}{.5em}{}
\theoremstyle{journal}
\newtheorem{assumption}{Assumption}[section]
\newtheorem{lemma}{Lemma}[section]
\newtheorem{proposition}{Proposition}[section]
\newtheorem{corollary}{Corollary}[section]
\renewcommand{\qedsymbol}{$\blacksquare$}
\renewenvironment{proof}[1][Proof]{\par\noindent\textbf{#1.}\enspace}{\hfill\qedsymbol\par}
\numberwithin{equation}{section}
\title{Why Good Experience Can End Joint Activity}
\author{Weikang Zhang\thanks{Email: \href{mailto:weikang.zhang@siu.edu}{weikang.zhang@siu.edu}}}
\date{}
\begin{document}
\maketitle
\begin{abstract}
\noindent Learning a partner's preferences through joint activity can make both partners worse off than receiving no feedback from the activity. We study a responder who values the partner's payoff but bears a cost when joining is attributed to personal interest. Both partners observe noisy evidence of whether the responder wanted the activity, available only after participation. Evidence of interest raises this cost, discouraging joining on unwanted days; joining only on wanted days would then be too revealing to sustain. With choices based on current payoffs, we derive a condition under which learning can end participation without ever making stopping beneficial to either partner. Feedback accuracy and preference persistence described by couples' diaries satisfy this condition at baseline payoffs. Once activity stops, no further evidence arrives and the belief that the responder generally wants the activity stays fixed, so waiting may never restore it; preference information available without participating can. A separate forward-looking model with highly persistent preferences shows why this remedy can also reduce joint welfare. When participation reveals preferences exactly, a responder may accept at a current loss to establish a lack of interest, making acceptance less revealing and so sustaining participation on wanted days. Free daily revelation outside the activity removes this incentive. At the model's baseline, it lowers the best joint value in monotone Markov equilibria with exact participation feedback, but raises it with feedback as weak as the diary's.


\end{abstract}

\section{Introduction}\label{sec:introduction}

Trying an activity together is a common way for partners to learn whether it is worth continuing, and an activity that ends after such trials is naturally read as one that was not. Yet partners in an ongoing collaboration may take part both because they like the activity and because they value doing something for each other. Participation can also matter for what it conveys about this willingness to respond to a partner. If experience makes personal interest more apparent, agreeing next time may no longer carry the same meaning, and the incentive to participate changes with it. Can good experience end activities that would still benefit both partners on average?

We study a new activity that competes with an existing arrangement both partners accept and value. One partner proposes it; the responder may agree even when they prefer the existing arrangement, because they also value the proposer's payoff. Participation in the existing arrangement is understood as responding to the partner. When joining the new activity is instead attributed to personal interest, the responder loses part of the value this understanding carries, whereas declining leaves it intact. We assume that when interest is fully revealed, this attribution cost outweighs the gain from joining.

Evidence of wanting the activity can thus undermine participation. As the belief that the responder wants the activity rises, joining becomes unprofitable on days they prefer the existing arrangement. Accepting only on the other days would fully reveal their preference and also be unprofitable. Participation when the responder wants the activity depends on participation when they do not. Above a participation threshold for this belief, no equilibrium of the current interaction implements the activity, even with counter-proposals and costless communication. Participation in both preference states ceases to benefit the responder on average only at a higher payoff threshold. Between the thresholds, the activity cannot take place, although participation in both states would benefit the proposer in each state and the responder on average.

We first analyze repeated encounters in which partners choose an equilibrium of the current interaction, taking no account of future learning. Participation produces noisy feedback, observed by both partners, about whether the responder wanted the activity that day. It informs public beliefs about today's preference and the responder's initially unknown long-run propensity to want the activity. Because opportunities themselves convey no preference information, evidence arrives only while participation remains feasible, and each report moves beliefs only so far. The participation threshold therefore also limits how far learning can go. We derive a condition on feedback informativeness and preference persistence under which learning never reaches the higher payoff threshold. Every opportunity lost to learning then retains positive expected value for both partners. Whenever participation is blocked with positive probability, both are strictly worse off than without feedback after the activity.

Daily diaries of couples describe how preferences vary and how well partners recognize them. In the main diary, each partner records their own sexual desire and judges the other's. Comparing judgments with the partner's own desire ratings on days without activity gives the model's feedback informativeness; variation in desire across days constrains preference persistence. At baseline payoffs, this informativeness satisfies the condition throughout the persistence range consistent with the diaries. Experience can stop participation, but cannot reach beliefs at which stopping would protect the responder.

The resulting loss can persist long after the experience that triggered it. During a stop, no further public preference evidence arrives. The judgment of the responder's long-run propensity stays fixed, and beliefs about their preference on later days drift toward it. Waiting restores participation only if that judgment lies below the participation threshold. In each of 36 simulated environments consistent with the diaries, stops that waiting cannot end account for at least 95.1\% of blocked opportunities over days 1--84. The main burden therefore comes from opportunities that time alone cannot restore.

Nor can more precise feedback from participation restore them: it supplies no new evidence to a pair that has stopped and can enlarge the overall loss. When evidence requires participation, a responder seeking to establish a lack of interest must propose taking part again. The proposer may read this request as interest, reinforcing the attribution that made participation costly. Preference evidence available without participation can restore some of these opportunities. We compare such outside feedback with feedback after participation, holding fixed the expected daily number of reports across pairs and how well each report distinguishes wanting from not wanting. In the main simulated environment, outside feedback reduces the joint payoff loss over days 1--84 from 14.8\% to 8.4\% of the no-report benchmark.

When both partners look ahead, this remedy can also have a cost. We study a separate economy in which the responder's long-run propensity is public and preferences persist longer than the diaries allow. With exact feedback after participation, a responder who does not want the activity may still accept it at a current loss: the report comes too late to lower that day's attribution cost, but it establishes a lack of interest and lowers future ones. Such acceptances make joining less revealing of personal interest, enabling more participation when the responder wants the activity.

Free daily revelation of preferences without participation removes this incentive. At this economy's baseline, it lowers the best joint value in Markov equilibria, even if additional public information is available before acceptance. With participation feedback as weak as the main diary's, unwanted participation provides too little evidence to compensate for a current loss at high beliefs. The same daily outside information then restores participation and raises the best joint value in monotone Markov equilibria. Outside evidence can therefore restore participation by supplying missing information, or reduce it by removing an incentive to participate.

These results bear on what the end of an activity reveals about its value. In a matching model in which match quality can be learned only by forming the match, a worker leaves exactly when the match rent reaches zero \citep{jovanovic1979}. Experience can also leave errors uncorrected: an option that yields evidence only when chosen can remain underestimated relative to an inferior alternative \citep{le_mens_denrell2011}, and forced or noninstrumental contact reduces that bias \citep{denrell2005}. Here evidence of personal interest lowers the return to the next acceptance, so an activity can end while participation would still benefit both partners on average.

Information can also halt joint activity in related models. With unanimity and perfectly negatively correlated types, one member's success can end collective experimentation by revealing the other to be a sure loser, who vetoes the risky action \citep{strulovici2010}; when only trade generates feedback and records are complete, a negative record can exclude a seller permanently \citep{kovbasyuk_spagnolo2024}. In Strulovici's model, stopping serves the member who imposes it; here, by refusing on unwanted days, the responder also loses the activity on wanted days. We establish when learning removes only opportunities that remain valuable to both partners on average. For the resulting persistent stops, recovery depends on evidence that remains available after participation ends.

Section~\ref{sec:literature} reviews related work, Sections~\ref{sec:static}--\ref{sec:forward} develop the analysis, and Sections~\ref{sec:robustness}--\ref{sec:conclusion} discuss its scope and conclude.

\section{Related literature}\label{sec:literature}

Learning can end cooperation in several related models. In \citet{khubulashvili2025}, success invites a partner's free riding, and the successful researcher chooses to work alone. Under public offers and the patience conditions of \citet{horner_vieille2009}, rejecting the first serious offer raises inferred seller quality but leaves only losing offers. In \citet{strulovici2010}, one voter's success can tell another that they are a sure loser, who imposes the safe action; under simple majority, voters may stop experimenting although the risky action has the higher immediate expected payoff, an outcome that unanimity rules out. In the bad-reputation model of \citet{ely_valimaki2003}, a patient agent's incentive to separate from a bad type hurts short-lived customers, who stop participating. Identity costs can also block take-up entirely when imitation is cheap \citep{hupkau_maniquet2018}. \citet{friedrichsen_koenig_schmacker2018} charge stigma only upon take-up as a simplification, noting that image costs of non-take-up would leave their results qualitatively unchanged. Our attribution cost is likewise paid only when the activity takes place, but this assumption matters: because declining is free, a responder who wants the activity can avoid the cost by staying out. Evidence that the responder wants the activity raises this cost, discouraging joining on unwanted days; joining only on wanted days would then be too revealing to be worthwhile. In Khubulashvili's and Strulovici's models, stopping serves whoever imposes it; here, by refusing on unwanted days, the responder loses the activity on wanted days. Between the participation and payoff thresholds, no equilibrium of the current interaction implements the activity, although participation in both states would benefit the proposer in each state and the responder on average.

When learning requires participation, a stop can last indefinitely, because no evidence arrives while it lasts. In \citet{kovbasyuk_spagnolo2024}, trade generates feedback, beliefs without trade drift toward a known mean, and, when that mean cannot support trade, a negative record under full disclosure excludes a seller permanently. In \citet{bardhi_guo_strulovici2026}, workers are learned about only when assigned, and, under breakdown learning, a lower initial rank can leave a worker behind indefinitely. In the alternative model of Section~6 of \citet{wiseman2008}, entrants learn about a fixed commitment identity together with a changing action state and may eventually all stay out. Here, with choices based on current payoffs, participation also limits how far learning can go: feedback follows only participation, which stops once the belief that the responder wants the activity today exceeds the participation threshold, and each observation moves beliefs only so far. Experience therefore cannot raise public beliefs about today's preference or the responder's enduring propensity above a bound set by the participation threshold, feedback informativeness and preference persistence (Proposition~\ref{cor:ceiling}). When this bound lies below the payoff threshold, as it does at baseline payoffs for the main diary's feedback informativeness and any persistence consistent with the diaries, every opportunity that experience blocks still has positive expected value for both partners. Both are then strictly worse off than without feedback whenever experience blocks participation with positive probability.

A stop can end in several ways. Negative records can expire in \citet{kovbasyuk_spagnolo2024}; in \citet{daley_green2012}, a seller optimally waits between two belief thresholds while independent news continues to arrive, and more precise news can reduce efficiency. In \citet{pallais2014}, where only employment generates performance information, a more precise evaluation lowers later employment for workers initially above the hiring threshold while raising every group's expected earnings. Patient informed sellers can sell at a loss to reveal high quality \citep{kovbasyuk_spagnolo2024}, and workers accept negative wages to be observed, although favored workers bid down too \citep{bardhi_guo_strulovici2026}. How an action is read also matters: \citet{daley_green2014} separate the interim belief after an action from the final belief after a grade, and a high type's attempt to prove itself can be read as coming from a lower type \citep{feltovich_harbaugh_to2002}. Here, since feedback follows participation, an excluded responder can show that they do not want the activity only by proposing it themselves, and the proposal can be read as interest, raising the attribution cost of the very participation that would produce the evidence. A stop that waiting cannot end therefore needs evidence that still arrives after participation has stopped, which more precise feedback from participation cannot supply.

Outside evidence can change the incentive to generate information through participation. Career incentives depend on today's behavior affecting future inferences \citep{holmstrom1999}, costly verification before purchase can crowd out a seller's reputational incentives \citep{knutsen2025}, and a good seller who is observed only when selling keeps selling at a short-run loss \citep{bar_isaac2003}. Social image already depends on who participates: participation brings honor and abstention stigma, and participation can be complementary across agents \citep{benabou2006}; undeserving welfare claimants indistinguishable from deserving ones lower the esteem of all claimants \citep{besley_coate1992}. In our forward-looking economy, the attribution cost is set by the composition of the acceptance pool, the probability that a responder who accepts wants the activity that day, and is fixed before the day's participation feedback arrives. With exact participation feedback, the responder can accept an unwanted activity at a loss to show that they do not want it today; this makes acceptance less revealing of interest and so supports participation on days they want the activity. Free, exact daily revelation outside the activity removes this incentive: every Markov equilibrium is then held each day to the participation constraint of a single encounter, whatever public signal precedes acceptance (Proposition~\ref{prop:full}), and at the baseline the best joint value falls. With participation feedback as weak as the main diary's, accepting at a loss reveals too little to pay off at high beliefs, and the same outside revelation instead raises the best joint value within monotone Markov equilibria (Proposition~\ref{prop:weak}). As in \citet{prat2005}, each comparison places a value attainable under one feedback arrangement above the best under the other. An initial public signal can lower or raise welfare depending on its precision and on whether the agent's own experimentation produces bad or good news \citep{halac_kremer2020}; here, too, the effect of outside evidence depends on what participation itself reveals, but it works through the composition of the acceptance pool.

\section{Motive attribution and joint participation}\label{sec:static}

When the return to participation depends on how a partner interprets one's motive, a joint activity can become infeasible before its average value is exhausted. The model identifies the beliefs at which this occurs and who bears the loss.

\subsection{The joint choice}\label{subsec:primitives}

Let $f$ denote the existing arrangement that both parties accept and $x$ a new option competing with it for the same occasion. Party~1, the proposer, makes the initial proposal; party~2, the responder, decides whether to accept. Both activities require mutual agreement, and either party may end the interaction unilaterally, yielding outcome $n$ and direct payoff $-\varepsilon_i<0$ to party $i$. Direct payoffs $u_i(c;z_i)$ of outcomes $c\in\{x,f,n\}$, with $z_1=s$ and $z_2=\theta$, are net of the resources used and measured relative to $f$.

Each party privately knows their own direct ranking of the two activities. The responder's state is $\theta\in\{H,L\}$: relative to $f$, choosing $x$ gives them $g>0$ in state $H$, where they directly prefer $x$, and $-b<0$ in state $L$, where they directly prefer $f$; these are two preference states of the same responder. The proposer's state is $s\in\{h,l\}$: choosing $x$ gives them $v>0$ in state $h$ and $-w<0$ in state $l$. All four combinations have positive probability. Write $\pi_h=\Pr(H\mid h)$ for the probability that the responder prefers $x$, conditional on the proposer also preferring it, and $\rho_\theta=\Pr(h\mid\theta)$. Each party also derives utility from the other's direct payoff, the responder with weight $\alpha>0$ and the proposer with weight $\alpha_1\ge0$, each the same in both of that party's states, so a responder who prefers $f$ may join $x$ for what it gives the proposer.

The responder also cares how participation is understood. Under $f$, participation is understood as a response to the partner, and this understanding is part of the shared value that $f$ carries. When the responder directly prefers $x$, their own gain from $x$ is a reason to join, so participation in $x$ can instead be read as following their own interests.\footnote{\citet{benabou2006} analyze how external rewards affect reputational payoffs through inferences about motives.} Let $t$ record everything the parties jointly observe up to the implementation of $x$, and $q_s(t)=\Pr_1(H\mid t,s)$ the resulting posterior of a proposer in state $s$ that the responder directly prefers $x$. We assume that the proposer attributes participation in $x$ to the responder's own interests with probability $q_s(t)$, and that such an attribution costs the responder a component $\kappa>0$ of the shared value of $f$. Joining $x$ therefore entails an expected attribution cost $\kappa q_s(t)$ relative to $f$. This cost is positive even when both preference states join, and it leaves the responder's weight on the proposer's direct payoff unchanged:
\begin{equation}\label{eq:utility}
U_1(c;s,\theta)=u_1(c;s)+\alpha_1u_2(c;\theta),\qquad
U_2(c,t;\theta,s)=u_2(c;\theta)+\alpha u_1(c;s)-\kappa\mathbf 1\{c=x\}\,q_s(t).
\end{equation}
The cost is fixed by the posterior at implementation; later information does not change it. We also assume that declining $x$ costs nothing, since the understanding belongs to $f$, which remains available.

When the proposer is in state $h$, the responder derives utility $A=\alpha v$ from the proposer's payoff. Before the attribution cost, the responder's payoffs from $x$ in states $L$ and $H$, and the proposer's, are
\begin{equation}\label{eq:dvalues}
d_L=A-b,\quad d_H=A+g,\quad \Delta=d_H-d_L=g+b;\qquad
v_L=v-\alpha_1b,\quad v_H=v+\alpha_1g=v_L+\alpha_1\Delta.
\end{equation}

\begin{assumption}[Participation region]\label{ass:region}
$d_L>0$, $d_H<\kappa$ and $0\le\alpha_1<\min\{v/b,\,w/g\}$.
\end{assumption}
\noindent The condition $d_L>0$ lets a responder in state $L$ join when the attribution cost is low. The condition $d_H<\kappa$ means that the understanding is worth more than the net gain from $x$ to a responder who prefers it, so participation that fully revealed that preference would not pay. If instead $\kappa<d_H$, a responder in state $H$ would join even when fully revealed, and information could not prevent participation. The restriction on $\alpha_1$ ensures $v_L>0$ and $-w+\alpha_1g<0$: whatever the responder's state, the proposer gains from implementing $x$ in state $h$ and does not want it in state $l$.

Before choosing, the parties can engage in a fixed number $J\ge2$ of rounds of costless public communication; the proposer then offers $x$ or $f$. The responder can accept, reject an $x$ offer and counteroffer $f$, or answer an $f$ offer by counteroffering $x$, which the proposer can accept or reject by offering $f$ again. Once rejected, $x$ cannot be proposed again; there are no transfers. An equilibrium consists of behavior strategies and beliefs. Beliefs obey Bayes' rule at every information set and implementation history reached with positive probability, and continuation choices are optimal at every information set. Deviations are evaluated holding the other party's strategy and the attribution posteriors fixed. The equilibrium is consistent if one sequence of fully mixed behavior strategies yields its strategies and beliefs in the limit. Appendix~\ref{subsec:s1_construction} completes the action rules and gives the consistent constructions.

Either party can always secure the zero payoff of $f$ through later voluntary choices, so in every equilibrium a proposer in state $l$ never implements $x$, and a responder who accepts an offer of $x$ on the equilibrium path is certain that the proposer is in state $h$ (Lemma~\ref{lem:outside}). Implementation thus identifies the proposer's direct preference but not necessarily the responder's.

\subsection{When joint participation is feasible, and what an implementation is worth}\label{subsec:participation}

Fix the proposer's state at $h$. Since the responder can secure zero, participation in state $L$ at a history with implementation posterior $q$ requires $d_L-\kappa q\ge0$, that is, $q\le c_*\equiv d_L/\kappa\in(0,1)$. This posterior, however, is endogenous: it depends on how often each preference state joins. Let $a_\theta=\Pr(x\mid\theta,h)$ be the probability of implementing $x$ in responder state $\theta$, $V_\theta=\mathbb E[U_2\mid\theta,h]$ the corresponding expected utility, and $M_H=\pi_ha_H$ and $M_L=(1-\pi_h)a_L$ the probabilities, conditional on $h$, that the responder is in state $H$ or $L$ and $x$ is implemented.

\begin{lemma}[Participation constraints and strategy imitation]\label{lem:restrictions}
Under Assumption~\ref{ass:region}, every equilibrium satisfies
\begin{equation}\label{eq:restrictions}
M_L\geq\frac{1-c_*}{c_*}M_H,
\qquad
\Delta a_L\leq V_H-V_L\leq\Delta a_H.
\end{equation}
In particular, $a_H\geq a_L$.
\end{lemma}

The first inequality says that participation in state $H$ must be accompanied by participation in state $L$. A history at which only $H$ implements $x$ would reveal $H$ and give them $d_H-\kappa<0$; at a history shared with $L$, participation in state $L$ requires a posterior of at most $c_*$. Averaged over implementations, the posterior $M_H/(M_H+M_L)$ is therefore at most $c_*$, which is the first inequality. This constraint alone would still allow implementation at priors $\pi_h$ above $c_*$, because a responder in state $H$ gains from implementation at any posterior below $d_H/\kappa>c_*$. Rationing participation in state $H$ to hold the posterior at $c_*$, with $L$ always joining, would then give implementation probability $(1-\pi_h)/(1-c_*)$ conditional on $h$. The second pair rules this out. It compares each state's payoff with that from following the other state's complete strategy, which yields the same public histories and attribution costs. Since $H$ gains $\Delta$ more than $L$ from each implementation on the same history, $H$ implements at least as often as $L$. Hence the posterior averaged over implementations is at least the prior, and equal participation is the best dilution available.\footnote{\citet[Proposition~1]{hupkau_maniquet2018} show that when imitation is cheap and identity costs are large, nobody takes up a conditional benefit.} Appendix~\ref{app:restrictions} gives the proof.

\begin{proposition}[The threshold for joint participation]\label{prop:threshold}
Suppose Assumption~\ref{ass:region} holds and no exogenous signal about $\theta$ arrives after communication begins. A consistent equilibrium with positive probability of implementing $x$ exists if and only if $\pi_h\le c_*$, or equivalently $A\ge b+\kappa\pi_h$. When $\pi_h>c_*$, no equilibrium, consistent or not, implements $x$. When $\pi_h<c_*$, a consistent equilibrium exists in which the responder accepts $x$ in both preference states and strictly prefers acceptance to $f$ in each.
\end{proposition}

\begin{proof}
If $x$ is implemented with positive probability, the first inequality of Lemma~\ref{lem:restrictions} rules out $M_L=0$, and the lemma gives $a_H\ge a_L>0$ and
\[
\frac{\pi_h}{1-\pi_h}
\leq\frac{c_*}{1-c_*}\frac{a_L}{a_H}
\leq\frac{c_*}{1-c_*},
\]
so $\pi_h\le c_*$. Conversely, if $\pi_h\le c_*$, let communication convey no private information, let the proposer offer $x$ in state $h$ and $f$ in state $l$, and let the responder accept $x$ in both preference states. The posterior at implementation is $\pi_h$, and the responder obtains $d_H-\kappa\pi_h>0$ and $d_L-\kappa\pi_h\ge0$. The proposer obtains $v_L+\alpha_1\Delta\pi_h>0$ in state $h$; in state $l$, deviating to an $x$ offer gives at most $-w+\alpha_1g<0$. Appendix~\ref{subsec:s1_construction} specifies play at the remaining nodes, off-path beliefs and consistency.
\end{proof}

When the prior exceeds $c_*$, the first inequality requires $L$ to participate more often than $H$, while imitation requires $H$ to participate at least as often as $L$. The two restrictions can then hold only if $x$ is never implemented. With two preference states, the highest implementation probability therefore falls from one to zero at the participation threshold $c_*$. Since the proposer in state $h$ gains from implementation in both responder states, the participation blocked here is participation the proposer wants.

Beyond the payoffs, the argument uses only that both preference states have the same messages and actions, that the proposer sees only the public history (hence the hypothesis of no exogenous signal about $\theta$), and that $f$ remains available until implementation. Whatever the fixed number of rounds $J$ and whoever names $x$, the threshold is unchanged, since Lemma~\ref{lem:restrictions} covers accepted offers and counteroffers.

Satisfying the threshold does not ensure implementation. Besides joint acceptance by both preference states, at any prior $\pi_h=p\in(0,c_*]$ there is a mixed-acceptance equilibrium, in which the responder accepts $x$ for sure in state $H$ and with probability
\begin{equation}\label{eq:ell}
\ell_*(p)\equiv\frac{p(1-c_*)}{(1-p)c_*}
\end{equation}
in state $L$; the posterior after acceptance is exactly $c_*$, $L$ is indifferent, $H$ obtains $\Delta$, and the implementation probability given $h$ is $p/c_*$. A zero-participation equilibrium exists at every prior $p\in(0,1)$, with off-path beliefs that attribute an unexpected move toward $x$ to state $H$.

In every equilibrium, summing over implementing histories given $h$ and using Bayes' rule gives
\begin{equation}\label{eq:accounting}
\begin{gathered}
\mathbb E[\kappa\mathbf 1\{x\}\,q_h(t)\mid h]=\kappa M_H,\\
\mathbb E[U_2\mid h]=(d_H-\kappa)M_H+d_LM_L,\qquad
\mathbb E[U_1\mid h]=v_HM_H+v_LM_L.
\end{gathered}
\end{equation}
By the first identity, the expected total attribution cost depends only on the mass of implementations in state $H$, as if each of them paid the full $\kappa$ and each implementation in state $L$ paid nothing; participation in both states only shifts part of the cost from $H$ to $L$ at each history. The identities therefore give each party a price for an implementation in each responder state: holding the other mass fixed, a unit of $M_L$ gives the proposer $v_L$ and the responder, on average, $d_L$, so both gain; a unit of $M_H$ gives the proposer $v_H\ge v_L$ and the responder $d_H-\kappa<0$, so their interests conflict. An implementation in state $H$ itself yields the responder $d_H-\kappa q>0$ at $q\le c_*$; its price is nonetheless negative, because it also raises the posterior, and hence the cost, of every implementation on histories it shares with $L$.

The same prices determine the value of blocked participation at a shared belief $p$. If the proposer is in state $h$ and both preference states participate, the responder's average payoff is
\begin{equation}\label{eq:phat}
(1-p)d_L-p(\kappa-d_H)=d_L-(\kappa-\Delta)p,
\end{equation}
positive exactly when $p$ is below the responder's payoff threshold $\widehat p\equiv d_L/(\kappa-\Delta)$, where $c_*<\widehat p<1$. The thresholds differ because in state $L$ the cost $\kappa p$ must be covered by $d_L$ alone, whereas on average the extra gain $\Delta$ in state $H$ covers part of it; without public information, participation in state $L$ alone is unavailable, since $H$ would imitate it. Let $\Phi(p)\equiv[d_L-(\kappa-\Delta)p]\mathbf 1\{p\le c_*\}$ denote the responder's average payoff where participation is feasible; at the participation threshold itself, $\Phi(c_*)=c_*\Delta>0$. When $c_*<\pi_h<\widehat p$, therefore, no equilibrium implements $x$, yet participation in both states would give the proposer, and the responder on average, a positive expected payoff. This is a counterfactual average at the shared belief, neither an equilibrium payoff attainable above $c_*$ nor a gain in each state, since $d_L-\kappa p<0$ in state $L$. The following sections, where preference states change from day to day, value blocked participation by this average.

\subsection{What public information blocks}\label{subsec:information}

Before communication begins, both parties observe a signal $Z$ that takes finitely or countably many values. Its distribution $K_\theta$ depends only on the responder's state. After $z$, a proposer in state $h$ assigns probability $p_h(z)=\Pr(H\mid h,Z=z)$ to $H$, with $\mathbb E[p_h(Z)\mid h]=\pi_h$, and play follows the original rules with the updated prior. Party $i$'s maximal attainable payoff $\overline U_i(Z)$ is the largest expected payoff, conditional on $h$, that the party obtains in any equilibrium under $Z$, with $\overline U_2$ averaged over the responder's preference states; $\varnothing$ denotes no signal and $F$ full disclosure of the responder's state.

\begin{proposition}[What public information blocks]\label{prop:info_value}
Suppose Assumption~\ref{ass:region} holds.
\begin{enumerate}
\item[(i)] After each signal value $z$ with positive probability, some equilibrium implements $x$ with positive probability if and only if $p_h(z)\le c_*$.
\item[(ii)] One consistent equilibrium has joint acceptance after every signal value with $p_h(z)\le c_*$. After every signal value, it simultaneously attains the highest implementation probability, the proposer's highest payoff, and the responder's highest payoff in each preference state with positive probability, among all equilibria after that value. Hence $\overline U_2(Z)=\mathbb E[\Phi(p_h(Z))\mid h]$.
\item[(iii)] Under any signal and equilibrium, the responder's average payoff is at most its value under full disclosure, $\overline U_2(F)=(1-\pi_h)d_L$.
\item[(iv)] Let $\pi_h\le c_*$ and $C=\{p_h(Z)>c_*\}$. Relative to no information, maximal attainable payoffs change by
\begin{equation}\label{eq:info_value}
\begin{aligned}
\overline U_2(Z)-\overline U_2(\varnothing)
&=(\kappa-\Delta)\,\mathbb E\bigl[(p_h(Z)-\widehat p)\,\mathbf 1\{p_h(Z)>c_*\}\mid h\bigr],\\
\overline U_1(Z)-\overline U_1(\varnothing)
&=-\mathbb E\bigl[\bigl(v_L+\alpha_1\Delta p_h(Z)\bigr)\mathbf 1\{p_h(Z)>c_*\}\mid h\bigr].
\end{aligned}
\end{equation}
If $\Pr(C\mid h)>0$ and $\mathbb E[p_h(Z)\mid h,C]<\widehat p$, every equilibrium under $Z$ is strictly worse for both parties, the responder on average, than joint acceptance without information.
\end{enumerate}
\end{proposition}

Appendix~\ref{app:information} gives the proof. For a given signal, the equilibrium in part (ii) attains all these maxima at once, so selecting it involves no trade-off between the parties or between the responder's preference states. Wherever participation remains feasible, the maxima are linear in the posterior, whose mean is $\pi_h$. Information therefore changes maximal attainable payoffs only through the signal realizations that prevent participation. Every blocked posterior costs the proposer; blocked posteriors in $(c_*,\widehat p)$ also hurt the responder, whereas those above $\widehat p$ spare the responder participation that does not pay on average.

Full disclosure is best for the responder on average because it removes exactly the implementations in state $H$, whose price is negative; yet when $\pi_h\le c_*$ it costs the responder in state $H$ the payoff $d_H-\kappa\pi_h>0$ and the proposer $\pi_hv_H$. That the cost is paid only upon implementation also matters: if declining $x$, by rejecting it or by ending the interaction, carried the same cost, whether participation stops at $c_*$ would depend on how an unexpected refusal is read, and full disclosure would no longer exclude participation in state $H$.

What blocks participation can thus be favorable information that the responder likes $x$, not unfavorable information about the activity. When the shared belief lies in $(c_*,\widehat p)$, participation is already infeasible although its average value is not exhausted. Section~\ref{sec:dynamic} asks whether joint experience carries the shared belief into this interval, and whether evidence can still arrive once the activity has stopped.

\section{Learning from joint activity}\label{sec:dynamic}

Section~\ref{sec:static} takes information as given. Here, public information about the responder's preference comes from the pair's joint experience and changes from day to day. A stop after good experience invites three natural inferences: the activity is no longer worth doing; stopping at least protects the responder, as full disclosure does; and waiting will restore participation. We give conditions under which each fails.

\subsection{Preferences, beliefs and feedback}\label{subsec:dynamic_environment}

Time is measured in days, $j=1,2,\ldots$. On each day an opportunity for the joint choice may arise, and both parties observe whether it does. When it does, a public draw selects a stage equilibrium, an equilibrium of that day's interaction under the complete action rules of Section~\ref{subsec:primitives}. A report about the responder's preference may arrive at the end of the day. Payoffs are measured relative to that day's $f$; a trial of $x$ changes neither the resources available on other days nor the shared value that $f$ carries.

The responder knows their preference on that day, $\theta_j\in\{H,L\}$, but must learn from their own preference history how often they tend to prefer $x$. Their persistent type $\pi\in(0,1)$, drawn once at the outset from a distribution with mean $\bar\pi$, is the probability that a newly drawn preference is $H$. The first day's preference is $H$ with probability $\pi$; on each later day the previous preference persists with probability $\gamma\in[0,1)$ and is otherwise drawn anew:
\begin{equation}\label{eq:preference_process}
\Pr(\theta_{j+1}=H\mid\pi,\theta_1,\dots,\theta_j)
=\gamma\mathbf 1\{\theta_j=H\}+(1-\gamma)\pi.
\end{equation}
Across days, the responder's weight on the proposer's payoff and concern for how participation is understood stay the same. We begin with the case in which neither party observes $\pi$.

The proposer prefers $x$ (state $h$) with probability $\eta$ each day, independently across days and of all other variables. Neither choices nor learning change these processes or the process generating opportunities. Opportunities carry no information about the responder's preference or type. The results of this section maintain this condition; Section~\ref{sec:robustness} considers opportunities that reveal information about the type.

Before the day's choice, the forecast $P_j$ is the probability of state $H$ given the information both parties observe. We assume $P_1=\bar\pi\le c_*$, so participation is feasible at the outset. The proposer's own preference carries no information about the responder's, so the proposer in state $h$ uses $P_j$ as the prior. The participation threshold applies with $\pi_h=P_j$: when $P_j>c_*$, no stage equilibrium implements $x$, whatever the responder privately knows about past preferences (Appendix~\ref{app:prescribed}). Implementation imposes the attribution cost $\kappa q_j^{\mathrm{imp}}$ on the responder, where $q_j^{\mathrm{imp}}$ is the proposer's posterior at implementation. At the end of the day, the parties may jointly observe a binary report about the responder's preference that day. It is positive with probability $k_H$ in state $H$ and $k_L\le k_H$ in state $L$, independently of everything else conditional on that preference. Unlike the signal in Section~\ref{subsec:information}, it arrives after the choice and leaves that day's attribution cost unchanged.

After updating on the day's public actions and any report, $q_j^{\mathrm{end}}$ denotes the public posterior probability of state $H$ that day, and $m_j$ the public posterior mean of $\pi$, called the type judgment. The responder's own beliefs also reflect their private history. The transition law gives
\begin{equation}\label{eq:forecast}
P_{j+1}=\gamma q_j^{\mathrm{end}}+(1-\gamma)m_j.
\end{equation}
The first term carries today's preference into tomorrow; the second corresponds to a new draw according to the type. The type judgment still matters after today's preference has faded.

Feedback arrangements differ in when this report can be obtained. Under feedback after joint activity, or experience, it arrives with probability $\zeta$ only on days when $x$ is implemented. Under feedback independent of participation, or outside feedback, it arrives each day with a probability fixed in advance. Section~\ref{sec:design} sets this probability to match the expected number of reports across the population each day under experience. The no-report benchmark $\varnothing$ has no postchoice reports but otherwise the same environment; a rejection under mixed acceptance still reveals $L$. A further comparison makes $\pi$ public from the first day.

All arrangements select stage equilibria by the same rule at a given forecast, isolating differences in public information. This prescribed stage behavior puts no weight on how today's choices affect future beliefs; it is a behavioral assumption, not an equilibrium for parties who look ahead (Sections~\ref{sec:forward} and~\ref{sec:robustness}). At an opportunity with $P_j\le c_*$, the rule selects positive participation with probability $\lambda$. Conditional on this selection, it chooses joint acceptance with probability $1-\chi$ and mixed acceptance with probability $\chi$: under joint acceptance both preference states accept, and under mixed acceptance $H$ accepts and $L$ accepts with probability $\ell_*(P_j)$ from~\eqref{eq:ell}. With the remaining probability $1-\lambda$, and always when $P_j>c_*$, participation is zero. The draws are public. New public information about preferences thus comes only from actions under mixed acceptance and from postchoice reports. When $\chi=0$, so that participation always takes the form of joint acceptance (pooling proposals only), it comes only from reports.

\subsection{The value of a blocked opportunity}\label{subsec:exclusion_payoffs}

Whether feedback helps either partner depends on the opportunities it prevents. A qualified opportunity is a day with an opportunity, $O_j=1$, on which the proposer prefers $x$, $s_j=h$; only such days can implement $x$, and each carries the opportunity weight $G_j=\lambda\mathbf 1\{O_j=1,\ s_j=h\}$. If $P_j>c_*$, the attribution cost prevents participation and the opportunity is blocked.

At a qualified opportunity with $P\le c_*$, averaging over the responder's preference, joint acceptance gives the proposer $v_L+\alpha_1\Delta P$ and the responder $d_L-(\kappa-\Delta)P$ as in~\eqref{eq:phat}; mixed acceptance implements $x$ with probability $P/c_*$ at posterior $c_*$, giving the proposer $(P/c_*)(v_L+\alpha_1\Delta c_*)$ and the responder $\Delta$ in state $H$ and zero in state $L$. Weighting by the selection probabilities,
\begin{equation}\label{eq:weights}
\phi_1(P)=v_L\Bigl[1-\chi+\chi\frac{P}{c_*}\Bigr]+\alpha_1\Delta P,\qquad
\phi_2(P)=(1-\chi)\bigl[d_L-(\kappa-\Delta)P\bigr]+\chi\Delta P.
\end{equation}
Party $i$'s expected payoff at a qualified opportunity is $G_j\phi_i(P_j)$ if $P_j\le c_*$ and zero otherwise. To value blocked opportunities, we extend the weights $\phi_i$ in~\eqref{eq:weights} linearly above $c_*$, where they are not attainable equilibrium payoffs. For a regime $Z$, let $P_j^Z$ be the forecast and $U_{i,j}^Z$ the day's utility from~\eqref{eq:utility}, zero without an opportunity. Expected payoff over a finite window $\mathcal W$ is $\mathcal U_i(Z)=\mathbb E[\sum_{j\in\mathcal W}U_{i,j}^Z]$, averaging over daily preferences and the type. The comparisons concern these expectations, not individual pairs or paths.

\begin{proposition}[The value of feedback]\label{prop:feedback_value}
Suppose Assumption~\ref{ass:region} holds. Fix a feedback regime $Z$ and its benchmark $\varnothing$ in the same environment, and suppose that (a) they share the same exogenous type, preference and opportunity processes, the occurrence of an opportunity is observed by both parties, and the proposer's preference is drawn independently each day and of all other variables; (b) $P_j$ is the Bayesian posterior probability of $\theta_j=H$ given the public history before the day's choice; and (c) both regimes use prescribed stage behavior with the same $\lambda$ and $\chi$. Then, for every finite window $\mathcal W$ and $i\in\{1,2\}$,
\begin{equation}\label{eq:feedback_value}
\begin{aligned}
\mathcal U_i(Z)-\mathcal U_i(\varnothing)
={}&-\mathbb E\Bigl[\sum_{j\in\mathcal W}G_j\mathbf 1\{P_j^Z>c_*\}\phi_i(P_j^Z)\Bigr]\\
&+\mathbb E\Bigl[\sum_{j\in\mathcal W}G_j\mathbf 1\{P_j^\varnothing>c_*\}\phi_i(P_j^\varnothing)\Bigr].
\end{aligned}
\end{equation}
\end{proposition}

Wherever participation is feasible, expected payoffs are linear in the forecast, whose mean is fixed by the exogenous processes. As in Proposition~\ref{prop:info_value}(iv), the arrangements therefore differ only through what each blocks, valued at that day's forecast (Appendix~\ref{app:feedback_value}). A positive weight means that exclusion costs the party; a negative one means that it saves a net burden.

With a public type, the benchmark blocks the pairs with $\pi>c_*$ (Corollary~\ref{cor:known_type}). With an unknown type and $\bar\pi\le c_*$, it blocks nothing (Lemma~\ref{lem:no_report}): acceptance under mixed acceptance raises the posterior only to $c_*$, and after a rejection the evidence of $L$ limits how far a later acceptance can raise the type judgment. These limits hold after any sequence of acceptances and rejections. Learning from the responder's choices alone thus never makes the activity infeasible; a stop requires postchoice reports.

\begin{corollary}[Who can gain]\label{cor:who_gains}
Suppose the conditions of Proposition~\ref{prop:feedback_value} hold, the type is unknown, opportunities carry no information about the type or preferences, and $\bar\pi\le c_*$. The benchmark $\varnothing$ then blocks no opportunity (Lemma~\ref{lem:no_report}). For every feedback regime $Z$ and finite window $\mathcal W$:
\begin{enumerate}
\item[(i)] $\mathcal U_1(Z)\le\mathcal U_1(\varnothing)$.
\item[(ii)] The responder's gain is bounded by
\[
\mathcal U_2(Z)-\mathcal U_2(\varnothing)
\le(1-\chi)(\kappa-\Delta)\,
\mathbb E\Bigl[\sum_{j\in\mathcal W}G_j(P_j^Z-\widehat p)
\mathbf 1\{P_j^Z>\widehat p\}\Bigr].
\]
\item[(iii)] If $\chi>(\kappa-d_H)/(\kappa-d_L)$, then $\mathcal U_2(Z)\le\mathcal U_2(\varnothing)$.
\item[(iv)] If $v_H\ge\kappa-d_H$, then $\mathcal U_1(Z)+\mathcal U_2(Z)\le\mathcal U_1(\varnothing)+\mathcal U_2(\varnothing)$.
\end{enumerate}
\end{corollary}

\begin{proof}
By Lemma~\ref{lem:no_report}, the benchmark term vanishes, and~\eqref{eq:feedback_value} becomes
\begin{equation}\label{eq:unknown_gain}
\mathcal U_i(Z)-\mathcal U_i(\varnothing)
=-\mathbb E\Bigl[\sum_{j\in\mathcal W}G_j\mathbf 1\{P_j^Z>c_*\}\phi_i(P_j^Z)\Bigr].
\end{equation}
It therefore suffices to sign the weights on $(c_*,1]$. There $\phi_1\ge v_L>0$, which gives (i). Also $\phi_2(P)=(1-\chi)(\kappa-\Delta)(\widehat p-P)+\chi\Delta P\ge(1-\chi)(\kappa-\Delta)(\widehat p-P)$, which is nonnegative for $P\le\widehat p$, so $-\phi_2(P)\le(1-\chi)(\kappa-\Delta)(P-\widehat p)\mathbf 1\{P>\widehat p\}$, which gives (ii). Since $\phi_2$ and $\phi_1+\phi_2$ are affine, their endpoint values determine their signs:
\begin{gather*}
\phi_2(c_*)=c_*\Delta>0,\qquad\phi_2(1)=(\kappa-d_L)(\chi-\tau),\\
(\phi_1+\phi_2)(c_*)>0,\qquad(\phi_1+\phi_2)(1)=(1-\chi)(v_H+d_H-\kappa)+\chi\Bigl(\frac{v_L}{c_*}+\alpha_1\Delta+\Delta\Bigr),
\end{gather*}
where $\tau=(\kappa-d_H)/(\kappa-d_L)$. Both endpoint values of $\phi_2$ are positive if $\chi>\tau$, which gives (iii), and both endpoint values of $\phi_1+\phi_2$ are nonnegative if $v_H\ge\kappa-d_H$, which gives (iv).
\end{proof}

Under joint acceptance alone, the weights are the prices in~\eqref{eq:accounting}: a lost $L$ day costs both partners, while a lost $H$ day costs the proposer $v_H$ and, on average, spares the responder $\kappa-d_H$. Exclusion in the lower band $(c_*,\widehat p]$ thus harms both, the responder only weakly at $\widehat p$, while exclusion in the upper band $(\widehat p,1]$ protects the responder (Figure~\ref{fig:where_ends}A). If $v_H\ge\kappa-d_H$, even this protection is no larger than the proposer's loss, so exclusion does not raise the pair's combined payoff (Corollary~\ref{cor:who_gains}(iv)). Mixed acceptance gives the responder $\Delta$ in state $H$ and leaves no net burden to save in state $L$. When $\chi>\tau=(\kappa-d_H)/(\kappa-d_L)$, as in part (iii), the responder's weight $\phi_2$ is therefore positive at every blocked forecast, wherever learning stops. At the baseline payoffs of Section~\ref{sec:design}, $\tau=1/3$ lies below the maintained mixture $\chi=1/2$; every positive $\chi$ in the numerical analysis likewise exceeds its environment's $\tau$. Otherwise, as with pooling proposals only, whether stopping protects the responder depends on how high experience takes the forecast.

\subsection{How high experience can take the forecast}\label{subsec:experience_ceiling}

Experience reports arrive only after implementation, when the posterior on today's preference is at most $c_*$, so one positive report can raise it only so far. With persistence and an unknown type, that limit is not enough: the forecast in~\eqref{eq:forecast} also depends on a type judgment that can accumulate across reports. The participation threshold limits this accumulation. By~\eqref{eq:forecast}, the forecast before the day's choice is at least $1-\gamma$ times the type judgment, and implementation requires this forecast to be at most $c_*$. Joint acceptance reveals nothing, and acceptance under mixed acceptance raises the posterior on today's preference to $c_*$ and the type judgment by at most the same factor. Before every report, therefore, the type judgment is at most today's posterior divided by $1-\gamma$, hence at most $c_*/(1-\gamma)$. One positive report can then raise neither today's posterior nor the type judgment beyond the ceiling $\overline P$ in~\eqref{eq:ceiling}, and negative reports and rejections weakly lower both relative to no such observation (Lemma~\ref{lem:covariance}). Thereafter, either the activity stops and experience supplies no further report, or the forecast returns to $c_*$ or below and the next report starts within the same bounds. Favorable experience can thus cut off its own source.

\begin{proposition}[Exclusion below the payoff threshold]\label{cor:ceiling}
Suppose the conditions of Corollary~\ref{cor:who_gains} hold, $Z$ is feedback after joint activity, $k_L>0$ and $c_*\le1-\gamma$, and let $\Lambda=k_H/k_L$. For any $\lambda$ and $\chi$:
\begin{enumerate}
\item[(i)] On every day and after every public history of positive probability, the forecast $P_j^Z$ and the type judgment, before and after the day's choice, are at most
\begin{equation}\label{eq:ceiling}
\overline P(c_*,\Lambda,\gamma)
=\frac{c_*\bigl[\Lambda+\gamma/(1-\gamma)\bigr]}{1+c_*(\Lambda-1)}\le1.
\end{equation}
\item[(ii)] If $\overline P(c_*,\Lambda,\gamma)\le\widehat p$, no blocked opportunity lies in the upper band and $\mathcal U_i(Z)\le\mathcal U_i(\varnothing)$ for $i\in\{1,2\}$ and every finite window $\mathcal W$. If $\overline P(c_*,\Lambda,\gamma)<\widehat p$, both inequalities are strict if and only if $\mathbb E[\sum_{j\in\mathcal W}G_j\mathbf 1\{P_j^Z>c_*\}]>0$; if that expectation is zero, both are equalities.
\end{enumerate}
\end{proposition}

With the ceiling below $\widehat p$, every blocked opportunity still has positive value for both parties, without requiring $v_H\ge\kappa-d_H$. Experience makes the activity infeasible once the forecast passes $c_*$; stopping saves the responder a net burden only above $\widehat p$, which the forecast never reaches. Full disclosure's benefit to the responder in Section~\ref{subsec:information} therefore does not carry over to information obtained through participation. Outside feedback also arrives during a stop, so the ceiling does not apply and blocked forecasts can exceed $\widehat p$ (Section~\ref{subsec:onset}).

When $\gamma=0$, the ceiling is the posterior after one positive report at prior $c_*$; in general, $c_*\le1-\gamma$ keeps it at most one. It is the same for every type distribution with mean at most $c_*$ and does not depend on $\lambda$, $\chi$, report delivery or the window. Since the argument for (i) in Appendix~\ref{app:ceiling} never uses the probability with which a stage equilibrium is selected, part (i) holds for any selection among the three stage equilibria based on the public history, whereas (ii) compares arrangements with the same $\lambda$ and $\chi$.

Rearranging~\eqref{eq:ceiling}, the ceiling is at most the payoff threshold if and only if
\begin{equation}\label{eq:ceiling_condition}
\Lambda+\frac{\gamma/(1-\gamma)}{1-\widehat p}
\le\frac{\widehat p/(1-\widehat p)}{c_*/(1-c_*)}
=\frac{\kappa-d_L}{\kappa-d_H}=\frac1\tau.
\end{equation}
The right side is the ratio of the odds at the two thresholds, determined by payoffs. The left side is one report's likelihood ratio plus a persistence correction. The diary-shaped report used in Section~\ref{sec:design} satisfies this condition strictly at the baseline payoffs throughout the diary-compatible range of persistence.

\subsection{After participation stops}\label{subsec:after_stop}

Under experience, excluded days provide no new public preference information. The public type judgment freezes at its value when the stop begins, although the responder still learns privately, and the forecast moves toward it as the preference at the start of the stop fades. Whether waiting restores feasibility therefore depends on what the pair has learned about the type; under outside feedback, reports can still arrive during a stop.

An exclusion spell, or stop, is a maximal run of consecutive days with $P_j>c_*$. Its first day is $j_0$, and $m$ is the type judgment before the choice on that day; if the type is known from the start, $m=\pi$.

\begin{corollary}[The path of an exclusion spell]\label{cor:spell_path}
Suppose the conditions of Proposition~\ref{prop:feedback_value} hold for feedback after joint activity, opportunities carry no information about the type or preferences, and $0<\gamma<1$. For every $j\ge j_0$, if days $j_0,\ldots,j-1$ belong to the spell, the type judgment before the choice on day $j$ remains $m$ and
\begin{equation}\label{eq:spell_path}
P_j=m+\gamma^{\,j-j_0}(P_{j_0}-m).
\end{equation}
Hence:
\begin{enumerate}
\item[(i)] If $m<c_*$, the spell lasts exactly $\bigl\lceil\ln[(P_{j_0}-m)/(c_*-m)]/\ln(1/\gamma)\bigr\rceil$ days and then ends by itself.
\item[(ii)] If $m\ge c_*$, the forecast exceeds $c_*$ on every day from $j_0$ onward, so the spell does not end by itself.
\item[(iii)] If $m<\widehat p<P_{j_0}$, the forecast exceeds $\widehat p$ on exactly the first $\bigl\lceil\ln[(P_{j_0}-m)/(\widehat p-m)]/\ln(1/\gamma)\bigr\rceil$ days of the spell; if also $m\ge c_*$, it lies in $(c_*,\widehat p]$ on every subsequent day.
\item[(iv)] If $m\ge\widehat p$ and $P_{j_0}\ge\widehat p$, the forecast is at least $\widehat p$ on every day.
\item[(v)] If the type is unknown and $\bar\pi\le c_*$, the day before $j_0$ ends with a positive report, or $c_*<m\le c_*/(1-\gamma)$.
\end{enumerate}
\end{corollary}

Waiting ends a stop with $m<c_*$ but not one with $m\ge c_*$, including $m=c_*$, and by (i) the wait can be arbitrarily long when $m$ is just below $c_*$. By (v), a stop can begin without a positive report only if $c_*<m\le c_*/(1-\gamma)$, a range that vanishes as $\gamma\to0$ and in which a stop does not end by itself. Under joint acceptance alone, protection means a negative responder weight in~\eqref{eq:weights}. When $c_*\le m<\widehat p<P_{j_0}$, part (iii) shows that protection expires during a persistent stop: the forecast falls to the payoff threshold or below while remaining above the participation threshold, so stopping no longer saves the responder a net burden. In (iv), the stop continues to protect the responder, only weakly at $\widehat p$. The duration observed so far therefore reveals neither whether a stop will end by itself nor whom it benefits.

Under prescribed stage behavior, when the experience ceiling lies below $\widehat p$, a lasting stop is no evidence of exhausted value and no protection for the responder. Learning blocks only opportunities that remain valuable to both partners; whenever the expected blocked opportunity weight is positive, both are strictly worse off than without reports. Waiting does not end a stop whose frozen type judgment is at or above $c_*$, so evidence capable of ending it must still reach the pair after participation has ceased (Section~\ref{sec:evidence}). The two thresholds and participation-dependent evidence are not enough to rule out protective stops: stronger reports or more persistent preferences can produce them, and greater type dispersion can sustain that protection (Section~\ref{sec:learning_stops}).

\section{Data and numerical design}\label{sec:design}

With unknown types and uninformative opportunities, condition~\eqref{eq:ceiling_condition} guarantees that experience never reaches the band where stopping protects the responder, whatever the report history. The diaries calibrate report discrimination and constrain preference persistence, which enter its left side; the payoffs fix its right side. When it holds, Proposition~\ref{cor:ceiling} establishes that experience can only lower each partner's payoff; simulations quantify the loss, tracing where stops fall and how many opportunities they block.

\subsection{What the diaries measure}\label{subsec:data}

The main source is Study~3 of \citet{vaccarino2025}: 121 couples reported their own sexual desire for the partner, their judgment of the partner's desire for them, and sexual activity for 21 days. The first two measures use the questionnaire's 1--7 scale, recording agreement with a statement of strong desire that day, from strongly disagree to strongly agree \citep{vaccarino2025data}. Self-reports of 6 or 7 proxy daily preference $H$, and partners' judgments proxy reports about it; the diaries record neither model proposals nor partners' payoffs. A 30-day diary of 133 couples \citep{vowels2018,vowels2018data} provides an independent description of preference variation.

Consecutive high-desire days can arise because some people are persistently more likely to have high desire, or because yesterday's preference carries over. Type dispersion $R=\operatorname{Var}(\pi)/[\bar\pi(1-\bar\pi)]$, the share of the variance of daily preference explained by the type, captures the former and persistence $\gamma$ the latter. In the main diary, the probability of high desire is 0.34 higher after a high self-report than after a low one; in the model this difference is $R+(1-R)\gamma$. Attributing this difference entirely to short-run persistence would predict an across-person standard deviation in high-day shares about half the observed value, and almost no correlation between individuals' shares in the two halves of the diary. At the baseline setting, $R=0.3$ and $\gamma=0.05$, the model gives $0.30+0.70\times0.05=0.335\approx0.34$: stable differences account for most of the dependence, with carry-over contributing about 0.035. The observed dependence of 0.34 therefore cannot be read directly as persistence $\gamma$.

A setting $(R,\gamma)$ is diary-compatible when the model's values of these three statistics all lie within their whole-couple bootstrap intervals for one column of Table~\ref{tab:states}; this constrains, rather than identifies, the process. The baseline $(R,\gamma)=(0.3,0.05)$ is compatible with the main diary. The largest compatible persistence on a grid of $(R,\gamma)$ values is $0.235$, from the independent diary: yesterday's preference carries over on fewer than one day in four; higher persistence is counterfactual (Appendix~\ref{subsec:s4_states}). Each pair draws a fixed type from a Beta distribution with mean $0.4$ and dispersion $R$. This mean, near the main diary's high-day share of $0.39$, lies below the baseline participation threshold of 50\%, so participation is feasible at the outset when the type is unknown.

Report discrimination comes from days when both members report no sexual activity, so judgments cannot rest on an activity both observed. With each judge's median over all valid matched days as their usual level, about 26\% of judgments exceed it when the partner's self-report is 6 or 7, and 17\% when it is 1 to 5. A positive report raises the odds of high desire by about half; a negative report, at or below the usual level, lowers them by about a tenth. A single report is weak evidence. Simulations draw public reports with these probabilities given the day's preference; we call them diary-shaped reports. With a judgment of 6 or 7 counted as positive for every judge, positive reports have likelihood ratios of 2.25 on no-activity days and 1.44 on activity days, so measuring discrimination on activity days instead would make condition~\eqref{eq:ceiling_condition} easier to satisfy (Appendix~\ref{subsec:s4_signals}).

Strong and near-perfect reports, correct with probability 90\% and 99\% under either preference, form a Blackwell order with the diary-shaped report for the precision comparisons; six further empirical shapes and a weak reference, which mixes diary-shaped and uninformative reports equally, show whether conclusions depend on how the report is constructed (Table~\ref{tab:signals}).

Transitions between joint desire states in the main diary provide a three-state clock that makes opportunities cluster in time (Appendix~\ref{subsec:s4_clock}). Baseline opportunities depend only on this clock. Although its states come from desire self-reports, we maintain that the clock evolves independently of the type and daily preference, so an opportunity itself supplies no preference evidence. Section~\ref{sec:robustness} lets arrivals also carry information about the type, holding the average opportunity rate fixed.

\subsection{Payoffs, the experience ceiling and comparisons}\label{subsec:comparisons}

At baseline, the proposer's direct gain or loss is one unit and the responder's is half a unit ($v=w=1$, $g=b=1/2$). The proposer gives equal weight to the responder's direct payoff and their own ($\alpha_1=1$); the responder's utility from the proposer's gain is $A=2$, which Section~\ref{sec:robustness} varies, and the maximum attribution cost is $\kappa=3$. These values satisfy Assumption~\ref{ass:region}, with $d_L=3/2>0$ and $d_H=5/2<\kappa$: valuing the proposer's gain can support participation, while motive attribution can end it.

At these baseline payoffs, substituting $c_*=1/2$ and $\widehat p=3/4$ into~\eqref{eq:ceiling_condition} gives $\Lambda+4\gamma/(1-\gamma)\le3$. The two thresholds correspond to odds of one to one and three to one that the responder wants the activity today. Experience reports arrive only while participation is feasible, so each starts from odds of at most one to one. A positive diary-shaped report, with $(k_H,k_L)=(0.264,0.173)$, multiplies them by $\Lambda=264/173\approx1.53$. Only preference carry-over could close the remaining gap: the ceiling is at most the payoff threshold whenever persistence is at most $255/947\approx0.27$. This critical value exceeds the largest diary-compatible persistence, $0.235$ (Appendix~\ref{subsec:s4_states}). Thus, at the baseline payoffs and every diary-compatible persistence, the diary-shaped report satisfies~\eqref{eq:ceiling_condition} strictly: the forecast never reaches the upper band, where stopping can protect the responder. Appendix~\ref{subsec:s4_signals} shows how this margin depends on sampling error and report construction. The strong report, $(k_H,k_L)=(0.9,0.1)$, instead gives $\Lambda=9$ and violates~\eqref{eq:ceiling_condition} at every persistence, so for this report the ceiling alone cannot rule out forecasts in the upper band.

Comparisons of report precision and type knowledge use pooling proposals only ($\chi=0$), where the payoff threshold separates blocked opportunities that hurt the responder from protective ones (Section~\ref{subsec:exclusion_payoffs}). In the main environment, the maintained mixture selects joint and mixed acceptance equally ($\chi=1/2$), so acceptance itself is evidence; because $\chi$ exceeds $\tau=1/3$, every blocked opportunity there has positive average value for the responder (Corollary~\ref{cor:who_gains}(iii)).

The main environment uses diary-shaped reports, the baseline preference process, an unknown type and uninformative opportunities. We set the probability of selecting a positive-participation equilibrium at a feasible opportunity to $\lambda=75\%$ and report delivery after implementation to $\zeta=80\%$. Further environments set $\lambda$ to 50\% or 100\% and $\zeta$ to 50\% or 20\%; all enter the complete comparisons of Appendix~\ref{subsec:s6_census}, and Section~\ref{sec:evidence} reports how delivery changes the loss. The proposer's daily probability of preferring the new activity is $\eta=60\%$ in every environment.

Each feedback arrangement is compared with its environment's no-report benchmark. Section~\ref{subsec:type_knowledge} separately values type revelation from the outset by comparing known-type and unknown-type no-report environments. Outside feedback gives all pairs the same daily delivery probability: $\zeta$ times that day's mean implementation frequency in a separate simulation of experience in the same environment, matching experience's expected daily report count. Its reports have the same conditional probabilities as experience reports, whether or not the pair participates. With these probabilities and the expected daily report count fixed, this comparison changes the timing of evidence: reports can reach pairs without their own implementation, including blocked pairs; individual report dates and contents still differ.

\subsection{Histories, windows and precision}\label{subsec:precision}

Each environment is simulated eight times with 1,024 pairs over 84 days. Its arrangements share common random numbers, so within a replication each pair has the same type, preferences and opportunities under every arrangement. Payoffs are per 100 pair-days, counting every pair and day; with the mean opportunity probability set to $7/20$, about 21 of these pair-days bring an opportunity the proposer wants. Histories start on day 1 and beliefs are never reset; comparisons cover days 1--84. Three periods, days 1--21, 22--30 and 31--84, split where the two diaries end, trace how effects evolve. With uninformative opportunities, stops with a type judgment of at least $c_*$ do not end by themselves under experience, so magnitudes depend on the common horizon. An onset is an exclusion spell's first day, with type judgment $m$ as in Corollary~\ref{cor:spell_path}; blocked opportunities are weighted by $G_j$ (Section~\ref{subsec:exclusion_payoffs}), and the day-84 stock of excluded pairs is counted per 100 pairs.

Each comparison has a 95\% Monte Carlo interval from eight paired replication differences, conditional on the parameter values and fixed outside-feedback schedules. A difference whose interval contains zero is not detected at this precision, which for the joint payoff of experience in days 1--84 is about $\pm0.09$ per 100 pair-days with the diary-shaped report under pooling proposals only and $\pm0.25$ in the main environment. A ratio with a loss or cost in the denominator is reported only when that denominator is detected positive. Appendix~\ref{sec:s5} gives interval and ratio calculations, integration precision and implementation checks.

\section{Where learning stops and who bears it}\label{sec:learning_stops}

At the baseline payoffs and diary-compatible persistence, Sections~\ref{sec:dynamic} and~\ref{sec:design} establish that diary-shaped experience cannot reach the region where stopping protects the responder. That bound leaves open where learning actually stops, which stops carry the blocked opportunities and how much each partner loses. We answer these questions with simulated histories that start from the report probabilities measured in the diaries, a diary-compatible preference process and the baseline payoffs, and then vary each.

\subsection{Where learning stops}\label{subsec:onset}

Consider a pair using pooling proposals only. When participation is feasible, the responder accepts whether or not they want the activity, so acceptance reveals nothing. Only the report that may follow brings new evidence, and it can change whether participation is feasible next time. Negative reports lower forecasts relative to no report (Lemma~\ref{lem:covariance}). In all 53 computed experience environments with unknown types and uninformative opportunities, every stop in days 1--84 begins the day after a positive report. The activity becomes infeasible just as the partners learn that the responder is more likely to want it.

\begin{figure}[tbp]
\centering
\includegraphics[width=.98\linewidth]{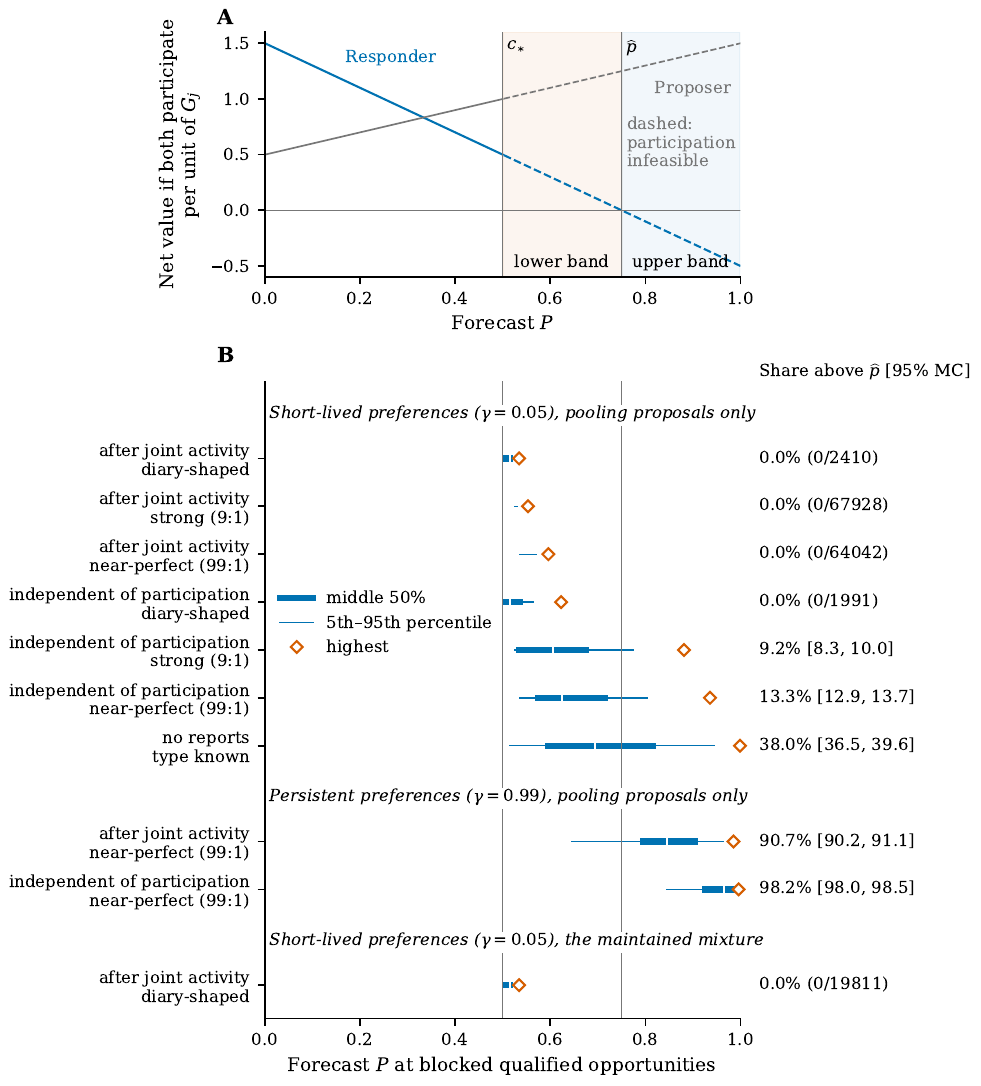}
\caption{Where blocked opportunities lie and what they are worth}\label{fig:where_ends}
\par\smallskip\begin{minipage}{\linewidth}\footnotesize\emph{Notes:} The horizontal axis is the public forecast $P_j$ of the responder directly preferring the activity that day; vertical lines mark the participation threshold $c_*$ and the responder's payoff threshold $\widehat p$. Panel A gives average net value if both preference states participate, per unit of $G_j$, as in~\eqref{eq:weights}; negative values mean attribution costs saved exceed participation benefits forgone. Panel B gives the distribution of forecasts at blocked qualified opportunities in days 1--84: medians (white ticks), interquartile ranges, 5th--95th percentiles, sample maxima (diamonds) and shares above $\widehat p$. Row labels give the feedback source, after joint activity (experience) or independent of participation (outside feedback), and the report: diary-shaped reports are as discriminating as partners' judgments in the main diary; strong and near-perfect reports are correct with probability 90\% and 99\%. In $0/n$, $n$ counts blocked qualified opportunities across eight replications; Proposition~\ref{cor:ceiling} guarantees the zero share for baseline diary-shaped experience. Unless indicated otherwise, $\gamma=0.05$, $A=2$, $R=0.3$, the type is unknown and proposals are pooling only. Share intervals are pointwise 95\% Monte Carlo intervals conditional on fixed outside-feedback schedules.\end{minipage}
\end{figure}

Diary-shaped reports take the forecast only a little beyond the participation threshold. With pooling proposals only, the mean forecast at blocked opportunities is 51.4\% and the maximum is 53.4\%, just above the 50\% participation threshold and well below the 75\% payoff threshold (Figure~\ref{fig:where_ends}B, top row). A positive report at the prior raises the type judgment only from 40\% to about 43\%; at baseline persistence $\gamma=0.05$, this judgment receives 95\% of the weight in tomorrow's forecast. Positive reports must accumulate, and forecasts cross the threshold by only a few percentage points. The activity thus stops where participation under both preferences would still give each partner a positive expected net payoff conditional on public information (Figure~\ref{fig:where_ends}A). Proposition~\ref{cor:ceiling} extends this conclusion to every report history: the forecast ceiling is below 62.5\% at baseline persistence and below the payoff threshold throughout the diary-compatible range.

An accurate report identifies today's preference, but one day leaves the persistent propensity uncertain. From the prior, learning with certainty that the responder wants the activity on day 1 raises the type judgment to
\[
\bar\pi+R(1-\bar\pi)=0.4+0.6R.
\]
At baseline dispersion $R=0.3$, this is only 58\%. A first positive near-perfect report gives a type judgment of 57.6\% and a next-day forecast of 59.6\%, also the highest blocked forecast in this environment's complete simulated histories. Even a perfectly accurate first report leaves tomorrow's forecast below 70\% under every diary-compatible preference process. Today's good news suffices to make the next activity infeasible, well before it makes stopping beneficial to the responder.

Blocked forecasts lie below the payoff threshold in every computed diary-compatible experience environment with unknown types and uninformative opportunities. Proposition~\ref{cor:ceiling} guarantees this for all empirical report shapes and the weak reference; complete simulated histories show the same for strong and near-perfect reports, whose ceilings lie above the payoff threshold (Appendix~\ref{subsec:s6_census}). During a stop, experience supplies no further report. If the same near-perfect reports arrive independently of participation, excluded pairs can continue learning: at the baseline preference process, 13.3\% of blocked opportunities lie above the payoff threshold, where stopping protects the responder (Figure~\ref{fig:where_ends}B). Near-perfect experience also protects the responder at high persistence (Figure~\ref{fig:where_ends}B, $\gamma=0.99$); Section~\ref{subsec:protection} examines how long this protection lasts.

\subsection{Who bears the cost of stopping}\label{subsec:denominators}

Whether waiting restores feasibility depends on the type judgment $m$ when the stop begins. Experience no longer supplies reports, so $m$ stays fixed while the influence of today's preference fades and the forecast falls toward it (Corollary~\ref{cor:spell_path}). Waiting restores feasibility only if $m<c_*$; completed spells of this kind have median lengths of one or two days in compatible environments. If $m\ge c_*$, participation requires new evidence.

In every diary-compatible experience environment, stops that cannot end on their own carry at least 95.1\% of blocked opportunities in days 1--84, even where they make up only 44.1\% of onsets (Table~\ref{tab:spell_burden}). With diary-shaped reports and pooling proposals only, they account for 67.1\% of onsets but 96.7\% of blocked opportunities: the other stops end quickly, whereas these block every later opportunity in the window. No stop in the diary-compatible environments begins with $m\ge\widehat p$: every stop that cannot end on its own lies between the two thresholds. Stops that eventually end can bear substantial losses when highly persistent preferences delay recovery, but those environments lie outside the compatible range.

\begin{table}[tbp]
\centering\small
\caption{Stops between the thresholds: frequency and blocked opportunities}\label{tab:spell_burden}
\begin{tabularx}{\linewidth}{@{}>{\raggedright\arraybackslash}Xrrr@{}}
\toprule
Persistence band & Environments & \shortstack{Share of\\onsets (\%)} & \shortstack{Share of blocked\\opportunities (\%)}\\
\midrule
Diary-compatible & 36 & 44.1--100.0 & 95.1--100.0\\
Other, $\gamma=0.18$--$0.50$ & 4 & 14.9--54.8 & 70.5--98.7\\
$\gamma=0.70$--$0.90$ & 7 & 2.8--46.8 & 4.4--91.0\\
$\gamma=0.95$--$0.99$ & 6 & 0--80.4 & 0--91.4\\
\bottomrule
\end{tabularx}
\par\smallskip\begin{minipage}{\linewidth}\footnotesize\emph{Notes:} Stops between the thresholds have onset type judgment $c_*\le m<\widehat p$, where $c_*$ is the participation threshold and $\widehat p$ the responder's payoff threshold. Shares are minima and maxima of environment-specific point estimates. Both shares use days 1--84; blocked opportunities are weighted by $G_j$. All 53 experience environments have unknown types and uninformative opportunities. Compatibility is determined jointly by type dispersion $R$ and persistence $\gamma$ (Appendix~\ref{subsec:s4_states}); the first row includes compatible combinations and the others group incompatible combinations by $\gamma$. Most compatible environments use the baseline process and differ in the report, the participation rule, payoffs, or the selection or delivery probability; the other three compatible processes use only diary-shaped reports under the maintained mixture. The smallest number of onsets in any environment is 15. The highest-persistence band includes the environments with $R=0.1$ and $0.7$ at $\gamma=0.95$.\end{minipage}
\end{table}

Participation rules determine the burden's scale. With diary-shaped reports, 1.7\% of all qualified opportunities in days 1--84 are blocked under pooling proposals only, versus 13.7\% under the maintained mixture, the main environment. Joint losses are respectively 1.6\% and 14.8\% of each rule's no-report benchmark (Table~\ref{tab:core_comparisons}A--B). Opportunity and loss shares are close because an opportunity just beyond the threshold is worth about as much as an average benchmark opportunity. On day 84, the two rules leave 5.2 $[4.7,5.8]$ and 31.4 $[30.0,32.8]$ of every 100 pairs excluded. Losses under the maintained mixture are larger because mixed acceptance raises the posterior on today's preference to $c_*$ and increases the type judgment; at the prior, this evidence is comparable to a positive diary-shaped report.

Both partners bear these losses. The responder loses participation benefits that remain positive after attribution costs; the proposer loses about twice as much per blocked opportunity and bears roughly two-thirds of the joint loss. These magnitudes rest on prescribed stage behavior: under mixed acceptance, a responder who does not want the activity is indifferent about accepting today, but at the sampled states of the main environment, rejecting becomes profitable once future payoffs are counted and later behavior is held fixed (Section~\ref{sec:robustness}).

Report shape also changes how often pairs reach the threshold. Under pooling proposals only, the other six empirical shapes leave mean blocked forecasts at 51--53\%, yet joint losses range from 5.2\% to 18.4\% of their respective benchmarks; the diary-shaped report produces the smallest loss. Reports constructed from activity days, closest in timing to feedback after a joint activity, produce a loss of 16.6\%. At the prior, roughly two in three of these reports are positive, compared with one in five diary-shaped reports. Their negative reports are also more discriminating, and each report moves the type judgment more than twice as far on average. The two reports cannot be ranked by precision (Appendix~\ref{subsec:s4_signals}); within the same window, the activity-day report's larger belief movements bring more pairs to the participation threshold.

Near-perfect reports block 44.2\% of qualified opportunities rather than 1.7\%, while the mean blocked forecast rises only from 51.4\% to 57.0\%. In this environment's simulated histories, a stop that cannot end on its own begins when positive reports first outnumber negative ones by one. By day 84, 53.3 of every 100 pairs are excluded, almost all in such stops. Near-perfect reports bring many more pairs to the threshold, but the opportunities they block remain valuable to both partners.

\begin{table}[tbp]
\centering\small
\caption{Payoff changes relative to the no-report benchmark}\label{tab:core_comparisons}
\begin{tabularx}{\linewidth}{@{}>{\raggedright\arraybackslash}Xrrr@{}}
\toprule
Comparison & Proposer & Responder & Joint\\
\midrule
\multicolumn{4}{@{}l}{\emph{A: Pooling proposals only, $\gamma=0.05$}}\\
Experience, diary-shaped & $-0.28$& $-0.13$& $-0.40$\\
 & $[-0.31,-0.25]$& $[-0.19,-0.06]$& $[-0.49,-0.31]$\\
Experience, strong & $-7.69$& $-3.13$& $-10.82$\\
 & $[-7.94,-7.44]$& $[-3.44,-2.82]$& $[-11.37,-10.27]$\\
Experience, near-perfect & $-7.48$& $-2.48$& $-9.95$\\
 & $[-7.75,-7.20]$& $[-2.85,-2.11]$& $[-10.59,-9.32]$\\
Type known from the start & $-6.80$& $-0.48$& $-7.28$\\
 & $[-7.13,-6.47]$& $[-0.89,-0.06]$& $[-8.01,-6.54]$\\
\addlinespace
\multicolumn{4}{@{}l}{\emph{B: Maintained mixture, diary-shaped, $\gamma=0.05$}}\\
Experience & $-2.19$& $-1.06$& $-3.25$\\
 & $[-2.33,-2.04]$& $[-1.17,-0.96]$& $[-3.50,-3.00]$\\
Outside minus experience & $0.95$& $0.46$& $1.41$\\
 & $[0.83,1.06]$& $[0.37,0.56]$& $[1.21,1.62]$\\
Outside & $-1.24$& $-0.60$& $-1.84$\\
 & $[-1.30,-1.18]$& $[-0.63,-0.57]$& $[-1.93,-1.75]$\\
\addlinespace
\multicolumn{4}{@{}l}{\emph{C: Pooling proposals only, $A=1.70$, $\gamma=0.05$}}\\
Type known from the start & $-8.33$& $0.74$& $-7.59$\\
 & $[-8.63,-8.02]$& $[0.39,1.08]$& $[-8.23,-6.96]$\\
Experience, near-perfect & $-8.78$& $-1.10$& $-9.88$\\
 & $[-9.05,-8.51]$& $[-1.39,-0.81]$& $[-10.43,-9.33]$\\
Experience, diary-shaped & $-7.51$& $-2.91$& $-10.41$\\
 & $[-7.73,-7.28]$& $[-3.06,-2.75]$& $[-10.79,-10.04]$\\
\bottomrule
\end{tabularx}
\par\smallskip\begin{minipage}{\linewidth}\footnotesize\emph{Notes:} Payoff changes are per 100 pair-days in days 1--84, relative to the same environment without reports except for Panel B's direct difference. Brackets give pointwise 95\% paired Monte Carlo intervals. Experience is feedback after participation; outside feedback arrives independently of participation and matches the population's expected daily report count. Intervals condition on fixed outside-feedback schedules. The type-known and experience rows in Panels A and C use the unknown-type no-report benchmark. Benchmark joint payoffs in Panels A--C are 25.18, 22.01 and 20.45. Joint payoff sums both utilities in~\eqref{eq:utility}: direct payoffs and utility from the partner's direct payoff, less the responder's attribution cost. Here $A$ is the responder's utility from the proposer's gain; the maintained mixture selects joint and mixed acceptance equally. Unstated parameters take their baseline values in Section~\ref{sec:design}.\end{minipage}
\end{table}

\subsection{Is being known what hurts?}\label{subsec:type_knowledge}

If becoming known were itself the problem, revealing the type from the outset should also hurt. Yet type revelation and experience can give opposite answers. With pooling proposals only and an unknown-type benchmark that blocks nothing, the responder's payoff change has the sign of the mean blocked forecast minus $\widehat p$ (equation~\eqref{eq:weights}). Under type revelation, blocked forecasts equal excluded pairs' true types (Figure~\ref{fig:where_ends}B). At baseline $A=2$, their mean is 71\%, below the 75\% payoff threshold, so type revelation also hurts the responder. At $A=1.70$, where the responder values the proposer's gain less and the two thresholds fall to 40\% and 60\%, the mean blocked forecast is 65.2\%, above the payoff threshold: revealing the type helps, while both diary-shaped and near-perfect experience remain harmful (Table~\ref{tab:core_comparisons}C).

Under the baseline prior, the effect of type revelation on the responder changes sign at $A\approx1.86$. Throughout $A\in[1.70,2]$ at $\gamma=0.05$, however, Proposition~\ref{cor:ceiling} keeps diary-shaped experience below the corresponding payoff threshold. At $A=1.75$, type revelation again benefits the responder while near-perfect experience hurts. Even at $A=1.70$, a perfectly accurate first positive report from the prior gives a next-day forecast of 60.1\% that makes stopping protective for only one day. The forecast then falls toward the 58\% type judgment, while participation remains infeasible. Knowing the type from the outset can therefore help even when learning through participation hurts.

Experience also excludes many pairs that could continue if the type were known. At baseline payoffs, with pooling proposals only and near-perfect reports, 37.9\% $[35.9,39.9]$ of pairs excluded on day 84 have true types no higher than the participation threshold. About three-quarters of excluded pairs stop after their first report from the prior is positive, when the posterior probability that the type is no higher than the threshold is still 38\% (37\% for a perfectly accurate first report). Except for the weak reference under pooling proposals only, this type share among excluded pairs ranges from 32.9\% to 51.2\% in the other diary-compatible experience environments. The weak reference leaves only nine pairs excluded on day 84 across eight replications, two of whose types are no higher than the threshold.

This mix of types need not reflect biased beliefs. Exclusion selects on public information, and Bayesian judgments can be unbiased on average even when choices determine which observations are collected \citep{le_mens_denrell2011}. In the main environment, among pairs still excluded on day 84 whose current stop began with a type judgment strictly above $c_*$, the mean true type is 50.8\% $[49.2,52.4]$ and the mean judgment at onset is 51.2\%. Average calibration does not distinguish individual pairs, and stopping cuts off the experience that could distinguish them further. Prolonged nonparticipation therefore does not show that the responder's type alone would make the activity infeasible.

With pooling proposals only, diary-shaped experience costs the pair less than type revelation over days 1--84 (Table~\ref{tab:core_comparisons}A), but in the long run it costs both the pair and the responder more. Under this report and the baseline preference process, pairs with types above $c_*$ almost surely enter a stop that never ends on its own. If feasible days continued indefinitely, reports would eventually pin down the type, and a type judgment settling above the threshold would make participation infeasible. Appendix~\ref{subsec:s6_census} gives the argument and the long-run bounds.

Pairs with types above $c_*$ make up 35.7\% of the population and eventually lose every future opportunity. Diary-shaped experience keeps forecasts below 62.5\%, so each blocked unit of $G_j$ costs at least 1.375 in joint payoff and 0.25 in responder payoff. The baseline opportunity weight is 15.75 per 100 pair-days, approximately 21 qualified opportunities times the participation-selection probability of 0.75. Long-run average expected losses therefore have lower limits of at least 7.7 and 1.4, exceeding the respective losses from type revelation, about 7.27 and 0.49 (Table~\ref{tab:core_comparisons}A gives their simulation estimates). The small loss from diary-shaped experience within 84 days reflects slower entry into these stops; once they begin, valuable opportunities continue to be lost.

\subsection{Can stopping protect the responder?}\label{subsec:protection}

\begin{figure}[tbp]
\centering
\includegraphics[width=\linewidth]{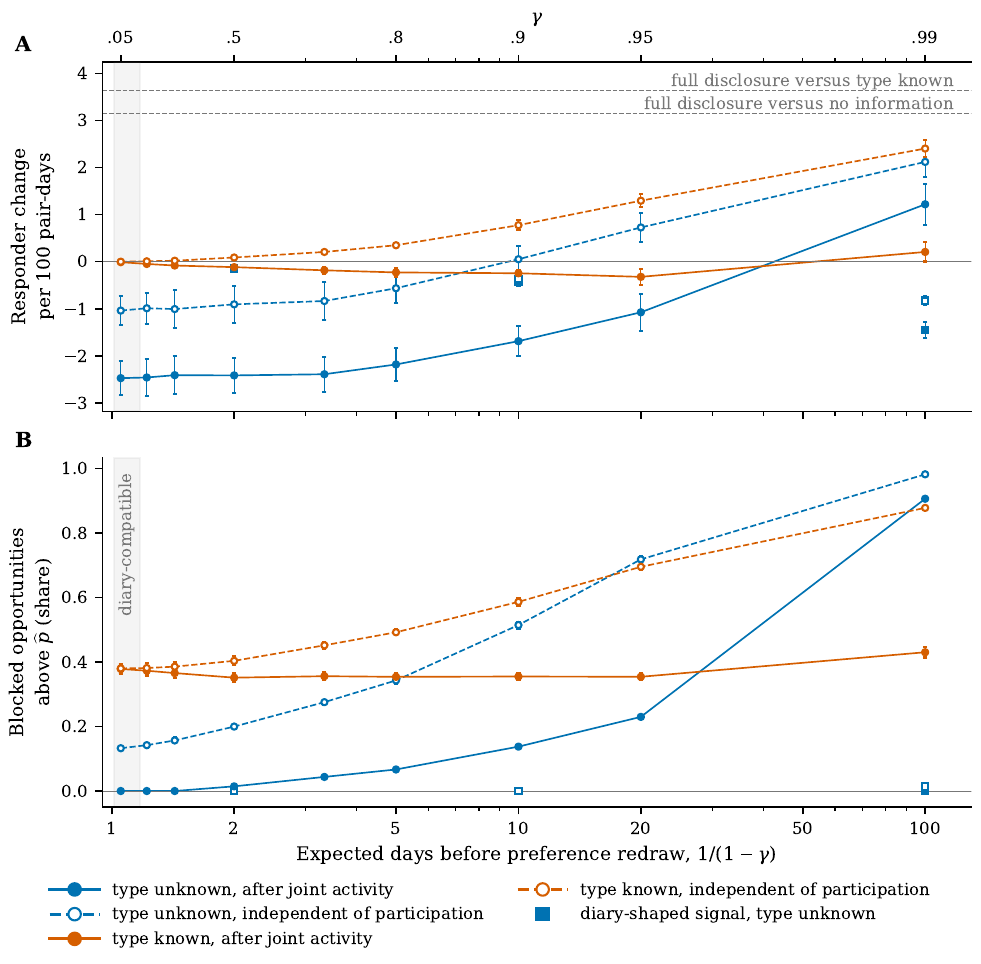}
\caption{Preference persistence and the value of feedback}\label{fig:persistence}
\par\smallskip\begin{minipage}{\linewidth}\footnotesize\emph{Notes:} Panel A gives the responder's payoff change relative to the same environment without reports, per 100 pair-days in days 1--84; Panel B gives the share of blocked forecasts above the responder's payoff threshold $\widehat p=3/4$ in that window. Circles are near-perfect reports; squares, plotted from $\gamma=0.5$, are diary-shaped reports with an unknown type; filled markers and solid lines are feedback after joint activity (experience), open markers and dashed lines feedback independent of participation (outside feedback). Known-type series value feedback relative to a known-type no-report benchmark, which already excludes pairs with $\pi>c_*$; their Panel B shares include those pairs' blocked opportunities. The type-known rows in Table~\ref{tab:core_comparisons} instead use an unknown-type benchmark. The logarithmic horizontal axis gives expected days before a preference redraw, $1/(1-\gamma)$, with $\gamma$ above; a redraw can return the same preference. Proposals are pooling only and $R=0.3$; shading marks the compatible range at this dispersion. Lines join computed points. Pointwise 95\% Monte Carlo intervals use paired differences for payoffs and levels for shares, conditional on fixed outside-feedback schedules. Gray dashed lines give the responder's static gain from full preference disclosure relative to no information or to type revelation, per opportunity and scaled by the common benchmark rate of $G_j$: approximate references, not attainable dynamic values or limits.\end{minipage}
\end{figure}

With pooling proposals only, a report that takes tomorrow's forecast above the payoff threshold can make stopping initially protective. At moderate dispersion, protection appears in the computed experience environments only at $\gamma\ge0.50$, outside the diary-compatible range (Figure~\ref{fig:persistence}B). With $R=0.3$ and $\gamma=0.95$, a first positive near-perfect report from the prior raises tomorrow's forecast to 96.5\%, but the type judgment only to 57.6\%. As today's preference fades, the forecast falls toward that judgment, passing below the 75\% payoff threshold after 16 days while staying above the 50\% participation threshold (Corollary~\ref{cor:spell_path}). Protection expires while exclusion continues.

For simulated pairs starting from the common prior on day 1, near-perfect experience at $R=0.3$ and $\gamma=0.95$ raises the responder's payoff relative to no reports by $0.844$ $[0.436,1.252]$ per 100 pair-days in days 1--21; the change in days 22--30 is $-0.146$ $[-0.756,0.464]$, not detected, and in days 31--84 it is $-1.979$ $[-2.375,-1.583]$. In each of these three periods, more than nine in ten new stops begin above the payoff threshold, yet only 23.0\% of blocked opportunities in days 1--84 lie above it (Figure~\ref{fig:persistence}B). New stops still protect the responder while older, continuing stops cause losses. When the same reports also arrive during stops, the late period instead shows a gain: near-perfect outside feedback raises the responder's payoff by $0.514$ $[0.163,0.865]$ in days 31--84.

Greater persistence extends protection: at $R=0.3$, near-perfect experience raises the responder's payoff in all three periods only at $\gamma=0.99$, with a gain of 1.22 $[0.78,1.66]$ over days 1--84. Diary-shaped experience lowers payoff in this window at every computed persistence (Figure~\ref{fig:persistence}A).

The formula in Section~\ref{subsec:onset} divides type dispersion into three ranges: after a perfectly accurate first positive report, the judgment lies below the participation threshold if $R<1/6$ and at or above the payoff threshold if $R\ge7/12$. If the first report triggers exclusion, a stop in the intermediate range cannot end on its own, and any initial protection must expire. The four diary-compatible $(R,\gamma)$ preference processes have dispersion between 0.255 and 0.37, all in this intermediate range, and insufficient persistence to produce initial protection.

With near-perfect reports at $\gamma=0.95$ and $R=0.1$, a first positive report from the prior gives a type judgment of 45.9\%: protection lasts only 11 days, but participation becomes feasible again on day 49 after onset, counting onset as day 0. Valuable participation is lost in between. At $R=0.3$, the judgment is 57.6\%, participation cannot resume on its own, and stops of this kind account for 77.5\% of blocked opportunities in days 1--84. At $R=0.7$, the first-report judgment is 81.0\%; about two-thirds of stops begin with type judgments at or above the payoff threshold and forecasts above it, and continue to protect the responder. The responder gains in all three periods, including $0.794$ $[0.226,1.363]$ in days 31--84. How long a stop lasts therefore does not reveal whether it protects or harms the responder.

Within the diary-compatible range, theory leaves open a gain relative to no reports for one partner or the pair in 36 comparisons from 24 environments, all at the baseline process ($R=0.3$, $\gamma=0.05$). No gain is detected in any of them. The largest upper endpoint of their pointwise 95\% Monte Carlo intervals is 0.062 per 100 pair-days, for the responder under near-perfect outside feedback at $A=1.70$. Even this upper endpoint is only 1.0\% of that environment's no-report responder payoff (Appendix~\ref{subsec:s6_census}). Outside feedback, which also reaches excluded pairs, still causes a loss relative to no reports in the main environment, though a smaller one than experience.
\FloatBarrier

\section{Evidence after the stop}\label{sec:evidence}

Persistent stops in diary-compatible environments need evidence of not wanting, which experience supplies only after participation. Section~\ref{sec:design}'s diaries calibrate report discrimination from partners' judgments on days without activity, but cannot show whether such judgments still become evidence both partners share once participation stops. Outside feedback represents the case in which they do: equally informative reports arrive regardless of participation.

Making experience reports more precise or more frequent brings no evidence to pairs that have already stopped, and it can enlarge the loss. Along the Blackwell chain in Lemma~\ref{lem:blackwell}, with short-lived preferences and pooling proposals only, we measure joint losses in units of the loss from type revelation (7.28 per 100 pair-days, Table~\ref{tab:core_comparisons}A). Experience's loss in days 1--84 rises from 0.06 with diary-shaped reports to 1.49 with strong reports, then falls to 1.37 $[1.30,1.43]$ with near-perfect reports. Both sharper reports cost the pair more than type revelation. Strong reports cost more than near-perfect ones because their false positives trap more low-type pairs and block participation at lower forecasts, where an opportunity is worth more to the pair. Among all pairs with true types no higher than $c_*$, 38.2\% are excluded on day 84 under strong reports and 31.4\% under near-perfect reports; mean blocked forecasts are 53.6\% and 57.0\%, respectively. Both exclude about nine in ten pairs with types above $c_*$.

In the main environment, raising the probability that a report follows a joint activity from 0.2 to 0.5 and 0.8 increases joint losses from 1.143 to 2.318 and 3.251 per 100 pair-days. Positive reports accumulate faster toward stopping, but no report reaches pairs already excluded.

Outside feedback supplies contrary evidence while participation is infeasible. With the same expected number of reports per day across the population, it reduces the main environment's joint loss from 3.251 to 1.840 per 100 pair-days, or from 14.8\% to 8.4\% of the no-report benchmark (Table~\ref{tab:core_comparisons}B). It also lowers the number of pairs excluded on day 84 by 13.9 $[12.6,15.1]$ per 100, from 31.4 under experience: 6.7 fewer have ever been excluded, and 7.2 more have been excluded but are no longer blocked. Diverting reports to excluded pairs leaves participating pairs with fewer, changing entry as well as exit. Among all pairs with true types no higher than $c_*$, the day-84 excluded share falls from 23.4\% to 12.1\%; among all pairs with types above $c_*$, it falls from 45.8\% to 27.4\%. The reduction reaches both pairs that type revelation would leave participating and those it would exclude from the outset.

Under pooling proposals only, outside feedback removes 30--41\% of the joint loss for the other six empirical shapes. Across 64 within-environment comparisons, inside and outside the compatible range, outside feedback detectably raises the joint payoff relative to experience in 57; the other seven differences are not detected, including three negative point estimates (Appendix~\ref{subsec:s6_census} lists the seven). For diary-shaped reports with pooling proposals only, experience costs the pair just 0.40, and the improvement of $+0.084$ $[-0.012,0.180]$ is not detected. In the computed unknown-type environments, outside feedback removes at most about half the loss. Its negative reports can end stops, but its positive reports still take participating pairs across the threshold.

Release need not last. Consider negative reports delivered to excluded pairs in days 1--83 in the main environment, grouped by the type judgment just before arrival. For judgments at or above the participation threshold but below the payoff threshold, 32.1\% $[29.7,34.5]$ restore feasibility the next day. A negative diary-shaped report lowers the judgment by less than one percentage point, so release requires a judgment already near the participation threshold. Among next-day releases with complete 21-day follow-up (reports through day 63), 61.2\% $[58.5,64.0]$ remain feasible throughout. If the type judgment remains above the threshold after the report, the forecast can rise back across it as the reported day's preference fades. Even with the judgment below the threshold, later reports or acceptances can take it back above.

When feedback is available only after participation, the responder must make the activity happen to obtain contrary evidence; counter-proposing is their only unilateral way on an excluded day. The prescribed reading interprets the counter-proposal as wanting with certainty. A subsequent noisy report about the same day cannot move this posterior from one, and with later behavior held fixed, no profitable counter-proposal is detected.

Obtaining evidence can pay under a less certain reading, that is, when the partner assigns a probability below one to the responder wanting the activity that day. At the tested readings from 0.90 to one, no profitable counter-proposal is detected for a responder who does not want the activity, either with diary-shaped reports, even at $\gamma=0.99$, or with near-perfect reports at $\gamma=0.05$. Detected gains require near-perfect reports and $\gamma\ge0.50$, or strong reports and $\gamma=0.99$, as well as a patient responder. The responder pays a current cost for evidence that may improve future participation. A responder who wants the activity gains in most sampled states only under weaker readings: either $d_H/\kappa$, where today's payoff is zero, or a reading under which the counter-proposal conveys no information (Appendix~\ref{subsec:s7_counter}).

Under prescribed stage behavior, no responder accepts in order to show that they do not want the activity today, so these comparisons cannot tell whether outside evidence would displace such acceptance. Section~\ref{sec:forward} examines this choice when both partners look ahead.
\FloatBarrier

\section{When both partners look ahead}\label{sec:forward}

Evidence that arrives without participation can end stops that waiting cannot (Section~\ref{sec:evidence}). With forward-looking choices, participation can produce contrary evidence too. A responder who prefers the existing arrangement today may accept the new activity at a current loss, because an exact participation report then shows that they do not want it today and, as preferences persist, can lower future attribution costs. This self-identification makes acceptance less revealing of personal interest and so supports participation on more days when the responder wants the activity. We study it in an economy in which the responder's type is public and preferences persist longer than the diaries allow. At this economy's baseline, introducing free, exact daily outside reports has opposite effects: with exact participation reports, the best joint value over monotone Markov equilibria falls even with arbitrary public information before acceptance; with weak, diary-shaped reports, it rises.

\subsection{A forward-looking economy}\label{subsec:fwd_economy}

With the responder's type public, evidence from participation and from outside reports concerns only today's preference. The partners meet daily and discount by $\delta\in(0,1)$. The proposer always directly prefers the activity and receives a positive payoff from implementation. The responder privately observes today's preference $\theta_j\in\{H,L\}$. It persists with probability $\gamma\in(0,1)$; otherwise it is redrawn and is $H$ with probability $\pi\in(0,1)$, the responder's public type. Tomorrow's probabilities of $H$, conditional on $H$ and $L$ today, are $p_{\mathrm{hi}}=\gamma+(1-\gamma)\pi$ and $p_{\mathrm{lo}}=(1-\gamma)\pi$. The public forecast $p$ is the probability of $H$ before today's choices. A posterior $r$ gives tomorrow's forecast $p_{\mathrm{lo}}+\gamma r$, so subsequent forecasts lie in $I=[p_{\mathrm{lo}},p_{\mathrm{hi}}]$.

Each day the responder accepts or declines; after acceptance the proposer chooses whether to implement, possibly at random. Implementation gives the proposer $v_{\theta_j}$ from~\eqref{eq:dvalues} and the responder $d_{\theta_j}-\kappa q$, where $q$ is the posterior after acceptance, the probability of $H$ in the \emph{acceptance pool} at the current forecast; otherwise both receive zero. Under Assumption~\ref{ass:region}, preference $L$ has a nonnegative current payoff only at $q\le c_*$, and preference $H$ only at $q\le c^H=d_H/\kappa\in(c_*,1)$. Implementation produces a participation report that is positive with probabilities $k_H$ and $k_L<k_H$ under $H$ and $L$: either the exact report $(1,0)$ or the diary-shaped report $(0.264,0.173)$ in Table~\ref{tab:signals}. Independently, an outside report publicly reveals today's preference at day's end with probability $\omega\in[0,1]$, whether or not activity occurred. Reports arrive after the day's attribution cost has been paid. Unless otherwise stated, there is no public signal before acceptance.

This section and Appendices~\ref{app:forward}--\ref{sec:s3} maintain $p_{\mathrm{lo}}<c_*<\pi<c^H<p_{\mathrm{hi}}$. Without new evidence, a forecast above $c_*$ moves toward $\pi$ and never returns below $c_*$, as in the stops that waiting cannot end (Corollary~\ref{cor:spell_path}(ii)). Revelation of $L$ sends tomorrow's forecast below $c_*$, and revelation of $H$ above $c^H$.

Payoffs are those of Section~\ref{subsec:comparisons}: $v_H=3/2$, $v_L=1/2$, $d_H=5/2$, $d_L=3/2$ and $\kappa=3$. Valuing the partner's payoff gives $L$ a reason to participate; full revelation of $H$ still yields a current loss. The thresholds are $c_*=1/2$ and $c^H=5/6$, with payoff difference $\Delta=d_H-d_L=1$. The public type $\pi=3/5$ lies between them. Both report strengths share $\gamma=9/10$ and $\delta=19/20$, giving $p_{\mathrm{lo}}=3/50$ and $p_{\mathrm{hi}}=24/25$, so the two cases differ only in what participation reveals. With exact participation reports and no outside reports, this persistence and patience allow future identification gains to offset today's loss (Proposition~\ref{prop:evidence}), as in the settings where Section~\ref{sec:evidence} detects gains from obtaining evidence. Only the diary-shaped report's probabilities come from the diaries; $\gamma=9/10$ exceeds the highest diary-compatible persistence, $0.235$, in the grid of Appendix~\ref{subsec:s4_states}.

Stationary Markov strategies and interpretations of unexpected choices depend on public history only through the current forecast; the responder also conditions on today's private preference. Let $\mathcal V_H(p)$ and $\mathcal V_L(p)$ be the responder's discounted values conditional on today's preference, $\mathcal V_2(p)=p\mathcal V_H(p)+(1-p)\mathcal V_L(p)$ their average, and $\mathcal V_1(p)$ the proposer's value. Equilibrium choices are optimal and on-path beliefs follow Bayes' rule. Local screening, a forecast-by-forecast version of the D1 criterion, attributes an unexpected choice to one preference if every interpretation weakly benefiting the other strictly benefits it, but not conversely (Appendix~\ref{app:forward}). An equilibrium is \emph{monotone} if both responder values are nonincreasing in $p$.

\subsection{Self-identification supports the acceptance pool}\label{subsec:fwd_identification}

In the equilibrium below, self-identification lowers tomorrow's attribution cost by $\mathcal D=\kappa(c^H-p_{\mathrm{lo}})$. Let $\omega^*=1-\Delta/(\delta\mathcal D)$ be the outside-report frequency at which this saving just offsets today's loss $\Delta$.

\begin{proposition}[Evidence-seeking equilibrium]\label{prop:evidence}
Let the report be exact. Consider the profile in which preference $L$ accepts at every forecast; preference $H$ accepts at $p\le c^H$ and with probability $c^H(1-p)/[(1-c^H)p]$ at $p>c^H$; the proposer implements every acceptance; and a decline at $p\le c^H$ is read as $H$. This profile is a Markov equilibrium if and only if $\omega\le\omega^*$. On its path every day ends with today's preference public, and its values do not depend on $\omega$.
\end{proposition}

Call this profile the evidence-seeking equilibrium, ES. Above $c^H$, $H$ accepts just often enough to keep the acceptance pool's posterior at $c^H$, where its current payoff is zero. At this posterior, $L$ loses $\Delta$ today by accepting, but the report identifies the day as $L$ and sends tomorrow's forecast to $p_{\mathrm{lo}}$; a decline is read as $H$. Outside reports identify $L$ even after a decline, so the private return $(1-\omega)\delta\mathcal D$ to acceptance must cover $\Delta$. At the baseline, $\omega^*=301/551\approx0.546$.

Preference $L$ accepts for its own future benefit, but this makes acceptance less revealing of $H$. Let $M_\theta(p)$ be the probability of preference $\theta$ and implementation at forecast $p$. Above $c^H$, each implemented $L$ day supports $M_H/M_L=c^H/(1-c^H)=d_H/(\kappa-d_H)$ implemented $H$ days in the acceptance pool. With only current payoffs, this ratio is at most $d_L/(\kappa-d_L)$. At the baseline high forecast, one implemented $L$ day supports five $H$ days in the same pool; static equilibrium permits at most one (Figure~\ref{fig:forward}A).\footnote{Preference $H$'s acceptance probability has the same form as the bad seller's trading probability in \citet[Proposition~2 and Section~4.2]{bar_isaac2003}: $1-p$ and $1-c^H$ here correspond to reputation and its threshold there. The bad seller's trade is inefficient in that model, whereas the mixing preference here wants the activity.}

\subsection{What exact outside reports remove}\label{subsec:fwd_full}

At $\omega=1$, tomorrow's forecast depends only on today's preference, so today's choices no longer affect continuation values. The static participation constraint therefore applies: $L$ accepts only at posteriors no higher than $c_*$. To ask whether information before acceptance can restore the $H$ days that self-identification supported, we allow any public signal conditional on the forecast and today's preference, presuming that preference can be measured beforehand. The \emph{pooling signal} $\star$ is uninformative at $p\le c_*$. Above $c_*$, as in the binary split of \citet{kamenica2011}, it pools every $L$ day with just enough $H$ days to give posterior $c_*$ and reveals the rest. \emph{Disclosure before acceptance}, $F$, reveals today's preference. Without a signal, the \emph{waiting equilibrium}, WT, accepts at $p\le c_*$ and waits for an outside report above it.

\begin{proposition}[Exact outside reports]\label{prop:full}
Let $\omega=1$. In every Markov equilibrium, whatever the report after implementation and the public signal before acceptance, the proposer implements every acceptance and $M_H(p)\le c_*M_L(p)/(1-c_*)$ at every forecast $p\in[0,1]$. Hence, from every initial forecast and for all weights $w_1,w_2\ge0$, the highest value of $w_1\mathcal V_1+w_2\mathcal V_2$ is attained by the pooling signal if $w_1v_H\ge w_2(\kappa-d_H)$ and by disclosure otherwise. Without a signal before acceptance, WT attains the highest value for every pair of weights.
\end{proposition}

\begin{corollary}[The days that cannot be restored]\label{cor:gap}
With exact participation reports, ES at any $\omega\le\omega^*$ and $\star$ at $\omega=1$ implement the same $L$ days, but ES implements more $H$ days from every initial forecast in $I$. Writing $N_\theta$ for the discounted expected number of implemented days with preference $\theta$, from $p_{\mathrm{lo}}$,
\begin{equation}\label{eq:gap}
N_H^{\mathrm{ES}}-N_H^\star=n_{\mathrm{hi}}(1-p_{\mathrm{hi}})\Bigl[\frac{c^H}{1-c^H}-\frac{c_*}{1-c_*}\Bigr],\qquad n_{\mathrm{hi}}=\frac{\delta p_{\mathrm{lo}}}{(1-\delta)(1-\delta\gamma)}.
\end{equation}
Thus the joint value of ES exceeds the supremum of joint value over Markov equilibria at $\omega=1$, allowing any public signal before acceptance, if and only if $v_H+d_H>\kappa$.
\end{corollary}

\begin{figure}[tp]
\centering
\includegraphics[width=\linewidth]{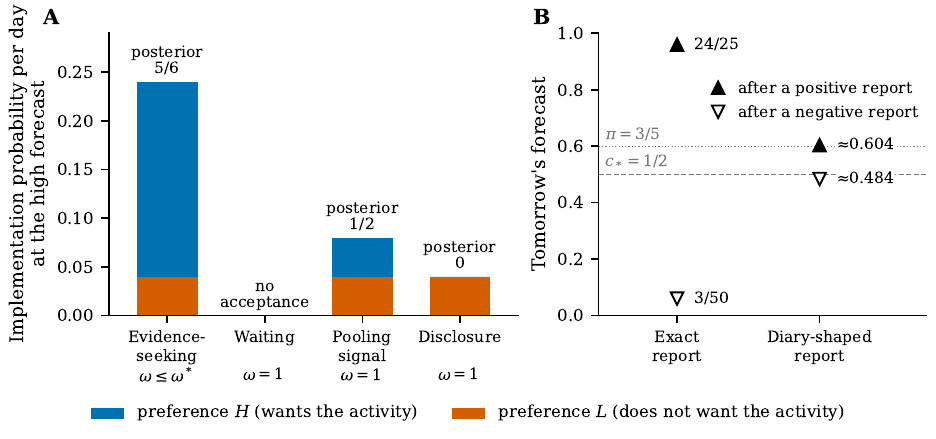}
\caption{Self-identification and the composition of the acceptance pool}\label{fig:forward}
\par\smallskip\begin{minipage}{0.97\linewidth}\footnotesize\emph{Notes:} Baseline of Section~\ref{subsec:fwd_economy}. Panel A: probabilities $M_H$ and $M_L$ of implementation with each preference on a day at the high forecast $p_{\mathrm{hi}}=24/25$; labels give the acceptance pool's posterior. Exact outside reports arrive with daily probability $\omega$; $\omega^*$ is the largest probability supporting the evidence-seeking equilibrium under exact participation reports. Panel B: tomorrow's forecast after an acceptance implemented at posterior $c_*=1/2$, the participation threshold, without an outside report. Here $\pi=3/5$ is the probability of $H$ after a preference redraw.\end{minipage}
\end{figure}

Self-identification entails a current loss: at the baseline, implementing an $L$ day at posterior $c^H$ gives joint payoff $v_L-\Delta=-1/2$. Outside reports eliminate this loss, and the pooling signal preserves every $L$ implementation, now at posterior $c_*$. In the job-market signaling model of \citet[pp.~126--128]{daley_green2014}, a more informative symmetric binary test weakly reduces wasteful education in the D1 equilibrium and weakly raises efficiency. Here the best joint value falls instead, through the composition of the acceptance pool. The extra attribution cost on $L$ days at posterior $c^H$ equals the saving on the $H$ days in the same pool, so the pool's expected attribution cost is $\kappa$ per implemented $H$ day (Lemma~\ref{lem:fwd_accounting}). What changes is how many $H$ days each $L$ day supports: at the high forecast, the pooling signal pairs each $L$ day with one $H$ day rather than five (Figure~\ref{fig:forward}A). With all $L$ days preserved, each missing $H$ day costs the proposer $v_H$ and raises the responder's average value by $\kappa-d_H$; their difference determines the joint change.

At the baseline, ES implements about $2.30$ discounted $H$ days from $p_{\mathrm{lo}}$, versus $1.04$ under the pooling signal. Each missing $H$ day costs $v_H+d_H-\kappa=1$ jointly, so the joint values also differ by $912/725\approx1.26$. Since ES is attainable at $\omega=0$, this gap bounds the fall in the best joint value from below. By Proposition~\ref{prop:full}, ES also exceeds every weighted Markov equilibrium value at $\omega=1$, allowing any public signal before acceptance, whenever the responder's weight is less than $v_H/(\kappa-d_H)=3$ times the proposer's. Within Markov equilibria, neither selection nor information design restores these $H$ days.

\subsection{When participation reveals little}\label{subsec:fwd_weak}

Diary-shaped reports reveal little: after an acceptance implemented at posterior $c_*=1/2$, the participation threshold, both positive and negative reports leave tomorrow's forecast near $1/2$ (Figure~\ref{fig:forward}B).

\begin{samepage}
\begin{proposition}[Diary-shaped reports]\label{prop:weak}
Let the report after implementation be diary-shaped, $\omega=0$, and the remaining parameters take their baseline values. In every monotone Markov equilibrium, no further implementation occurs once the forecast reaches $29757/50000\approx0.595$. From $p_{\mathrm{lo}}$, $N_H\le5$, $\mathcal V_2\le10.67$ and $\mathcal V_1+\mathcal V_2\le22.56$; apart from the joint-value bound, these conclusions hold whatever the proposer's payoffs and choices. Exact daily outside reports raise the supremum of $\mathcal V_1+\mathcal V_2$ over monotone Markov equilibria from $p_{\mathrm{lo}}$ by at least $0.99$, and that of $\mathcal V_2$ by at least $6.08$.
\end{proposition}
\end{samepage}

\begin{table}[tp]
\centering
\caption{Values and bounds at two report strengths}\label{tab:forward}
\vspace{0.35em}
\small
\begin{tabular}{@{}llcccl@{}}
\toprule
& & & \multicolumn{2}{c}{$\omega=1$} & \\
\cmidrule(lr){4-5}
Report after implementation & Value & $\omega=0$ & no signal & any signal & Exact outside reports\\
\midrule
Exact & $\mathcal V_1+\mathcal V_2$ & $\ge25.74$ & 23.548 & 24.491 & lower the best value\\
 & $\mathcal V_2$ & $\ge16.43$ & 16.750 & 17.586 & direction unresolved\\
\addlinespace
Diary-shaped & $\mathcal V_1+\mathcal V_2$ & $\le22.56$ & 23.548 & 24.491 & raise the best value\\
 & $\mathcal V_2$ & $\le10.67$ & 16.750 & 17.586 & raise the best value\\
\bottomrule
\end{tabular}
\par\smallskip\begin{minipage}{0.97\linewidth}\footnotesize\emph{Notes:} Baseline of Section~\ref{subsec:fwd_economy}; infinite-horizon discounted values from initial forecast $p_{\mathrm{lo}}=3/50$, computed in exact arithmetic; $\mathcal V_1$ is the proposer's value and $\mathcal V_2$ the responder's, averaged over today's preference. Exact outside reports arrive with daily probability $\omega$. At $\omega=0$, without a signal before acceptance, the evidence-seeking equilibrium gives lower bounds and Appendix~\ref{sec:s3} gives upper bounds on the best monotone Markov values, rounded outward. At $\omega=1$, outside reports fix tomorrow's forecast, so values do not depend on the report after implementation. They are the highest values over Markov equilibria (Proposition~\ref{prop:full}), rounded to three decimals: without a signal, those of the waiting equilibrium; with any signal, those of the pooling signal for joint value and of full disclosure for $\mathcal V_2$. Signals require preference to be measurable before acceptance. The last column gives the effect of exact daily outside reports on the best value.\end{minipage}
\end{table}

Diary-shaped evidence cannot sustain the self-identification of ES. Activity ends for good once the forecast reaches about $0.595$, just below the public type $\pi$. At such forecasts the acceptance pool's posterior exceeds $c_*$, so $L$ loses today, and the continuation gain from a diary-shaped report cannot compensate. Acceptance by $H$ alone would reveal $H$ and likewise prevent implementation. Without implementation, no new report arrives; even if choices convey information, the responder's values remain zero at every later forecast, so implementation never resumes (Appendix~\ref{sec:s3}, Step~5). After implementation at posterior $1/2$, a positive diary-shaped report takes tomorrow's forecast to about $0.604$, above $\pi$ and inside this region (Figure~\ref{fig:forward}B). In these equilibria, forward-looking choices do not end a stop at such forecasts; exact outside reports instead return the forecast directly to $p_{\mathrm{lo}}$ after an $L$ day, restoring activity in WT.

The value bounds rest on the responder's incentives alone. Below $\pi$, implementation requires both preferences in the acceptance pool, at a posterior no lower than the forecast. With both preferences accepting for sure, $L$'s value must cover its discounted continuation at the same forecast; otherwise local screening would attribute a decline to $L$, and declining would pay. Together with monotonicity, these incentives bound the value of a day without implementation. Appendix~\ref{sec:s3} substitutes this bound into acceptance payoffs and iterates the resulting map in exact arithmetic on a grid of forecasts; each iterate is a valid upper bound, and $29757/50000$ is the first grid point from which the bounds are zero.

{\clubpenalty10000
Table~\ref{tab:forward} sets the two report strengths side by side within monotone Markov equilibria, a class that includes ES and WT (Appendix~\ref{app:fwd_class}). With exact participation reports, an attainable joint value at $\omega=0$ exceeds the highest value at $\omega=1$ even with arbitrary signals before acceptance (Corollary~\ref{cor:gap}). With diary-shaped reports, the upper bound at $\omega=0$ lies below WT at $\omega=1$ (Proposition~\ref{prop:weak}); only this comparison relies on monotonicity and local screening. With exact participation reports the proposer bears the loss, and with diary-shaped reports the responder's best average value rises; in each case the other partner's direction is unresolved.\par}

Between these two report strengths, the effect of outside reports on the best joint value remains open. These comparisons concern Markov equilibria: with exact participation reports, history-dependent punishment by stopping activity can sustain full participation. At the baseline, this attains $920/29\approx31.72$ from $p_{\mathrm{lo}}$ at every $\omega$, the highest joint value of any strategy profile (Appendix~\ref{app:fwd_class}).

In these comparisons, whether free outside reports raise or lower the best joint value depends on what participation itself reveals. When participation reveals today's preference exactly, outside reports displace self-identification and the additional $H$ days it supports, a cost that comparisons under prescribed stage behavior cannot record. When participation reveals as little as the diary-shaped report, outside reports supply evidence that joint activity cannot provide at high forecasts, and the comparison points the same way as in Section~\ref{sec:evidence}.

\section{Robustness and scope}\label{sec:robustness}

The diaries measure feedback informativeness and constrain persistence; payoffs, the form of the attribution cost, prescribed stage behavior and uninformative opportunities are assumptions. We ask how three conclusions of Sections~\ref{sec:dynamic}--\ref{sec:evidence} depend on these inputs: opportunities blocked by experience keep positive expected value for both partners, stops that waiting cannot end carry almost all of them, and outside feedback reduces the loss.

Payoffs set the two thresholds that enter~\eqref{eq:ceiling_condition}. Varying payoffs with $c_*$ fixed leaves every simulated history unchanged; blocked forecasts then stay below $\widehat p$ whenever it exceeds the ceiling. With diary-shaped experience at $\gamma=0.05$, $\widehat p$ exceeds the ceiling at every simulated value of $A$. Exclusion itself depends strongly on $c_*$: under the maintained mixture, raising $A$ from $1.70$ to $2.30$, and with it $c_*$ from $0.40$ to $0.60$, cuts day-84 exclusion from 79.5 to 10.8 pairs per 100. At both levels, stops that waiting cannot end still carry almost all blocked opportunities (Table~\ref{tab:spell_burden}).

Within the diary-compatible range, raising persistence from $0.05$ to $0.13$ changes experience's joint payoff effect under the maintained mixture by $+0.012$ $[-0.025,0.050]$ per 100 pair-days in days 1--84, against an effect of $-3.25$ at $\gamma=0.05$ (Table~\ref{tab:core_comparisons}B); the corresponding change in day-84 exclusion is also undetected. Outside that range, high persistence or type dispersion can make stopping protective (Section~\ref{subsec:protection}). Under the baseline preference process, diary-shaped experience with pooling proposals only costs the pair and the responder more than type revelation in the long run (Section~\ref{subsec:type_knowledge}).

The proposer's weight $\alpha_1$ on the responder's direct payoff changes no simulated history and enters each payoff comparison linearly, so its whole range $0\le\alpha_1<2$ under Assumption~\ref{ass:region} can be checked (Appendix~\ref{subsec:s6_census}). Among the 36 diary-compatible comparisons in which theory leaves open a gain (Section~\ref{subsec:protection}), no gain is detected in days 1--84, 1--21 or 1--30 at any such weight. For every such $\alpha_1$, strong and near-perfect experience yield detectably lower responder and joint payoffs than type revelation in days 1--84 (Table~\ref{tab:core_comparisons}A), but the proposer's payoff is detectably lower only for $\alpha_1$ below about $1.78$ with strong reports and $1.77$ with near-perfect reports. None of the 64 outside-minus-experience joint differences in Section~\ref{sec:evidence} is detected negative anywhere in this range. The experience loss in Table~\ref{tab:core_comparisons}B and its reduction through outside feedback hold for each partner.

The mechanism needs two features of the attribution cost. Attribution is costly even when correct; otherwise, participation in state $H$ would not need participation in state $L$. Declining $x$ carries no such cost; charging it there would remove $H$'s reason to avoid participation (Section~\ref{subsec:information}).

The ceiling holds under any selection among the three stage equilibria based on the public history (Section~\ref{subsec:experience_ceiling}), but stage equilibria put no weight on how today's choices affect future beliefs. With prescribed behavior fixed from the next day through day 84 and future days discounted by $\delta=0.95$, $L$ profitably rejects at all 24 sampled mixed-acceptance states of the main environment under each feedback source. Since $L$ is indifferent today, the gain comes from continuation. Under experience, at joint-acceptance states where one positive report would take tomorrow's forecast above $c_*$ and a refusal is read as $L$, $H$ profitably refuses at 15/24 sampled states in the main environment and at 23/24 under pooling proposals only (Appendix~\ref{subsec:s7_feasible}). Both deviations give up participation today to appear less interested in the activity. No proposer deviation has a positive estimate in 8,976 comparisons. During exclusion, counter-proposals can be profitable when they are less strongly interpreted as evidence of $H$ (Appendix~\ref{subsec:s7_counter}).

With uninformative opportunities, stops under experience that begin at $m\ge c_*$ need new evidence to end. Type-dependent opportunities supply it. In two variants of the main environment, a type one standard deviation higher doubles or halves the odds of an opportunity, independently of daily preferences (Appendix~\ref{subsec:s4_clock}); there, 78--80\% of experience spells beginning at $m>c_*$ end during days 1--84. Opportunities also inform the no-report benchmark, which then excludes 35--36 pairs per 100 on day 84. No difference in day-84 exclusion among experience, outside feedback and no reports is detected, and at $\alpha_1=1$ neither feedback source detectably changes any payoff; lower proposer weights can produce detected proposer losses. Because opportunities now move beliefs, the ceiling no longer applies: about 17--36\% of experience's blocked opportunity weight in days 1--84 lies above $\widehat p$. Under the maintained mixture, where $\chi>\tau$, these opportunities still have positive average value for both partners.

\section{Conclusion}\label{sec:conclusion}

Joint experience can leave both partners worse off than they would have been without feedback about the responder's preferences. In our model, evidence that the responder wants the activity raises the cost of joining, first making participation for the partner's sake unprofitable. Joining only when interested would then reveal too much. Learning can consequently end the activity and cut off further preference evidence before stopping becomes beneficial to either partner. With choices based on current payoffs, we establish a condition under which this remains true after every history: participation limits how far experience can move beliefs, so every opportunity that learning removes still has positive average value for both partners. At baseline payoffs, this condition holds for the main diary's feedback informativeness throughout the range of preference persistence consistent with the diaries.

The losses need not disappear as preferences change. Once activity stops, the belief that the responder generally wants it remains fixed, even as the influence of a particular day's preference fades. In simulations consistent with the diaries, stops that waiting cannot end account for almost all opportunities lost over days 1--84. With more persistent preferences and sharper feedback, stopping can initially protect the responder yet continue after that protection expires. Neither the end of an activity nor the length of its interruption therefore establishes that participation has lost its value.

Preference information that reaches partners without requiring participation can recover some of these opportunities, holding accuracy and the expected amount of feedback across pairs fixed. Our separate forward-looking economy reveals a cost of this remedy. With long-run propensities public and preferences more persistent than in the diaries, accepting an unwanted activity can pay because it provides exact evidence of a lack of interest and reduces future attribution costs. These acceptances make joining less revealing of personal interest, sustaining participation on wanted days. At this economy's baseline, free daily revelation outside the activity removes that incentive and lowers the best joint value in monotone Markov equilibria. With participation feedback as weak as the diary's, the same outside information instead restores activity and raises that value. Outside information can thus reopen an activity, or remove an incentive that keeps it going.

\section*{Data availability}

All data used in this paper are publicly available on OSF \citep{vaccarino2025data,vowels2018data,columbus2021data}. Replication code is available from the author upon request.

\bibliographystyle{apalike}
\bibliography{references}
\appendix
\counterwithin{figure}{section}
\counterwithin{table}{section}
\section{Static results: proofs and equilibrium constructions}\label{app:static}

Assumption~\ref{ass:region} holds throughout, and conditional statements refer only to signal values with positive probability.

\subsection{The existing arrangement and what an accepted offer reveals}\label{app:outside}

\begin{lemma}\label{lem:outside}
In every equilibrium, (i) either party can secure the zero payoff of $f$ through subsequent voluntary choices, the outcome $n$ does not occur on the equilibrium path, and a proposer in state $l$ does not implement $x$; (ii) a responder who, at a history reached with positive probability, accepts an initial offer of $x$ with positive probability assigns probability one to the proposer's state $h$.
\end{lemma}

\begin{proof}
(i) Once $x$ has been rejected it cannot be proposed again, so the subsequent $f$ offer can only be accepted, giving both parties zero, or answered by ending the interaction, giving the proposer $-\varepsilon_1-\alpha_1\varepsilon_2<0$ and the responder $-\varepsilon_2-\alpha\varepsilon_1<0$; it is therefore accepted. Accepting an $f$ offer, or rejecting an $x$ offer and offering $f$, thus secures zero at every node, strictly more than ending the interaction. A counteroffer of $x$ gives the proposer at least $v_L>0$ in state $h$ and at most $-w+\alpha_1g<0$ in state $l$, whatever their belief about $\theta$, so only state $h$ accepts it; an initial $x$ offer in state $l$ yields at most $-w+\alpha_1g<0$ if accepted and zero if rejected, so it is strictly worse than offering $f$ whenever acceptance has positive probability. Thus $\Pr(x\mid l)=0$.

(ii) Let $t_0$ be a public history that ends with an initial offer of $x$ and includes the signal value, if there is one, and let the responder in state $\theta$ reach $t_0$ with positive probability and accept with probability $\beta>0$. The responder's acceptance depends only on $\theta$ and $t_0$, so $\Pr(l,\theta,t_0)\beta\le\Pr(x,l)=0$ by (i), and $\Pr(l\mid\theta,t_0)=0$.
\end{proof}

\subsection{Participation constraints and strategy imitation}\label{app:restrictions}

\begin{proof}[Proof of Lemma~\ref{lem:restrictions}]
Take any equilibrium and condition on the proposer's state $s=h$. When a public signal is present, also fix a signal value $z$ with $0<p_h(z)<1$. Below, $a_\theta$, $V_\theta$ and $M_\theta$ are conditional on $h,z$, omitting $z$ without a signal. For each complete public history $t$ that implements $x$ with positive probability, write
\[
M_\theta(t)=\Pr(\theta,t,x\mid h,z),\qquad
q_h(t)=\frac{M_H(t)}{M_H(t)+M_L(t)},
\]
where the second equality follows from Bayes' rule.

\emph{Participation.} There are two routes to implementation. If the responder accepts an offer of $x$, they could instead have rejected and counteroffered $f$ for zero; by Lemma~\ref{lem:outside}, a responder accepting $x$ with positive probability on the equilibrium path is certain that the proposer is in state $h$, so acceptance requires $d_\theta-\kappa q_h(t)\ge0$. If the proposer accepts the responder's counteroffer of $x$, which only state $h$ does (Lemma~\ref{lem:outside}), the responder could instead have accepted the preceding $f$ offer for zero. The proposer's acceptance and its message depend only on the proposer's state and the public history, so, conditional on $h$, they leave the posterior unchanged; since a failed counteroffer returns to $f$, a successful one cannot yield a negative payoff. This route therefore also requires $d_\theta-\kappa q_h(t)\ge0$. A history with $M_H(t)>0=M_L(t)$ would have $q_h(t)=1$, giving the responder in state $H$ the payoff $d_H-\kappa<0$, which violates this requirement. Every history at which $H$ implements $x$ with positive probability must therefore also be reached with positive probability under $L$, where participation requires
\[
q_h(t)\le c_*=\frac{d_L}{\kappa},
\qquad\text{or equivalently}\qquad
M_L(t)\ge\frac{1-c_*}{c_*}M_H(t).
\]
The latter inequality also holds when $M_H(t)=0$. Summing over all implementing histories, with $M_\theta=\sum_tM_\theta(t)$, gives the first restriction in~\eqref{eq:restrictions}.

\emph{Imitation.} Fix the proposer's strategy and the attribution posteriors. In either state, the responder can follow all the messages, actions and random choices of the other state's strategy. Conditional on the proposer's state and signal $z$, this imitation reproduces the imitated strategy's distribution of public histories. The proposer observes neither the responder's private state nor any further exogenous signal about it, so responses to the same histories and the posteriors $q_s(t)$ remain unchanged, as do the responder's attribution costs. Under $h$, imitation preserves the implementation probability and attribution cost; only the direct payoff per implementation changes: imitating $L$ in state $H$ adds $\Delta$, while imitating $H$ in state $L$ subtracts $\Delta$. Under $l$, both strategies implement only $f$ and yield zero (Lemma~\ref{lem:outside}). Since the signal distribution depends only on the responder's preference state, $\Pr(h\mid\theta,z)=\rho_\theta>0$. Equilibrium play must be at least as good as imitation at each of the responder's first information sets; averaging over the proposer's opening message (Appendix~\ref{subsec:s1_construction}) and dividing by $\rho_\theta$ gives $V_H\ge V_L+\Delta a_L$ and $V_L\ge V_H-\Delta a_H$, the second pair of restrictions in~\eqref{eq:restrictions}; since $\Delta>0$, also $a_H\ge a_L$.
\end{proof}

\subsection{What public information blocks}\label{app:information}

\begin{proof}[Proof of Proposition~\ref{prop:info_value}]
Fix a signal value $z$ with positive probability and write $p=p_h(z)$. (i) For $0<p<1$, all four combinations of states keep positive probability after $z$, and Lemma~\ref{lem:restrictions} and Proposition~\ref{prop:threshold} applied to the interaction after $z$ give the feasibility condition $p\le c_*$; at $p\in\{0,1\}$ it follows from $d_L>0>d_H-\kappa$ and the endpoint constructions of Appendix~\ref{subsec:s1_construction}.

(ii) For $0<p\le c_*$, take any equilibrium after $z$, with $a_\theta$ and $V_\theta$ conditional on $h,z$. By~\eqref{eq:accounting}, $\mathbb E[U_2\mid h,z]=pV_H+(1-p)V_L=p(d_H-\kappa)a_H+(1-p)d_La_L$. Using $V_H-V_L\le\Delta a_H$, $a_L\le a_H$ and $d_L>0$ in the first line, and $V_H-V_L\ge\Delta a_L$, $a_H\ge a_L$ and $d_H<\kappa$ in the second,
\[
\begin{aligned}
V_H&=\mathbb E[U_2\mid h,z]+(1-p)(V_H-V_L)\le a_H(d_H-\kappa p),\\
V_L&=\mathbb E[U_2\mid h,z]-p(V_H-V_L)\le a_L(d_L-\kappa p),
\end{aligned}
\]
and both bounds are at most $d_\theta-\kappa p$ because $d_\theta-\kappa p\ge0$; averaging gives $\Phi(p)$. At $p=0$ only state $L$ remains, and~\eqref{eq:accounting} gives $V_L=d_La_L\le d_L=\Phi(0)$ directly. The proposer's payoff is $pv_Ha_H+(1-p)v_La_L\le pv_H+(1-p)v_L$. Joint acceptance attains all these bounds and an implementation probability of one. For $p>c_*$ every equilibrium gives zero. By Appendix~\ref{subsec:s1_construction}, selecting joint acceptance when $p_h(z)\le c_*$ and zero participation otherwise is one consistent equilibrium attaining all these bounds after every signal value; averaging over signal values conditional on $h$, $\overline U_2(Z)=\mathbb E[\Phi(p_h(Z))\mid h]$ and $\overline U_1(Z)=\mathbb E[(v_L+\alpha_1\Delta p_h(Z))\mathbf 1\{p_h(Z)\le c_*\}\mid h]$.

(iii) For any signal and any of its equilibria, \eqref{eq:accounting} gives $\mathbb E[U_2\mid h]=(d_H-\kappa)M_H+d_LM_L\le(1-\pi_h)d_L$, since $d_H<\kappa$ and $M_L\le1-\pi_h$; full disclosure attains this bound, with $L$ participating at posterior zero and $H$ not participating.

(iv) With $\pi_h\le c_*$, no information gives
\[
\overline U_1(\varnothing)=v_L+\alpha_1\Delta\pi_h,\qquad
\overline U_2(\varnothing)=d_L-(\kappa-\Delta)\pi_h.
\]
Since $\mathbb E[p_h(Z)\mid h]=\pi_h$ and $d_L=(\kappa-\Delta)\widehat p$, subtracting these from the expressions in (ii) gives~\eqref{eq:info_value}. The proposer's change is strictly negative when $\Pr(C\mid h)>0$, since $v_L+\alpha_1\Delta p>0$, and the responder's equals
\[
-(\kappa-\Delta)\Pr(C\mid h)\bigl[\widehat p-\mathbb E[p_h(Z)\mid h,C]\bigr],
\]
also strictly negative under the stated condition. Every equilibrium under $Z$ gives each party at most their maximal payoff, while joint acceptance without information attains $\overline U_i(\varnothing)$, so the strict comparison follows. Finally, full disclosure gives the proposer $(1-\pi_h)v_L=\overline U_1(\varnothing)-\pi_hv_H$.
\end{proof}

\subsection{Action rules and consistent constructions}\label{subsec:s1_construction}

In each of the $J$ rounds the proposer speaks first and the responder replies, and every action may also carry a message; messages are costless finite strings over a fixed finite alphabet, so the set $\mathcal M$ of messages is countable. At each choice-stage decision node the player may end the interaction; a counteroffer of $f$, and a renewed offer of $f$, can only be accepted or answered by ending the interaction. A player's information set consists of their own preference state, the public signal and all preceding public messages and actions; the complete public history $t$ at implementation includes the final action and its message, and the attribution posterior in~\eqref{eq:utility} is computed from it.

Let $p$ be the prior on $H$ of a proposer in state $h$: $p=\pi_h$ without a signal and $p=p_h(z)$ after signal $z$; first let $0<p<1$. All constructions use the same message strategy: at every speech turn and for every message attached to an action, messages are drawn independently of states and history from one distribution $\nu$ with $\nu(m)>0$ for all $m\in\mathcal M$, and later actions do not depend on message content, so messages reveal nothing and no deviation can gain by changing its messages. In all constructions every $f$ offer is accepted, a counteroffer of $x$ is accepted in state $h$ and answered by a new $f$ offer in state $l$, and nobody ends the interaction. Every construction also assigns attribution posterior one to each history at which $x$ is implemented after a zero-probability move by the responder toward $x$, so such a deviation would give the responder $d_\theta-\kappa<0$ if $x$ were implemented, while a failed counteroffer returns to $f$.

\emph{Joint acceptance, $0<p\le c_*$.} The proposer offers $x$ in state $h$ and $f$ in state $l$, and the responder accepts $x$ in both states; on-path optimality is shown in the proof of Proposition~\ref{prop:threshold}, and a counteroffer of $x$ after the on-path $f$ offer, which comes from $l$, would be rejected. \emph{Mixed acceptance, $0<p\le c_*$.} This construction differs only in that the responder in state $L$ accepts with probability $\ell_*(p)$ and otherwise rejects and counteroffers $f$; payoffs are as in Section~\ref{subsec:participation}, and a proposer in state $h$ still obtains $pv_H+(1-p)\ell_*(p)v_L>0$ from offering $x$. \emph{Zero participation, $0<p<1$.} The proposer offers $f$ in both states and the responder accepts; after an off-path $x$ offer, the responder believes the proposer is in state $h$, rejects and counteroffers $f$, so offering $x$ gains nothing. At $p=0$ joint acceptance by $L$ alone applies, and at $p=1$ zero participation, with attribution posteriors identically zero and one, respectively.

\emph{Consistency.} Let $\epsilon\downarrow0$, keep the message distribution at $\nu$ and the positive target probabilities of choice-stage actions as weights, give zero-probability actions the weights below, and normalize at each information set. A zero-probability initial offer by the proposer has weight $\epsilon$ in state $h$ and $\epsilon^2$ in state $l$; a zero-probability move by the responder toward $x$, accepting or counteroffering it, has weight $\epsilon$ in state $H$ and $\epsilon^2$ in state $L$, and a zero-probability rejection of $x$ the reverse; every other zero-probability action has weight $\epsilon^3$. In the limit an initial $x$ offer is attributed to state $h$, an initial $f$ offer reveals $l$ under joint and mixed acceptance and leaves the prior $\rho_\theta$ under zero participation, a zero-probability move toward $x$ is attributed to $H$, and a zero-probability rejection of $x$ to $L$; positive prior weights and the common message distribution do not change these limits, so all constructions are consistent equilibria.

One $\epsilon$ sequence and one message distribution serve all information sets and priors, so the constructions apply jointly across signal values. Action probabilities depend on $p$ only through the threshold comparison and $\ell_*(p)$, so the strategies are Borel measurable in $p$, as Appendix~\ref{app:prescribed} requires for selecting stage equilibria across the forecasts of the dynamic environment.

\section{Proofs for learning from joint activity}\label{app:dynamic}

We use the environment and notation of Section~\ref{sec:dynamic}. Assumption~\ref{ass:region} holds and $0\le\gamma<1$. The condition that opportunities reveal no information about the type or preferences is used only in the forecast recursion and results that require it.

\subsection{Public histories and belief updating}\label{app:histories}

Let $\mathcal I_j$ be the public history on day $j$ after observing whether an opportunity arises and before drawing the day's stage equilibrium: the opportunity records for days $1,\ldots,j$ and the selected stage equilibria, messages, actions and delivered postchoice signals of earlier days, together with $\pi$ when the type is public. It excludes the proposer's private preference $s_j$. Messages are finite strings over a finite alphabet (Appendix~\ref{subsec:s1_construction}), so for each finite $j$ the record $D_j$ other than the public type takes countably many values. With an unknown type, $\mathcal I_j=D_j$, and $P_j=\Pr(\theta_j=H\mid\mathcal I_j)$ is an ordinary conditional probability at every history of positive probability.

Write $\mathbb E_j$ and $\mathbb E_j^+$ for expectations given the prechoice and end-of-day public histories. Thus $P_j=\mathbb E_j[\mathbf 1\{\theta_j=H\}]$, $q_j^{\mathrm{end}}=\mathbb E_j^+[\mathbf 1\{\theta_j=H\}]$, $m_j=\mathbb E_j^+[\pi]$, and $\mathbb E_j[\pi]$ is the prechoice type judgment. To track the day's preference and the type jointly, let $\mu_j^\theta(B)=\Pr(\pi\in B,\theta_j=\theta\mid\mathcal I_j)$. Their sum $\mu_j$ is the type posterior, and $\mu_j^H$ has mass $P_j$; a superscript $+$ marks end-of-day measures. When $\mu^\theta$ has positive mass, $m^\theta$ is its normalized mean, the type mean conditional on $\theta_j=\theta$.

Under prescribed stage behavior, the likelihood of each later public observation that day, given the preceding public history, depends on the type and the responder's preference history only through $\theta_j$: only actions under mixed acceptance and the sign of a signal differ across the two states, and delivery is uninformative given the day's actions. With likelihoods $e_H,e_L$ and probability $q$ of $H$ just before an observation of positive probability, Bayes' rule multiplies $\mu^\theta$ by $e_\theta/[e_Hq+e_L(1-q)]$. Within a day, therefore, $m^H$ and $m^L$ stay fixed wherever defined and only $q$ moves; the type judgment is $qm^H+(1-q)m^L$, with only the defined term when $q\in\{0,1\}$. If the next day's opportunity is uninformative, equation~\eqref{eq:preference_process} gives
\begin{equation}\label{eq:prediction_step}
\begin{aligned}
\mu_{j+1}^H(d\pi)&=\gamma\mu_j^{H,+}(d\pi)+(1-\gamma)\pi\,\mu_j^+(d\pi),\\
\mu_{j+1}^L(d\pi)&=\gamma\mu_j^{L,+}(d\pi)+(1-\gamma)(1-\pi)\,\mu_j^+(d\pi).
\end{aligned}
\end{equation}
Adding gives $\mu_{j+1}=\mu_j^+$, so the type posterior is unchanged overnight, and the mass of the first equation gives~\eqref{eq:forecast}. Since $\pi\in(0,1)$ and $\gamma<1$, every conditional type mean and every prechoice forecast lie in $(0,1)$.

When the type is public, $\mathcal I_j=(\pi,D_j)$, and we apply these updates separately at each type using a point-mass prior. Then $m_j=\pi$ and $P_{j+1}=\gamma q_j^{\mathrm{end}}+(1-\gamma)\pi$. Statements by history hold for every type and every discrete record of positive probability conditional on that type. The type variance and the covariance below are zero.

\subsection{Prescribed stage behavior and the participation threshold}\label{app:prescribed}

Fix the public history on a day with an opportunity, with prescribed stage behavior on earlier days. Consider any equilibrium of the current interaction that puts no weight on future payoffs; stage strategies may depend on the private past, and no exogenous signal about the responder's preference arrives before implementation. To show that, as in Proposition~\ref{prop:threshold}, no such equilibrium implements $x$ when $P_j>c_*$, we verify that the responder's knowledge of earlier preferences does not prevent imitation of a complete strategy. Prescribed past play reveals the proposer's state under joint or mixed acceptance and nothing under zero participation. Since the proposer's preferences are independent across days and of everything else, their private history adds no information about the responder beyond the public history. Today's proposer in state $h$ therefore has prior $P_j$.

Let $a_\theta$ be the probability of implementing $x$ given this history, $s_j=h$ and $\theta_j=\theta$, and $V_\theta$ the responder's expected payoff under the same conditioning, both averaged over earlier private preferences. Fix the proposer's strategy and the attribution posteriors. After any private past of positive probability, a responder currently in state $H$ can ignore their actual past, privately draw an entire earlier preference sequence from its conditional distribution given the public history and current state $L$, and follow the complete stage strategy for that past and state $L$, including all messages, counteroffers and random choices; a responder in state $L$ can likewise imitate $H$. Only finitely many preference sequences exist over finitely many days, so these draws are defined after every private history of positive probability. The proposer sees neither the actual nor the sampled sequence, so imitation reproduces the distribution of public stage histories under the imitated strategy, with the same implementation probability and attribution costs; $H$ earns $\Delta$ more than $L$ whenever $x$ is implemented, and under $s_j=l$ both strategies yield the zero payoff of $f$. Averaging sequential optimality over the proposer's opening message and the responder's actual private past and dividing by $\eta$, as in Appendix~\ref{app:restrictions}, gives $V_H\ge V_L+\Delta a_L$ and $V_L\ge V_H-\Delta a_H$, hence $a_H\ge a_L$. Since the distribution of the private past may depend on the current preference, the imitation draws the entire past under the other current preference rather than comparing two current preferences after the same past.

Along both implementation routes of Appendix~\ref{app:restrictions}, the responder could choose $f$ and obtain zero, and a history reached only by $H$ would give posterior one and payoff $d_H-\kappa<0$. Every implementing history reached under $H$ is therefore also reached under $L$, whose participation requires a posterior no greater than $c_*$; summing over the countably many such histories gives $(1-P_j)a_L\ge[(1-c_*)/c_*]P_ja_H$. With $a_H\ge a_L$, this fails at $P_j>c_*$ unless $a_H=a_L=0$, so private preference history leaves the participation threshold unchanged. For the selected equilibria, Appendix~\ref{subsec:s1_construction} constructs zero participation at every prior $0<p<1$ and joint and mixed acceptance at $p\le c_*$, sharing a full-support message distribution and one sequence of fully mixed perturbations. These strategies do not use private pasts and remain sequentially optimal after every private history of positive probability, and a public lottery with the probabilities of Section~\ref{subsec:dynamic_environment} gives prescribed stage behavior. This behavior depends on the forecast only through $\mathbf 1\{P\le c_*\}$ and $\ell_*(P)$, so it is measurable in the forecast and the draw (Appendix~\ref{subsec:s1_construction}); since the transition and signal probabilities are also Borel, the type-by-type quantities of Appendix~\ref{app:histories} are measurable in $\pi$ when the type is public and, with bounded payoffs over a finite window, can be integrated over the type distribution. The construction requires strategy and belief limits at each prior, not convergence uniform across priors.

\subsection{The value of feedback}\label{app:feedback_value}

\begin{proof}[Proof of Proposition~\ref{prop:feedback_value}]
Fix a regime and day $j$, omitting the regime superscript. Payoffs are zero without an opportunity or with the proposer in state $l$. At $O_j=1,s_j=h$, condition (a) makes the proposer's preference independent of $(\theta_j,\mathcal I_j)$, so the probability of $H$ remains $P_j$, and prescribed behavior selects zero participation when $P_j>c_*$. The daily payoffs of Section~\ref{subsec:exclusion_payoffs} and the selection probabilities thus give, at every prechoice history of positive probability,
\begin{equation}\label{eq:daily_payoff}
\mathbb E[U_{i,j}\mid\mathcal I_j,s_j]
=G_j\phi_i(P_j)\mathbf 1\{P_j\le c_*\},\qquad i\in\{1,2\},
\end{equation}
with attribution costs fixed at the implementation posterior. Writing the indicator as $1-\mathbf 1\{P_j>c_*\}$ separates the blocked term from $\mathbb E[G_j\phi_i(P_j)]$. Since $\phi_i$ is affine, the latter equals $\phi_i(0)\mathbb E[G_j]+[\phi_i(1)-\phi_i(0)]\mathbb E[G_jP_j]$. Independence of $s_j$, public measurability of $O_j$ and condition (b) give $\mathbb E[G_jP_j]=\lambda\eta\,\mathbb E[O_jP_j]=\lambda\eta\,\mathbb E[O_j\mathbf 1\{\theta_j=H\}]=\mathbb E[G_j\mathbf 1\{\theta_j=H\}]$. This also holds with informative opportunities if their observation enters $\mathcal I_j$. Hence $\mathbb E[G_j\phi_i(P_j)]$ depends only on the joint distribution of $(O_j,s_j,\theta_j)$, which is the same under both regimes by (a), so it cancels day by day when expectations of~\eqref{eq:daily_payoff} are summed over $\mathcal W$ and subtracted, leaving~\eqref{eq:feedback_value}. With a public type, apply the argument at each type and then integrate over its distribution (Appendices~\ref{app:histories} and~\ref{app:prescribed}).
\end{proof}

The same proof applies to any two arrangements that differ only in public information and share these exogenous processes and prescribed stage behavior, including the no-report benchmarks with the type public from the start and with the type unknown.

\subsection{Who can gain}\label{app:who_gains}

\begin{lemma}[The no-report benchmark]\label{lem:no_report}
Suppose the type is unknown, opportunities carry no information about the type or preferences, $\bar\pi\le c_*$, and there are no postchoice signals. Under prescribed stage behavior, for any $\lambda,\chi\in[0,1]$, every day $j$ and every public history of positive probability satisfy $P_j\le c_*$ and $m_j\le c_*$.
\end{lemma}

\begin{proof}
Without postchoice signals, only acceptance or rejection under mixed acceptance carries preference information. Without either, $q_j^{\mathrm{end}}=P_j$ and $m_j=\mathbb E_j[\pi]=m_{j-1}$. Acceptance, with probability $P_j/c_*$ and likelihoods one and $\ell_*(P_j)$, gives $q_j^{\mathrm{end}}=c_*$ and $m_j=m^A\equiv c_*m^H+(1-c_*)m^L$; rejection, possible only if $P_j<c_*$, gives $q_j^{\mathrm{end}}=0$ and $m_j=m^L$, with $m^H,m^L$ taken before the choice. If $P_j\le c_*$ and $m_j\le c_*$, then $q_j^{\mathrm{end}}\in\{P_j,c_*,0\}$ and~\eqref{eq:forecast} give $P_{j+1}\le c_*$. Since $P_1=\bar\pi$, it suffices to show that $m_j\le c_*$ whenever $P_j\le c_*$ and $m_0,\ldots,m_{j-1}$ are at most $c_*$, with $m_0=\bar\pi$.

Define finite measures $\psi_j$ by letting $\psi_1$ be the initial type distribution, multiplying $\psi_j$ within the day by the factors applied to the $L$-component, and setting $\psi_{j+1}=\gamma\psi_j^++(1-\gamma)\mu_j^+$ overnight. Then $\mu_j^L(d\pi)=(1-\pi)\psi_j(d\pi)$ for every $j$: this holds on day 1, both sides are multiplied by the same constants within a day, and~\eqref{eq:prediction_step} gives $\mu_{j+1}^L=\gamma(1-\pi)\psi_j^++(1-\gamma)(1-\pi)\mu_j^+$. By construction, $\psi_j$ is a positive combination of the initial distribution and earlier end-of-day type posteriors, and by the induction hypothesis its mean $\bar\psi_j$ is at most $c_*$; let $\tilde\psi_j$ be $\psi_j$ normalized.

\emph{Rejection.} By Jensen's inequality, $m^L=\int\pi(1-\pi)\,d\tilde\psi_j/(1-\bar\psi_j)\le(\bar\psi_j-\bar\psi_j^2)/(1-\bar\psi_j)=\bar\psi_j\le c_*$.

\emph{Acceptance.} The identity $(P_j/c_*)m^A+(1-P_j/c_*)m^L=\mathbb E_j[\pi]$ shows that $m^A\le c_*$ if and only if
\begin{equation}\label{eq:acceptance_condition}
c_*(\mathbb E_j[\pi]-P_j)\le(c_*-P_j)\,m^L,
\end{equation}
whose right side is nonnegative. If no acceptance or rejection under mixed acceptance has occurred, $(\pi,\theta_j)$ keeps its initial joint distribution and $\mathbb E_j[\pi]=P_j=\bar\pi$. Otherwise let $j'<j$ be the last day with such an outcome and $\upsilon=1-\gamma^{\,j-j'}\in(0,1]$. If day $j'$ ended in acceptance, iterating~\eqref{eq:prediction_step} gives $P_j=(1-\upsilon)c_*+\upsilon m_{j'}\ge m_{j'}=\mathbb E_j[\pi]$. If it ended in rejection, then $\mu_{j'}^{H,+}=0$; with $y=\int\pi(1-\pi)\,\mu_{j'}^+(d\pi)$, iterating gives $\mu_j^H=\upsilon\pi\mu_{j'}^+$ and $\mu_j^L=[(1-\upsilon)+\upsilon(1-\pi)]\mu_{j'}^+$, so $P_j=\upsilon m_{j'}$, $\mathbb E_j[\pi]=m_{j'}$ and $m^L=[(1-\upsilon)m_{j'}+\upsilon y]/(1-\upsilon m_{j'})$, and~\eqref{eq:acceptance_condition} becomes $(1-c_*)(1-\upsilon)m_{j'}^2\le(c_*-\upsilon m_{j'})y$. As $c_*-\upsilon m_{j'}\ge(1-\upsilon)c_*$, it suffices that $(1-c_*)m_{j'}^2\le c_*y$. The rejection only normalized the $L$-component, so $\mu_{j'}^+=(1-\pi)\tilde\psi_{j'}/(1-\bar\psi_{j'})$, $m_{j'}=\int\pi(1-\pi)\,d\tilde\psi_{j'}/(1-\bar\psi_{j'})$ and $y=\int\pi(1-\pi)^2\,d\tilde\psi_{j'}/(1-\bar\psi_{j'})$. Cauchy--Schwarz applied to $\pi(1-\pi)=\sqrt\pi\cdot[\sqrt\pi\,(1-\pi)]$ gives $m_{j'}^2\le[\bar\psi_{j'}/(1-\bar\psi_{j'})]y\le[c_*/(1-c_*)]y$.
\end{proof}

\begin{corollary}[Who can gain when the type is known from the start]\label{cor:known_type}
Suppose the conditions of Proposition~\ref{prop:feedback_value} hold, the type $\pi$ is public from the first day, and, conditional on the type, opportunities provide no information about current or past preferences. For each public type, statements by history use the type-by-type updates of Appendix~\ref{app:histories}.
\begin{enumerate}
\item[(i)] In the no-report benchmark of the same environment, a qualified opportunity is blocked if and only if $\pi>c_*$, in which case $P_j^\varnothing=\pi$ on every day. Hence, for every feedback regime $Z$, finite window $\mathcal W$ and $i\in\{1,2\}$,
\begin{equation}\label{eq:known_gain}
\mathcal U_i(Z)-\mathcal U_i(\varnothing)
=\mathbb E\Bigl[\sum_{j\in\mathcal W}G_j
\bigl(\mathbf 1\{\pi>c_*\}\phi_i(\pi)
-\mathbf 1\{P_j^Z>c_*\}\phi_i(P_j^Z)\bigr)\Bigr].
\end{equation}
\item[(ii)] Under feedback after joint activity, pairs with $\pi>c_*$ remain blocked at $P_j^Z=\pi$; every other blocked opportunity has a forecast no greater than $\gamma+(1-\gamma)c_*$. Experience therefore does not raise the proposer's expected payoff. It does not raise the pair's combined expected payoff if $v_H\ge\kappa-d_H$, or the responder's expected payoff if $\chi>\tau$ or $\gamma+(1-\gamma)c_*\le\widehat p$, where $\tau=(\kappa-d_H)/(\kappa-d_L)$.
\end{enumerate}
\end{corollary}

\begin{proof}
With a public type, $m_j=\pi$. In the benchmark, pairs with $\pi>c_*$ are blocked from the first day and nothing reveals their preferences. For $\pi\le c_*$, each day ends with $q_j^{\mathrm{end}}\in\{P_j,c_*,0\}$ and~\eqref{eq:forecast} keeps the forecast at or below $c_*$. Substituting into~\eqref{eq:feedback_value} gives (i). For (ii), experience gives no signal to pairs with $\pi>c_*$ either, so their terms in~\eqref{eq:known_gain} cancel. For $\pi\le c_*$, $q_j^{\mathrm{end}}\le1$ bounds the next forecast by $\gamma+(1-\gamma)c_*$. The endpoint values in the proof of Corollary~\ref{cor:who_gains} determine the signs of the remaining weights. If $\gamma+(1-\gamma)c_*\le\widehat p$, the blocked forecasts lie in $(c_*,\widehat p]$, where $\phi_2\ge0$ for any $\chi$.
\end{proof}

Outside feedback also reaches pairs with $\pi>c_*$ and may lift the benchmark's own exclusions, so both terms in~\eqref{eq:known_gain} remain, and the responder may gain even if every blocked opportunity lies in the lower band. With $\bar\pi\le c_*$ and uninformative opportunities, revealing the type in the no-report benchmark blocks precisely the pairs with $\pi>c_*$, each weighted by $G_j\phi_i(\pi)$ (Appendix~\ref{app:feedback_value}, Lemma~\ref{lem:no_report} and (i)), so it helps the responder only through pairs with $\pi>\widehat p$.

\subsection{Exclusion below the payoff threshold}\label{app:ceiling}

On a day when $x$ is implemented, $\mathbb E_j^{\mathrm{imp}}$ denotes expectation given the public history after the stage actions and before any signal, $m_j^{\mathrm{imp}}=\mathbb E_j^{\mathrm{imp}}[\pi]$, and the posterior on the day's preference is $q_j^{\mathrm{imp}}$.

\begin{lemma}[Type and today's preference]\label{lem:covariance}
Suppose the conditions of Proposition~\ref{prop:feedback_value} hold and opportunities carry no information about the type or preferences. Under the feedback regime $Z$ and the no-report benchmark of the same environment, on every day $j$ and after every public history of positive probability, the joint posterior before the choice and after each subsequent public observation that day satisfies
\[
\operatorname{Cov}(\pi,\mathbf 1\{\theta_j=H\})\ge0.
\]
Thus a public observation moves the type judgment weakly in the same direction as the probability of $\theta_j=H$: one that does not raise this probability does not raise the type judgment, and one that raises it does not lower the type judgment.
\end{lemma}

\begin{proof}
For $0<q<1$, $\operatorname{Cov}=q(1-q)(m^H-m^L)$, and $\operatorname{Cov}=0$ if $q\in\{0,1\}$. Observations within a day leave $m^H$ and $m^L$ unchanged wherever defined and move only $q$ (Appendix~\ref{app:histories}), so they preserve the sign of the covariance and move $qm^H+(1-q)m^L$ weakly with $q$. Before the first choice, the covariance is the variance of the type, and~\eqref{eq:prediction_step} gives $\operatorname{Cov}_{j+1}=\gamma\operatorname{Cov}_j^++(1-\gamma)\operatorname{Var}_j^+(\pi)\ge0$ between days. Induction over days and observations completes the proof.
\end{proof}

\begin{proof}[Proof of Proposition~\ref{cor:ceiling}]
Write $P_j=P_j^Z$ and $\overline P=\overline P(c_*,\Lambda,\gamma)$; since $0<k_L\le k_H\le1$, $\Lambda\ge1$ is finite. Rearranging~\eqref{eq:ceiling}, $\overline P\le1$ if and only if $c_*\le1-\gamma$, and $c_*/(1-\gamma)\le\overline P$ if and only if $(\Lambda-1)(1-\gamma-c_*)\ge0$. Under the proposition's conditions, therefore,
\begin{equation}\label{eq:ceiling_constants}
\frac{\Lambda c_*}{1+c_*(\Lambda-1)}\le\overline P,\qquad
c_*\le\frac{c_*}{1-\gamma}\le\overline P\le1.
\end{equation}
Because tomorrow's forecast depends on both today's preference posterior and the type judgment, the induction controls both: at every prechoice history of positive probability,
\begin{equation}\label{eq:ceiling_induction}
P_j\le\overline P,\qquad \mathbb E_j[\pi]\le\overline P.
\end{equation}
This holds on day 1, where $P_1=\mathbb E_1[\pi]=\bar\pi\le c_*$. Given~\eqref{eq:ceiling_induction} on day $j$, it suffices by~\eqref{eq:forecast} and~\eqref{eq:prediction_step} to show $q_j^{\mathrm{end}}\le\overline P$ and $m_j\le\overline P$, since $P_{j+1}$ averages the two and $\mathbb E_{j+1}[\pi]=m_j$. Note that $(1-\gamma)\mathbb E_j[\pi]\le P_j$ on every day, by~\eqref{eq:forecast} after day 1. Only observations of positive probability matter.

\emph{No implementation.} Without a qualified opportunity or with zero participation, which includes every day with $P_j>c_*$, nothing reveals the responder's preference and experience supplies no signal, so $q_j^{\mathrm{end}}=P_j$ and $m_j=\mathbb E_j[\pi]$. After a rejection under mixed acceptance, $q_j^{\mathrm{end}}=0$ and Lemma~\ref{lem:covariance} gives $m_j\le\mathbb E_j[\pi]$.

\emph{Before a signal.} Implementation requires $P_j\le c_*$. Joint acceptance reveals nothing, so $q_j^{\mathrm{imp}}=P_j$ and $m_j^{\mathrm{imp}}=\mathbb E_j[\pi]$. Acceptance under mixed acceptance, with probability $P_j/c_*$, multiplies the $H$-component by $c_*/P_j$ and the $L$-component by $c_*\ell_*(P_j)/P_j\le c_*/P_j$. Thus $q_j^{\mathrm{imp}}=c_*$ and $m_j^{\mathrm{imp}}\le(c_*/P_j)\mathbb E_j[\pi]$. Since $\mathbb E_j[\pi]\le P_j/(1-\gamma)$, both forms of acceptance give
\begin{equation}\label{eq:report_lock}
q_j^{\mathrm{imp}}\le c_*,\qquad
m_j^{\mathrm{imp}}\le\frac{q_j^{\mathrm{imp}}}{1-\gamma}\le\frac{c_*}{1-\gamma}.
\end{equation}
Experience signals arrive only after implementation, so~\eqref{eq:report_lock} holds before every delivered signal. This is where experience enters the proof: outside feedback also arrives on days with $P_j>c_*$, when the posterior before the signal already exceeds $c_*$.

\emph{After a signal.} A positive signal multiplies the two components by $k_H$ and $k_L$, so
\[
\begin{aligned}
q_j^{\mathrm{end}}&=\frac{\Lambda q_j^{\mathrm{imp}}}{1+(\Lambda-1)q_j^{\mathrm{imp}}},\\
m_j&=\frac{(\Lambda-1)\mathbb E_j^{\mathrm{imp}}[\pi\mathbf 1\{\theta_j=H\}]+m_j^{\mathrm{imp}}}{1+(\Lambda-1)q_j^{\mathrm{imp}}}
\le\frac{q_j^{\mathrm{imp}}[\Lambda+\gamma/(1-\gamma)]}{1+(\Lambda-1)q_j^{\mathrm{imp}}},
\end{aligned}
\]
using $\mathbb E_j^{\mathrm{imp}}[\pi\mathbf 1\{\theta_j=H\}]\le q_j^{\mathrm{imp}}$ and~\eqref{eq:report_lock}. Both bounds increase with $q_j^{\mathrm{imp}}\le c_*$, so~\eqref{eq:ceiling_constants} puts them at or below $\overline P$. An undelivered signal changes nothing. A negative signal, with likelihoods $1-k_H\le1-k_L$, raises neither $q$ nor, by Lemma~\ref{lem:covariance}, the type judgment. After either outcome, $q_j^{\mathrm{end}}\le c_*$ and $m_j\le c_*/(1-\gamma)$, both at most $\overline P$ by~\eqref{eq:ceiling_constants}. This completes the induction for (i); no branch depends on the probability with which it is selected.

For (ii), if $\overline P\le\widehat p$, no forecast reaches the upper band, and Corollary~\ref{cor:who_gains}(i)--(ii) gives both weak inequalities. If $\overline P<\widehat p$, then on $[c_*,\overline P]$, for every $\chi\in[0,1]$,
\[
\begin{aligned}
\phi_1(P)&\ge v_L>0,\\
\phi_2(P)&=(1-\chi)(\kappa-\Delta)(\widehat p-P)+\chi\Delta P\\
&\ge(1-\chi)(\kappa-\Delta)(\widehat p-\overline P)+\chi\Delta c_*>0,
\end{aligned}
\]
so~\eqref{eq:unknown_gain} gives a strictly negative difference for each party exactly when the expected blocked opportunity weight is positive, and zero differences when it is zero.
\end{proof}

\subsection{The path of an exclusion spell}\label{app:spell_path}

\begin{proof}[Proof of Corollary~\ref{cor:spell_path}]
On an excluded day under experience, prescribed behavior neither implements $x$ nor reveals a preference, so no signal arrives, and opportunities are uninformative; hence $q_j^{\mathrm{end}}=P_j$ and $m_j=\mathbb E_j[\pi]$, with $\mathbb E_j[\pi]=\pi$ for each type when the type is public. Iterating~\eqref{eq:forecast} from $\mathbb E_{j_0}[\pi]=m$ while days $j_0,\ldots,j-1$ belong to the spell gives $\mathbb E_j[\pi]=m$ and~\eqref{eq:spell_path}; day $j$ need not be excluded, so the formula also gives the forecast on the first feasible day. For $m<p<P_{j_0}$, $P_j>p$ if and only if $j-j_0<\ln[(P_{j_0}-m)/(p-m)]/\ln(1/\gamma)$, a positive bound below which lie exactly $\lceil\ln[(P_{j_0}-m)/(p-m)]/\ln(1/\gamma)\rceil$ nonnegative integers, counted in calendar days from $j_0$.

Taking $p=c_*$ gives (i). For (ii), $P_j-c_*=(1-\gamma^{\,j-j_0})(m-c_*)+\gamma^{\,j-j_0}(P_{j_0}-c_*)$, whose first term is nonnegative and second strictly positive on every finite day, including when $m=c_*$. Taking $p=\widehat p$ gives the count in (iii); these days have $P_j>\widehat p>c_*$ and so belong to the spell, and a forecast equal to $\widehat p$ lies outside the upper band; if also $m\ge c_*$, later forecasts lie in $(c_*,\widehat p]$ by (ii). In (iv), each $P_j$ is a convex combination of $m$ and $P_{j_0}$. For (v), $P_1=\bar\pi\le c_*$ gives $j_0\ge2$ and $P_{j_0-1}\le c_*$. If day $j_0-1$ has no positive signal, then $q_{j_0-1}^{\mathrm{end}}\le c_*$, since it equals $P_{j_0-1}$ without implementation, zero after a rejection, and at most $q_{j_0-1}^{\mathrm{imp}}\le c_*$ after an undelivered or negative signal by~\eqref{eq:report_lock}. Also $m\le c_*/(1-\gamma)$: without implementation or after a rejection, Lemma~\ref{lem:covariance} gives $m\le\mathbb E_{j_0-1}[\pi]\le P_{j_0-1}/(1-\gamma)$, and after implementation $m\le m_{j_0-1}^{\mathrm{imp}}$, bounded by~\eqref{eq:report_lock}. If $m\le c_*$ as well,~\eqref{eq:forecast} gives $P_{j_0}\le c_*$, contradicting the definition of $j_0$; so without a positive signal, $c_*<m\le c_*/(1-\gamma)$.
\end{proof}

\section{Proofs for the forward-looking economy}\label{app:forward}

The economy and maintained ordering $p_{\mathrm{lo}}<c_*<\pi<c^H<p_{\mathrm{hi}}$ are given in Section~\ref{subsec:fwd_economy}. Exact reports reveal today's preference after every implementation, including implementation following a zero-probability choice.

Interpretations of zero-probability choices are restricted forecast by forecast. For such a choice at forecast $p$ and an interpretation $q\in[0,1]$, the probability then assigned to $H$, let $\mathrm{gain}_\theta(q)$ be preference $\theta$'s gain relative to its value under the profile, with the proposer responding optimally to $q$ and subsequent play following the profile; the optimal response is unique wherever used below. Let $\mathcal R_\theta^{\ge}$ and $\mathcal R_\theta^{>}$ be the sets of interpretations giving $\mathrm{gain}_\theta\ge0$ and $\mathrm{gain}_\theta>0$. \emph{Local screening} requires $q=1$ if $\mathcal R_L^{\ge}\subseteq\mathcal R_H^{>}$ and $\mathcal R_H^{\ge}\not\subseteq\mathcal R_L^{>}$, and $q=0$ in the symmetric case; otherwise the interpretation is unrestricted. For value functions and tomorrow's forecast $u$, the responder's expected values tomorrow, conditional on today's preference, are
\[
\overline{\mathcal V}_H(u)=p_{\mathrm{hi}}\mathcal V_H(u)+(1-p_{\mathrm{hi}})\mathcal V_L(u),\qquad
\overline{\mathcal V}_L(u)=p_{\mathrm{lo}}\mathcal V_H(u)+(1-p_{\mathrm{lo}})\mathcal V_L(u).
\]

\begin{lemma}[Payoffs from participation]\label{lem:fwd_accounting}
For every strategy profile, public signal before acceptance and initial forecast, $\mathcal V_1=v_HN_H+v_LN_L$, $\mathcal V_2=(d_H-\kappa)N_H+d_LN_L$, and hence $\mathcal V_1+\mathcal V_2=(v_H+d_H-\kappa)N_H+(v_L+d_L)N_L$.
\end{lemma}

\begin{proof}[Proof of Lemma~\ref{lem:fwd_accounting}]
Let $J_j$ indicate implementation. Conditional on public history after acceptance, including any signal before acceptance, $q_j=\Pr(\theta_j=H)$, and implementation depends only on that history and independent randomization. Thus $\mathbb E[q_jJ_j]=\mathbb E[\mathbf1\{\theta_j=H\}J_j]$: in expectation, attribution costs amount to $\kappa$ per implemented $H$ day. Discounting and summing the two payoffs gives the three identities.
\end{proof}

\paragraph{The revealed chain.} If every day ends with today's preference public, then $P_{j+1}=p_{\mathrm{hi}}$ when $\theta_j=H$ and $p_{\mathrm{lo}}$ otherwise, whatever the partners do. After the initial day the forecast follows an exogenous chain on $\{p_{\mathrm{lo}},p_{\mathrm{hi}}\}$. A quantity with daily flow $y$ satisfies, at every forecast,
\begin{equation}\label{eq:chain}
X(p)=y(p)+\delta\bigl[pX(p_{\mathrm{hi}})+(1-p)X(p_{\mathrm{lo}})\bigr].
\end{equation}
At $p_{\mathrm{hi}}$ and $p_{\mathrm{lo}}$ this is a linear system with determinant $(1-\delta)(1-\delta\gamma)>0$ and an entrywise positive inverse. Let $y_\theta(p)$ be preference $\theta$'s expected payoff on a day with forecast $p$. Applying~\eqref{eq:chain} to $\mathcal V_2$ with flow $py_H+(1-p)y_L$, write the endpoint values as $\mathcal C_H=\mathcal V_2(p_{\mathrm{hi}})=\overline{\mathcal V}_H(p_{\mathrm{hi}})$ and $\mathcal C_L=\mathcal V_2(p_{\mathrm{lo}})=\overline{\mathcal V}_L(p_{\mathrm{lo}})$, the responder's expected values from tomorrow on once today's preference $H$ or $L$ is public. Then $\mathcal V_\theta(p)=y_\theta(p)+\delta\mathcal C_\theta$. The proposer's value and the counts $N_\theta$ have flows $v_HM_H(p)+v_LM_L(p)$ and $M_\theta(p)$, respectively.

\subsection{The evidence-seeking equilibrium}\label{app:evidence}

\begin{proof}[Proof of Proposition~\ref{prop:evidence}]
\emph{Values.} On the path, a day with acceptance ends with a participation report; declines occur only at $p>c^H$ and only under preference $H$, revealing it. Thus every day ends with today's preference public, independently of $\omega$. The acceptance posterior is $\min\{p,c^H\}$, the implementation probability is $\iota(p)=\min\{1,(1-p)/(1-c^H)\}$, and $M_H(p)=\min\{p,c^H(1-p)/(1-c^H)\}$. Preference $H$ receives $d_H-\kappa\min\{p,c^H\}$ that day, zero at $p>c^H$ whether it accepts or not. Hence $y_\theta(p)=d_\theta-\kappa\min\{p,c^H\}$ and, for every forecast $u$, the value to $L$ of being identified rather than entering tomorrow at forecast $u$ is $\mathcal C_L-\overline{\mathcal V}_L(u)=\kappa[\min\{u,c^H\}-p_{\mathrm{lo}}]$; in particular, $\mathcal C_L-\overline{\mathcal V}_L(p_{\mathrm{hi}})=\mathcal D$. The flow of $\mathcal V_2$ is $d_L+p\Delta-\kappa\min\{p,c^H\}$. Subtracting~\eqref{eq:chain} at $p_{\mathrm{lo}}$ from that at $p_{\mathrm{hi}}$ gives $(1-\delta\gamma)(\mathcal C_H-\mathcal C_L)=\gamma\Delta-\mathcal D$, so at every forecast $\mathcal V_H-\mathcal V_L=(\Delta-\delta\mathcal D)/(1-\delta\gamma)$.

\emph{Responder.} A decline is read as $H$, on the path at $p>c^H$ and off it at $p\le c^H$. It gives preference $H$ the value $\delta\mathcal C_H$, whether or not an outside report arrives. Preference $L$ receives $\delta\{\omega\mathcal C_L+(1-\omega)\overline{\mathcal V}_L(p_{\mathrm{hi}})\}$, since an outside report identifies the day as $L$. The respective gains from acceptance are therefore $d_H-\kappa\min\{p,c^H\}\ge0$, zero exactly at $p\ge c^H$, where $H$ can mix, and $d_L-\kappa\min\{p,c^H\}+(1-\omega)\delta\mathcal D\ge(1-\omega)\delta\mathcal D-\Delta$, with equality at $p\ge c^H$. Thus $L$ accepts everywhere when $\omega\le\omega^*$ and strictly prefers to decline at $p_{\mathrm{hi}}$ when $\omega>\omega^*$. An off-path decline at $p\le c^H$, read as $q$, gives $\mathrm{gain}_\theta(q)=\delta\{\omega\mathcal C_\theta+(1-\omega)\overline{\mathcal V}_\theta(p_{\mathrm{lo}}+\gamma q)\}-\mathcal V_\theta(p)$. Since $\overline{\mathcal V}_H(u)-\overline{\mathcal V}_L(u)=\gamma[\mathcal V_H(u)-\mathcal V_L(u)]$ and $\delta(\mathcal C_H-\mathcal C_L)=\mathcal V_H-\mathcal V_L-\Delta$, at every interpretation $\mathrm{gain}_H(q)-\mathrm{gain}_L(q)=(1-\omega)\delta\mathcal D-\Delta$. This difference is nonnegative when $\omega\le\omega^*$. If positive, $\mathcal R_L^{\ge}\subseteq\mathcal R_H^{>}$, so local screening either attributes the decline to $H$ or leaves it unrestricted. If zero, the gain functions coincide and the interpretation is unrestricted. Under the profile's reading $q=1$, $\mathrm{gain}_L(1)\le\mathrm{gain}_H(1)=-(d_H-\kappa p)\le0$, so neither preference gains by declining.

\emph{Proposer.} Implementing an acceptance read as $q$ rather than declining to implement changes the discounted counts of implemented days and implemented $H$ days by $G(q)$ and $G_H(q)$. By Lemma~\ref{lem:fwd_accounting}, the proposer's gain is $v_LG(q)+(v_H-v_L)G_H(q)$. Without implementation, an outside report arrives with probability $\omega$; otherwise tomorrow's forecast is $u=p_{\mathrm{lo}}+\gamma q=(1-q)p_{\mathrm{lo}}+qp_{\mathrm{hi}}$, the mean forecast after implementation. By~\eqref{eq:chain}, each count equals its flow plus a term affine in the forecast, so with $t=\delta(1-\omega)$ and $\iota=M_H+M_L$ the two choices differ only through today's and tomorrow's flows:
\[
\begin{aligned}
G(q)&=1+t\bigl[(1-q)\iota(p_{\mathrm{lo}})+q\iota(p_{\mathrm{hi}})-\iota(u)\bigr],\\
G_H(q)&=q+t\bigl[(1-q)M_H(p_{\mathrm{lo}})+qM_H(p_{\mathrm{hi}})-M_H(u)\bigr].
\end{aligned}
\]
Since $\iota\le1=\iota(p_{\mathrm{lo}})$, $G(q)\ge1-tq[1-\iota(p_{\mathrm{hi}})]>0$. Since $M_H(u)\le u$ at every forecast, with equality at $p_{\mathrm{lo}}$, $G_H(q)\ge q[1-t(p_{\mathrm{hi}}-M_H(p_{\mathrm{hi}}))]\ge0$. With $v_L>0$, implementation is strictly optimal at every interpretation. Bounded payoffs and $\delta<1$ give the result by the one-shot deviation principle.
\end{proof}

\subsection{Exact outside reports}\label{app:full}

\begin{proof}[Proof of Proposition~\ref{prop:full}]
At $\omega=1$, including the initial day, tomorrow's forecast is $p_{\mathrm{hi}}$ under $H$ and $p_{\mathrm{lo}}$ under $L$. Markov continuation therefore gives each partner the same value after every choice. The proposer gains $(1-q)v_L+qv_H>0$ from implementing an acceptance read as $q$, so every acceptance is implemented. After signal realization $z$, preference $\theta$ accepts only if $d_\theta-\kappa q(z)\ge0$, where $q(z)$ is the acceptance posterior. If $L$ never accepted after $z$ but $H$ accepted with positive probability, then $q(z)=1$ and $H$ would receive $d_H-\kappa<0$. Hence whenever acceptance has positive probability, $L$ accepts with positive probability and $q(z)\le c_*$. The probabilities $m_H(z)$ and $m_L(z)$ of the respective preferences and acceptance satisfy $m_H(z)\le c_*m_L(z)/(1-c_*)$. Integrating over $z$ gives the stated bound; also $M_L(p)\le1-p$ and $M_H(p)\le p$.

By Lemma~\ref{lem:fwd_accounting}, a day with forecast $p$ contributes $[w_1v_H-w_2(\kappa-d_H)]M_H+(w_1v_L+w_2d_L)M_L$ to $w_1\mathcal V_1+w_2\mathcal V_2$. Since $w_1v_L+w_2d_L\ge0$, subject to the three bounds this is maximized by $M_L=1-p$ and $M_H=\min\{p,c_*(1-p)/(1-c_*)\}$ if $w_1v_H\ge w_2(\kappa-d_H)$, and by $M_L=1-p$, $M_H=0$ otherwise. The distribution of $P_j$ is the same in every Markov equilibrium, so the discounted sum of these daily maxima bounds value from every initial forecast.

The two signal rules attain these daily maxima at every $p\in[0,1]$. Under the pooling signal, both preferences accept at posterior $p$ when $p\le c_*$, gaining $d_\theta-\kappa p\ge0$. When $p>c_*$, the pooled message has posterior $c_*$, giving $L$ zero and $H$ a gain of $\Delta$, with $M_L=1-p$ and $M_H=c_*(1-p)/(1-c_*)<p$. At every interpretation, a decline gives each preference the negative of its gain from acceptance. Below posterior $c_*$, both sets $\mathcal R_\theta^{\ge}$ are empty and the interpretation is unrestricted. At posterior $c_*$, $\mathcal R_H^{\ge}=\varnothing$ whereas $\mathcal R_L^{\ge}=[0,1]\not\subseteq\mathcal R_H^{>}=\varnothing$, so a decline is read as $L$. After a message revealing $H$, an acceptance read as $q$ gives gains $d_\theta-\kappa q$, so $\mathcal R_L^{\ge}=[0,c_*]\subseteq[0,c^H)=\mathcal R_H^{>}$ while $\mathcal R_H^{\ge}=[0,c^H]\not\subseteq[0,c_*)=\mathcal R_L^{>}$. Acceptance is read as $H$, making both gains negative. Under disclosure before acceptance, $L$ receives $d_L>0$ at posterior zero, whereas $H$ would receive $d_H-\kappa<0$ from acceptance.

Without a signal, suppose acceptance has positive probability at forecast $p$. Then $L$ accepts with positive probability at some $q\le c_*<c^H$, where $H$ gains at least $\Delta>0$. Thus $H$ accepts for sure and $q\ge p$. Activity is impossible at $p>c_*$; at $p\le c_*$ every Markov equilibrium has either no activity or $M_H=p$, $M_L\le1-p$. WT has $M_H=p$ and $M_L=1-p$ there. Given $M_H=p$, each weighted daily payoff is nondecreasing in $M_L$ and nonnegative at $M_L=1-p$, so WT maximizes it. With optimal implementation and screening as above, these choices form a Markov equilibrium at $\omega=1$ for any participation report, with off-path acceptance read as $H$ and decline as $L$.
\end{proof}

\begin{proof}[Proof of Corollary~\ref{cor:gap}]
ES and the pooling signal at $\omega=1$ both end every day with today's preference public and implement every $L$ day. They therefore share the exogenous forecast chain and satisfy $N_L^{\mathrm{ES}}=N_L^\star$ from every initial forecast. Their daily probabilities of an implemented $H$ day coincide at $p\le c_*$, differ by $(p-c_*)/(1-c_*)$ at $c_*<p\le c^H$, and differ by $(1-p)[c^H/(1-c^H)-c_*/(1-c_*)]$ at $p>c^H$. From $p_{\mathrm{lo}}$, only $p_{\mathrm{lo}}$ and $p_{\mathrm{hi}}$ occur, with discounted expected time $n_{\mathrm{hi}}$ at the latter, giving~\eqref{eq:gap}. From any $p\in I$, the gap is today's difference plus $\delta$ times a positive combination of the strictly positive gaps from $p_{\mathrm{hi}}$ and $p_{\mathrm{lo}}$. The value comparison follows from Lemma~\ref{lem:fwd_accounting} and Proposition~\ref{prop:full}; if $v_H+d_H<\kappa$, disclosure attains the joint supremum, implementing every $L$ day and no $H$ days.
\end{proof}

\subsection{The comparison class and history dependence}\label{app:fwd_class}

Table~\ref{tab:forward} compares best values within monotone Markov equilibria, a class that contains ES and WT. In ES, $\mathcal V_\theta(p)=d_\theta-\kappa\min\{p,c^H\}+\delta\mathcal C_\theta$. In WT at $\omega=1$, $\mathcal V_\theta(p)=d_\theta-\kappa p+\delta\mathcal C_\theta$ for $p\le c_*$ and $\delta\mathcal C_\theta$ above, with $d_\theta-\kappa c_*\ge0$. Both are monotone. At $\omega=1$, the values of WT, $\star$ and $F$ do not depend on the participation report, since outside reports determine tomorrow's forecast. For every report and every $\omega\in[0,1]$, the zero-activity profile, with acceptance read as $H$, is a monotone Markov equilibrium, so the class is never empty. All continuation values are zero and the proposer implements every acceptance. An acceptance read as $q$ gives gains $d_\theta-\kappa q$; the screening argument in the proof of Proposition~\ref{prop:full} attributes it to $H$, making both gains negative. Declines are on the path.

\paragraph{History dependence.} With exact reports, consider full participation, the path on which both preferences accept and the proposer implements every day. By~\eqref{eq:chain} with $y_\theta(p)=d_\theta-\kappa p$, the responder's value $d_\theta-\kappa p+\delta\mathcal C_\theta$ is minimized on $I$ at $p_{\mathrm{hi}}$. At the baseline the minima are $437/725$ for $H$ and $8262/725$ for $L$. When these values are nonnegative, switching to the zero-activity equilibrium after any deviation sustains the path from every forecast in $I$. Every deviation is publicly observed and yields zero, while conforming gives the proposer at least $v_L>0$ today and a nonnegative continuation. From $p_{\mathrm{lo}}$, this path has $N_H=240/29$ and $N_L=340/29$. Full participation implements every day of each preference, so the positive baseline weights, $1$ on $N_H$ and $2$ on $N_L$ in Lemma~\ref{lem:fwd_accounting}, imply that it attains $920/29$ at every $\omega$, the highest joint value of any strategy profile.

\section{Upper bounds with diary-shaped reports}\label{sec:s3}

This appendix proves Proposition~\ref{prop:weak}; $\overline{\mathcal V}_\theta$ and local screening are as in Appendix~\ref{app:forward}. The proof has five steps: restrictions linking the two preferences' values (Step~1), the form of activity at forecasts up to $\pi$ (Step~2), an upper-bound operator (Step~3), its computation in exact arithmetic (Step~4), and the end of activity together with the bound on implemented $H$ days (Step~5).

After an acceptance implemented at posterior $q$, without an outside report, positive and negative reports give tomorrow's forecasts
\[
\tilde p^{+}(q)=p_{\mathrm{lo}}+\gamma\frac{qk_H}{qk_H+(1-q)k_L},\qquad
\tilde p^{-}(q)=p_{\mathrm{lo}}+\gamma\frac{q(1-k_H)}{q(1-k_H)+(1-q)(1-k_L)},
\]
both nondecreasing in $q$. Fix diary-shaped reports, $\omega=0$, the baseline parameters and a monotone Markov equilibrium. All forecasts, suprema and infima below are over $I$. Always declining and the largest daily payoff give $0\le\mathcal V_\theta\le d_H/(1-\delta)=50$. The proof applies local screening only to unexpected declines and allows the proposer to randomize. An acceptance implemented with probability $a$ at posterior $q$ gives preference $\theta$ the value $aA_\theta(q)+(1-a)\delta\overline{\mathcal V}_\theta(p_{\mathrm{lo}}+\gamma q)$, where
\[
A_\theta(q)=d_\theta-\kappa q+\delta\bigl[k_\theta\overline{\mathcal V}_\theta(\tilde p^{+}(q))+(1-k_\theta)\overline{\mathcal V}_\theta(\tilde p^{-}(q))\bigr]
\]
is nonincreasing. A decline with posterior $r$ gives $\delta\overline{\mathcal V}_\theta(p_{\mathrm{lo}}+\gamma r)$.

\emph{Step 1: restrictions linking the two values.} We establish $\delta\overline{\mathcal V}_\theta(p)\le\mathcal V_\theta(p)$: discounted continuation at the same public forecast cannot exceed current value. This allows Step~3 to bound current value by the payoff from implementation. The two preferences have different transition probabilities, requiring two separate inequalities.

(a) $\mathcal V_L\ge\tau'\mathcal V_H$, where $\tau'=\delta p_{\mathrm{lo}}/[1-\delta(1-p_{\mathrm{lo}})]=57/107$, equivalently $\delta\overline{\mathcal V}_L(p)\le\mathcal V_L(p)$. Let $z=\tau'\mathcal V_H-\mathcal V_L$ and $M=\sup_p\max\{z(p),0\}$. Suppose both preferences accept for sure at $p$. A decline then has zero probability and, when read as $r$, gives preference $\theta$ a gain $g_\theta(r)=\delta\overline{\mathcal V}_\theta(p_{\mathrm{lo}}+\gamma r)-\mathcal V_\theta(p)$, nonincreasing in $r$. At $r_p=(p-p_{\mathrm{lo}})/\gamma\in[0,1]$, tomorrow's forecast equals $p$ and $g_L(r_p)=[1-\delta(1-p_{\mathrm{lo}})]z(p)$. If $z(p)>0$, then $\mathcal V_L(p)<\mathcal V_H(p)$ and $g_H(r_p)<0$: every interpretation weakly benefiting $H$ lies below $r_p$, where $L$ strictly benefits, while at $r_p$ only $L$ benefits. Local screening therefore reads a decline as $r=0$, where $L$ strictly gains, a contradiction. Thus $z(p)\le0$. If $H$ accepts for sure but $L$ does not, a decline reveals $L$, so $\mathcal V_L(p)=\delta\overline{\mathcal V}_L(p_{\mathrm{lo}})$ and $\mathcal V_H(p)\le\mathcal V_H(p_{\mathrm{lo}})$. Since $\tau'-\delta p_{\mathrm{lo}}=\tau'\delta(1-p_{\mathrm{lo}})$, $z(p)\le\delta(1-p_{\mathrm{lo}})z(p_{\mathrm{lo}})\le(893/1000)M$. If $H$ declines with positive probability, let $u$ be tomorrow's forecast after a decline. Then $\mathcal V_H(p)=\delta\overline{\mathcal V}_H(u)$ and $\mathcal V_L(p)\ge\delta\overline{\mathcal V}_L(u)$, so $z(p)\le\delta\{\tau'\overline{\mathcal V}_H(u)-\overline{\mathcal V}_L(u)\}$, whose right-hand side equals
\[
-\delta(1-\tau')\bigl[p_{\mathrm{lo}}+\tau'(1-p_{\mathrm{hi}})\bigr]\mathcal V_H(u)+\delta\bigl[(1-p_{\mathrm{lo}})-\tau'(1-p_{\mathrm{hi}})\bigr]z(u)\le\tfrac{18677}{21400}M.
\]
Hence $M\le(893/1000)M$, giving $M=0$.

(b) $\mathcal V_H\ge\bar s\mathcal V_L$, where $\bar s=\delta(1-p_{\mathrm{hi}})/(1-\delta p_{\mathrm{hi}})=19/44$. Writing $w_{\bar s}=\mathcal V_H-\bar s\mathcal V_L$, the identity $\mathcal V_H-\delta\overline{\mathcal V}_H=(1-\delta p_{\mathrm{hi}})w_{\bar s}$ makes this equivalent to $\delta\overline{\mathcal V}_H(p)\le\mathcal V_H(p)$. For every forecast $u$ and $s\in[0,1)$,
\begin{equation}\label{eq:cone}
\overline{\mathcal V}_H(u)-s\,\overline{\mathcal V}_L(u)=(p_{\mathrm{hi}}-sp_{\mathrm{lo}})\,w_s(u)+(1-s)\bigl[(1-p_{\mathrm{hi}})+sp_{\mathrm{lo}}\bigr]\mathcal V_L(u),
\end{equation}
where $w_s=\mathcal V_H-s\mathcal V_L$. Weighting $H$'s expectation by $k_H$ and $L$'s by $k_L$, as after a positive report, changes the coefficient on $w_s(u)$ to $k_Hp_{\mathrm{hi}}-sk_Lp_{\mathrm{lo}}$ and that on $\mathcal V_L(u)$ to $k_H(sp_{\mathrm{hi}}+1-p_{\mathrm{hi}})-sk_L(sp_{\mathrm{lo}}+1-p_{\mathrm{lo}})$. After a negative report, replace $k_\theta$ by $1-k_\theta$. Let $m=\inf w_{\bar s}$. If $L$ declines with positive probability at $p$, $H$ can also decline. At $s=\bar s$, the coefficient on $\mathcal V_L$ in~\eqref{eq:cone} is $145/3872>0$, so $w_{\bar s}(p)\ge(7809/8800)\min\{m,0\}$, where $7809/8800=\delta(p_{\mathrm{hi}}-\bar sp_{\mathrm{lo}})$. If $L$ accepts for sure, the acceptance posterior satisfies $q\le p\le p_{\mathrm{hi}}$, and $H$ can accept. Today's difference $(d_H-\bar sd_L)-\kappa(1-\bar s)q$ is at least $19/88$. The coefficients on $w_{\bar s}$ are positive and sum to $\delta(p_{\mathrm{hi}}-\bar sp_{\mathrm{lo}})$; those on $\mathcal V_L$ are nonnegative except after a negative report, where the coefficient is $\delta$ times $-8049/774400$. To control this term, note that Lemma~\ref{lem:fwd_accounting} bounds $\mathcal V_2(p_{\mathrm{lo}})$ by $d_L$ times the discounted number of $L$ days from $p_{\mathrm{lo}}$, $340/29$, giving $510/29$. Since $\mathcal V_2(p_{\mathrm{lo}})\ge(1-p_{\mathrm{lo}})\mathcal V_L(p_{\mathrm{lo}})$, monotonicity implies $\mathcal V_L\le B=25500/1363$ everywhere, so
\[
w_{\bar s}(p)\ge\tfrac{7809}{8800}\min\{m,0\}+a\varepsilon,\qquad
\varepsilon=\tfrac{19}{88}-\delta\,\tfrac{8049}{774400}\,B=\tfrac{1316263}{42220288}>0.
\]
Thus $m\ge0$, and $w_{\bar s}(p)\ge a\varepsilon>0$ wherever $L$ accepts for sure and $a>0$.

\emph{Step 2: activity at forecasts up to $\pi$.} (a) No forecast has positive implementation probability while $L$ accepts for sure and $H$ does not. A decline there reveals $H$, so Step~1(b) and monotonicity give $\mathcal V_H(p)=\delta\overline{\mathcal V}_H(p_{\mathrm{hi}})\le\mathcal V_H(p_{\mathrm{hi}})\le\mathcal V_H(p)$. Equality implies $w_{\bar s}(p_{\mathrm{hi}})=0$, and hence $w_{\bar s}(p)=\mathcal V_H(p_{\mathrm{hi}})-\bar s\mathcal V_L(p)\le w_{\bar s}(p_{\mathrm{hi}})=0$, contradicting $w_{\bar s}(p)>0$.

(b) If $p\le\pi$ and nothing is implemented at $p$, both values there are zero. On-path choices yield nothing today. If they lead to different forecasts $u<u'$ tomorrow, a preference choosing the action leading to $u'$ requires $\overline{\mathcal V}_\theta(u')\ge\overline{\mathcal V}_\theta(u)$. Monotonicity and positive transition weights imply that both values are constant on $[u,u']$. Since $p$ is a mixture of the choice posteriors, $p_{\mathrm{lo}}+\gamma p\in[u,u']$. Thus in either case $\mathcal V_\theta(p)=\delta\overline{\mathcal V}_\theta(p_{\mathrm{lo}}+\gamma p)$. As $p\le\pi$, $p_{\mathrm{lo}}+\gamma p\ge p$, so $\mathcal V_\theta(p)\le\delta\overline{\mathcal V}_\theta(p)\le\delta\max\{\mathcal V_H(p),\mathcal V_L(p)\}$ for both preferences, forcing both values to zero.

(c) If both values are zero at some $p_0\le\pi$, nothing is implemented at $p_0$ or above. By monotonicity both values vanish on $[p_0,p_{\mathrm{hi}}]$. An acceptance implemented there at posterior $q<c^H$ would give $H$ positive value, so $q\ge c^H$. Since $\tilde p^{-}(c^H)=179121/225350>\pi$ and $p_{\mathrm{lo}}+\gamma c^H>\pi$, every forecast after such an acceptance exceeds $\pi$ and has zero values. A preference in the pool then obtains $a(d_\theta-\kappa q)<0$: both preferences if $q>c^H$, and $L$, which must be in the pool, if $q=c^H<1$.

(d) If $p\le\pi$ and implementation has positive probability, both preferences accept with positive probability and the acceptance posterior satisfies $q\ge p$. If $H$ accepts for sure, Bayes' rule gives $q\ge p$, while (a) excludes the case in which only $L$ accepts for sure. Otherwise both preferences decline with positive probability, so $\mathcal V_\theta(p)=\delta\overline{\mathcal V}_\theta(p_{\mathrm{lo}}+\gamma r)$ for both, with $r$ the decline posterior. If $p_{\mathrm{lo}}+\gamma r\ge p$, the argument in (b) would give zero values, contrary to (c). Hence $p_{\mathrm{lo}}+\gamma r<p\le p_{\mathrm{lo}}+\gamma p$, giving $r<p$ and $q>p$. Since $q>0$, $H$ is in the pool. If only $H$ were in it, $q=1$ and every forecast after acceptance would be $p_{\mathrm{hi}}$. Step~1(b) and monotonicity would then bound $H$'s payoff by $a(d_H-\kappa)+\delta\overline{\mathcal V}_H(p_{\mathrm{hi}})<\mathcal V_H(p_{\mathrm{hi}})\le\mathcal V_H(p)$.

\emph{Step 3: an upper-bound operator.} At a forecast $p\le\pi$ with implementation, Step~2(d) gives $\mathcal V_\theta(p)=aA_\theta(q)+(1-a)\delta\overline{\mathcal V}_\theta(p_{\mathrm{lo}}+\gamma q)$ for both preferences, with $a>0$ and $q\ge p$. Since $p_{\mathrm{lo}}+\gamma q\ge p$, monotonicity and Step~1 imply $\delta\overline{\mathcal V}_\theta(p_{\mathrm{lo}}+\gamma q)\le\delta\overline{\mathcal V}_\theta(p)\le\mathcal V_\theta(p)$, so $\mathcal V_\theta(p)\le A_\theta(q)\le A_\theta(p)$. Combining this with Step~2(b) and replacing forecasts above $\pi$ by $\pi$, which cannot lower values, gives $\mathcal V\le\mathcal S\mathcal V$ on $[p_{\mathrm{lo}},\pi]$. Here $\bar p^\pm=\min\{\tilde p^\pm,\pi\}$ and, for nonnegative $U=(U_H,U_L)$,
\begin{equation}\label{eq:operator}
(\mathcal SU)_\theta(p)=\max\Bigl\{0,\ d_\theta-\kappa p+\delta\bigl(k_\theta\overline U_\theta(\bar p^+(p))+(1-k_\theta)\overline U_\theta(\bar p^-(p))\bigr)\Bigr\},
\end{equation}
with $\overline U_\theta$ formed from $U$ as $\overline{\mathcal V}_\theta$ from $\mathcal V$. The operator $\mathcal S$ is monotone: a larger $U$ gives a larger $\mathcal SU$. The proposer's choices enter only through $a>0$, so the bound holds whatever the proposer does.

\emph{Step 4: a finite computation.} Let $N=10^4$ and $x_i=p_{\mathrm{lo}}+(\pi-p_{\mathrm{lo}})i/N$, $0\le i\le N$, with cells $[x_i,x_{i+1})$ for $i<N$ and $\{x_N\}=\{\pi\}$. Call $U$ a step upper bound if each $U_\theta$ is nonincreasing, constant on each cell and at least $\mathcal V_\theta$ on $[p_{\mathrm{lo}},\pi]$. The constant $50$ is a step upper bound. Given such a bound, $\mathcal V\le\mathcal S\mathcal V\le\mathcal SU$, and $\mathcal SU$ is nonincreasing because $\bar p^\pm$ are nondecreasing. Hence the function equal to $\min\{U_\theta(x_i),(\mathcal SU)_\theta(x_i)\}$ on each cell is again a step upper bound. By Step~1(a), replacing $U_H$ by $\min\{U_H,U_L/\tau'\}$ also preserves the bound. Evaluating $U$ at $\bar p^\pm(x_i)$ requires only locating its cell, which uses exact rational arithmetic. Bounds are integer multiples of $10^{-8}$; each iteration uses exact integer arithmetic and rounds upward, which preserves the bound. Starting from the constant $50$, 1000 iterations give $\mathcal V_H(p_{\mathrm{lo}})\le382447941/25000000$ and $\mathcal V_L(p_{\mathrm{lo}})\le1037120049/10^8$, hence $\mathcal V_2(p_{\mathrm{lo}})\le10666803519/10^9$. Every iteration gives a valid bound, without requiring convergence.

\emph{Step 5: implemented days with preference $H$.} (a) After 1000 iterations, both step upper bounds are zero on every cell from $x_{9910}=29757/50000$ onward. Thus $\mathcal V_H=\mathcal V_L=0$ on $[x_{9910},p_{\mathrm{hi}}]$. At a forecast $p$ where both values are zero, nothing is implemented: apply Step~2(c) at $p$ if $p\le\pi$, and at $x_{9910}$ otherwise. Every choice made there with positive probability yields nothing today and leads to a forecast $u$ satisfying $\delta\overline{\mathcal V}_\theta(u)=0$ for a preference $\theta$ making that choice. Both weights in $\overline{\mathcal V}_\theta$ are positive, so both values vanish at $u$. Therefore no further implementation occurs after any forecast with both values zero, in particular once the forecast reaches $x_{9910}$.

(b) Every implemented acceptance has a posterior below $x_{9910}$. Let $U$ be the final step upper bound, extended above $\pi$ by its value at $\pi$, and let $A_L^U$ be $A_L$ with $U$ replacing $\mathcal V$. It is nonincreasing and $A_L\le A_L^U$. After acceptance at posterior $x_{9910}$, a positive report leads above $\pi$, where both bounds are zero; a negative report leads to the cell starting at $x_{9446}$, where the bounds for $H$ and $L$ are $64689903/10^8$ and $34460976/10^8$. Their weights are $\delta(1-k_L)p_{\mathrm{lo}}=47139/10^6$ and $\delta(1-k_L)(1-p_{\mathrm{lo}})=738511/10^6$, giving $A_L^U(x_{9910})=3/2-3x_{9910}+(47139\cdot64689903+738511\cdot34460976)/10^{14}=-42772815747/10^{14}<0$. At a forecast $p$ with implementation, (a) gives $p<x_{9910}$, while Steps~2(d) and~3 give $q\ge p$ and $0\le\mathcal V_L(p)\le A_L(q)$ for the acceptance posterior $q$. Thus $q<x_{9910}$.

(c) To bound future implemented $H$ days, construct a concave function of the forecast and verify that it bounds current implementation plus discounted continuation after every choice. Let $\mathcal N$ be the continuous piecewise-linear function on $[0,x_{9910}]$ through
\[
(0,\tfrac{241}{50}),\ (\tfrac{3}{50},5),\ (\tfrac{17}{100},\tfrac{133}{25}),\ (\tfrac{7}{25},\tfrac{139}{25}),\ (\tfrac{19}{50},\tfrac{273}{50}),\ (\tfrac{49}{100},\tfrac{493}{100}),\ (x_{9910},\tfrac{67}{20}).
\]
Its slopes $3$, $32/11$, $24/11$, $-1$, $-53/11$ and $-79000/5257$ decrease, so $\mathcal N$ is concave and positive. Let $\mathcal N_0=\mathcal N$ on $[p_{\mathrm{lo}},x_{9910})$, $\mathcal N_0=0$ on $[x_{9910},p_{\mathrm{hi}}]$, and $k(q)=qk_H+(1-q)k_L$. For $0\le r\le x_{9910}$ and $p_{\mathrm{lo}}\le q\le x_{9910}$,
\begin{equation}\label{eq:potential}
\delta\mathcal N_0(p_{\mathrm{lo}}+\gamma r)\le\mathcal N(r),\qquad
q+\delta\bigl[k(q)\mathcal N_0(\tilde p^{+}(q))+(1-k(q))\mathcal N_0(\tilde p^{-}(q))\bigr]\le\mathcal N(q).
\end{equation}
Between consecutive nodes and their preimages under $r\mapsto p_{\mathrm{lo}}+\gamma r$ and $\tilde p^\pm$, both margins are affine: $k(q)\tilde p^+(q)$ and $(1-k(q))\tilde p^-(q)$ are affine in $q$. Evaluating the two inequalities at the respective 12 and 15 partition points, taking the larger value at the jump of $\mathcal N_0$, gives an exact minimum margin of $132409/114664000>0$. Both inequalities therefore hold throughout their domains.

(d) Let $N_H(p)$ be the discounted expected number of implemented $H$ days from forecast $p$, at most $20$. Let $e$ be the supremum of $N_H-\mathcal N$ over forecasts where some value is positive, taking $e=0$ if the supremum is negative or there are no such forecasts. By (a), $N_H(u)\le\mathcal N_0(u)+e$ at every forecast $u$. At a forecast $p$ where some value is positive, (a) gives $p<x_{9910}$, Step~2(b) implies positive implementation probability, and (b) gives $p\le q<x_{9910}$ for the acceptance posterior. Let $\sigma$ be the acceptance probability, $a$ the implementation probability and, if $\sigma<1$, $r\le p$ the decline posterior, so that $p=\sigma q+(1-\sigma)r$. Then
\begin{align*}
N_H(p)={}&\sigma\Bigl\{a\Bigl[q+\delta\bigl(k(q)N_H(\tilde p^{+}(q))+(1-k(q))N_H(\tilde p^{-}(q))\bigr)\Bigr]+(1-a)\delta N_H(p_{\mathrm{lo}}+\gamma q)\Bigr\}\\
&+(1-\sigma)\delta N_H(p_{\mathrm{lo}}+\gamma r).
\end{align*}
By~\eqref{eq:potential} and concavity, $N_H(p)\le\sigma\mathcal N(q)+(1-\sigma)\mathcal N(r)+\delta e\le\mathcal N(p)+\delta e$; if $\sigma=1$, the decline term is absent and $q=p$. Hence $e\le\delta e$, so $e=0$ and $N_H(p_{\mathrm{lo}})\le\mathcal N(p_{\mathrm{lo}})=5$, whatever the proposer does.

\emph{Joint value.} Eliminating $N_L$ from Lemma~\ref{lem:fwd_accounting} gives $\mathcal V_1+\mathcal V_2=[v_H+v_L(\kappa-d_H)/d_L]N_H+(1+v_L/d_L)\mathcal V_2$. At the baseline the coefficients are $5/3$ and $4/3$, so from $p_{\mathrm{lo}}$, $\mathcal V_1+\mathcal V_2\le(5/3)\cdot5+(4/3)\,10666803519/10^9=16916803519/750000000$. At $\omega=1$, WT implements exactly on days with forecast $p_{\mathrm{lo}}$, at posterior $p_{\mathrm{lo}}$. From $p_{\mathrm{lo}}$ there are $1/(1-\delta)-n_{\mathrm{hi}}=352/29$ such discounted days, giving $\mathcal V_2=12144/725$ and $\mathcal V_1+\mathcal V_2=17072/725$. The guaranteed gains in Proposition~\ref{prop:weak} are the amounts by which these values exceed the two upper bounds.

\section{Data}\label{sec:s4}

\subsection{Diary constraints on the preference process}\label{subsec:s4_states}

Averaging over types, the indicators of a high preference on days $j\ne j'$ have correlation $R+(1-R)\gamma^{|j-j'|}$, whose stable component $R$ does not decay with the lag. For each column of Table~\ref{tab:states}, the model statistics are computed from this correlation as if every day of the source's 21- or 30-day window were observed, with the column's high-report share as the type mean. Compatibility compares them with each marginal interval at the three decimals shown, not with a joint confidence region, on a grid of $R,\gamma\in[0,0.5]$ in steps of $0.005$. With no stable differences, matching the adjacent-day difference would need $\gamma=0.337$ and give a share standard deviation of 0.148 and a half-window correlation of 0.039, both far outside their intervals.

\begin{table}[htbp]
\centering
\caption{Diary statistics for the compatibility comparison}\label{tab:states}
\footnotesize
\setlength{\tabcolsep}{3pt}
\begin{tabular*}{\linewidth}{@{\extracolsep{\fill}}lccc@{}}
\toprule
 & 21-day diary, $\ge6$ & 21-day diary, $\ge5$ & 30-day diary, $>3$\\
\midrule
High-report share & 0.390 & 0.604 & 0.318\\
Personal-share SD & 0.282 [0.258, 0.302] & 0.285 [0.266, 0.303] & 0.292 [0.267, 0.314]\\
Half-window correlation & 0.752 [0.681, 0.813] & 0.780 [0.721, 0.832] & 0.869 [0.833, 0.899]\\
Adjacent-day frequency difference & 0.337 [0.276, 0.397] & 0.358 [0.308, 0.406] & 0.442 [0.388, 0.497]\\
\bottomrule
\end{tabular*}
\par\smallskip\begin{minipage}{0.97\linewidth}\footnotesize
\emph{Notes:} The high-report share supplies the source mean; the 30-day diary counts a self-report above 3 on its five-point scale as high. Individuals need at least ten valid reports; half-window correlations require six in each half. The samples contain about 230 people from 120 couples in the main diary (220 from 110 for half-window correlations) and 270 from 130 in the independent diary's single imputed release. Person and couple counts are separately rounded to tens. Brackets give 2.5th and 97.5th percentiles of 1,000 whole-couple bootstrap resamples, within that release for the independent diary.
\end{minipage}
\end{table}

Four compatible settings enter the analysis: $(R,\gamma)=(0.3,0.05)$ and $(0.255,0.18)$ fit both main-diary cutpoints, $(0.37,0.115)$ the independent diary, and $(0.3,0.13)$ all three columns. Only the diary-shaped report is computed at all four; other reports use the baseline process within the compatible range. On the grid, the largest compatible persistence is $0.215$ for the main diary and $0.235$ for the independent diary (at $R=0.34$). At $R=0.3$, as in Section~\ref{subsec:protection}, persistence from $0.015$ to $0.15$ is compatible with at least one main-diary cutpoint, and from $0.020$ to $0.135$ with the $\ge6$ cutpoint.

\FloatBarrier
\subsection{Report discrimination}\label{subsec:s4_signals}

Each judgment is matched to the partner's self-report on the same scheduled day, in both directions; $k_H$ and $k_L$ are the frequencies with which a judgment meets its high criterion after a high or low self-report. The diary-shaped report requires at least three high and three low self-report days in each direction. Two alternatives replace the median with the judge's mean or two-thirds quantile, the latter interpolating linearly at order position $1+2(n-1)/3$ among the $n$ sorted judgments. All cutpoints use every day with both the judgment and the judged person's self-report valid, before selection by activity. Only a judgment strictly above the cutpoint is positive.

\begin{table}[htbp]
\centering
\caption{Conditional report probabilities}\label{tab:signals}
\footnotesize
\setlength{\tabcolsep}{3pt}
\begin{tabularx}{\linewidth}{@{}>{\raggedright\arraybackslash}p{.25\linewidth}Xcc@{}}
\toprule
Report & Measurement or construction & $k_H$ & $k_L$\\
\midrule
Diary-shaped & Above the judge's median & 0.264 [0.216, 0.314] & 0.173 [0.142, 0.205]\\
Mean judgment & Above the judge's mean & 0.498 [0.430, 0.554] & 0.341 [0.291, 0.387]\\
Two-thirds-quantile judgment & Above the judge's two-thirds quantile & 0.162 [0.128, 0.204] & 0.072 [0.051, 0.095]\\
\midrule
No-activity days & Judgment at least 6; both report no activity & 0.384 [0.308, 0.456] & 0.171 [0.134, 0.210]\\
Activity days & Judgment at least 6; both report activity & 0.811 [0.769, 0.851] & 0.565 [0.498, 0.637]\\
\midrule
Nonconflict events & Actor's conflict rating at most 2 & 0.623 [0.562, 0.679] & 0.428 [0.343, 0.528]\\
Conflict events & Actor's conflict rating at least 4 & 0.698 [0.575, 0.802] & 0.385 [0.190, 0.600]\\
\midrule
Weak reference & Half diary-shaped, half uninformative reports & 0.2367 & 0.1912\\
Strong reference & Higher discrimination & 0.900 & 0.100\\
Near-perfect reference & Higher discrimination & 0.990 & 0.010\\
\bottomrule
\end{tabularx}
\par\smallskip\begin{minipage}{0.97\linewidth}\footnotesize
\emph{Notes:} $k_H$ and $k_L$ are positive-report probabilities under high and low preference, respectively. The first three rows use days when both members report no activity; their cutpoints use all valid matched days. Empirical probabilities enter simulations at three decimals. The weak reference's uninformative report is positive with probability $0.2094=0.4\times0.264+0.6\times0.173$ under either preference, leaving the positive-report frequency at prior $0.4$ unchanged. Brackets are bootstrap percentiles as in Table~\ref{tab:states}, holding cutpoints fixed; each resample's two probabilities are used jointly below. High and low records, rounded to tens: about 550 and 690 for the first three rows, 890 and 2,180 (fixed-cutpoint no-activity), 710 and 260 (activity), 1,020 and 140 (nonconflict events) and 60 and 30 (conflict events).
\end{minipage}
\end{table}

The additional empirical shapes come from alternative cutpoints and activity samples in the same diary and from the experience-sampling data on 139 couples \citep{columbus2021,columbus2021data}. In these data, an actor rates a behavior's effect on themselves and on the partner, and the partner separately rates its effect on the actor, each on a five-point scale from costly to beneficial. Records pair the two members' reports of the same event and enter if the actor rates the benefit to the partner at least 4; the actor's own benefit is high at 4 or above, and the report is positive when the partner rates the actor's benefit at least 4.

Sampling error and report construction determine the ceiling's margin. At baseline payoffs, condition~\eqref{eq:ceiling_condition} reads $\Lambda\le3-4\gamma/(1-\gamma)$. At $\gamma=0.05$ this is $\Lambda\le53/19\approx2.79$, which the diary-shaped report satisfies in all 1,000 whole-couple bootstrap resamples; every empirical shape and the weak reference also satisfy it at their point estimates. Using unrounded frequencies, one positive report at prior $0.4$ gives posterior $0.503$, with 2.5th and 97.5th bootstrap percentiles $0.449$ and $0.558$, corresponding to $\Lambda$ of about $1.22$--$1.89$. At the maximum compatible persistence values of $0.215$ and $0.235$, the largest admissible $\Lambda$ falls to $1.90$ and $1.77$, the latter below the upper end of this range. The sharper two-thirds-quantile and fixed-cutpoint no-activity shapes ($\Lambda\approx2.25$) allow persistence only up to about $0.16$, and the conflict-event shape to $0.23$. The guarantee near the compatible upper limits thus rests on point estimates and the median construction. When received before choice (Section~\ref{subsec:information}), a positive report raises the posterior above $c_*$ in 546 resamples: sampling uncertainty straddles the participation threshold, but its upper percentile $0.558$ remains far below $\widehat p$.

The precision comparisons in Section~\ref{sec:evidence} require the stronger report to contain all the information in the weaker one. For binary reports, one signal is a garbling of another if it can be generated from the original by randomization independent of the preference: reporting positive with probabilities $\xi_+$ and $\xi_-$ after positive and negative original signals gives $k_\theta'=\xi_+k_\theta+\xi_-(1-k_\theta)$, where $\theta\in\{H,L\}$ and $\xi_+,\xi_-\in[0,1]$. The original signal is then more informative in the Blackwell sense.

\begin{samepage}
\begin{lemma}[The Blackwell order of postchoice signals]\label{lem:blackwell}
For $0<k_L<k_H<1$ and $0<k_L'<k_H'<1$, $(k_H',k_L')$ is a garbling of $(k_H,k_L)$ if and only if
\[
\frac{k_H'}{k_L'}\le\frac{k_H}{k_L},\qquad
\frac{1-k_L'}{1-k_H'}\le\frac{1-k_L}{1-k_H}.
\]
The two conditions compare how well positive reports distinguish the high preference and negative reports distinguish the low preference. The diary-shaped signal $(0.264,0.173)$ is a garbling of the strong signal $(0.9,0.1)$, which is a garbling of the near-perfect signal $(0.99,0.01)$.
\end{lemma}
\end{samepage}

\begin{proof}
The two defining linear equations have determinant $k_H-k_L>0$ and the unique solution
\[
\xi_+=\frac{k_H'(1-k_L)-k_L'(1-k_H)}{k_H-k_L},\qquad
\xi_-=\frac{k_Hk_L'-k_Lk_H'}{k_H-k_L}.
\]
On the stated domain, $\xi_+>0$ and $\xi_-<1$ always hold. The solution is therefore feasible if and only if $\xi_-\ge0$ and $\xi_+\le1$. Cross-multiplying these conditions gives exactly the two likelihood-ratio inequalities in the statement. The strong signal is obtained from the near-perfect signal with $(\xi_+,\xi_-)=(89/98,9/98)$, and the diary-shaped signal from the strong signal with $(\xi_+,\xi_-)=(2203/8000,1293/8000)$. Both pairs of coefficients belong to $[0,1]^2$.
\end{proof}

The diary-shaped and activity-day reports differ in the evidence supplied by positive and negative messages. A positive diary-shaped report gives stronger evidence of state $H$, while a negative activity-day report gives stronger evidence of state $L$. Their likelihood ratios satisfy $0.264/0.173>0.811/0.565$ and $0.827/0.736<0.435/0.189$, respectively. Neither report therefore discriminates more strongly in both directions; by Lemma~\ref{lem:blackwell}, they are Blackwell-incomparable. The precision comparisons use only the chain in the lemma.

\subsection{The timing of opportunities}\label{subsec:s4_clock}

The hidden clock uses adjacent-day transitions in the main diary's joint desire states: both low if both self-reports are at most 3, both high if both exceed 4, other otherwise (probabilities in the replication materials); each pair starts at its stationary distribution. The log odds of an opportunity equal an intercept, plus a state offset of $-1$, $0$ or $1$ for both low, other and both high, plus $\beta_\pi$ times the standardized type $(\pi-\bar\pi)/\sqrt{R\bar\pi(1-\bar\pi)}$. Informative opportunities use $\beta_\pi=\pm\log2$, so that, at a given clock state, a type one standard deviation higher doubles or halves the odds of an opportunity; the baseline has $\beta_\pi=0$, and the intercept sets the mean to $7/20$.

\section{Numerical methods, diagnostics and complete comparisons}\label{sec:s5}

\subsection{Filtering and integration}\label{subsec:s5_filter}

The public belief is the joint distribution of the type and today's preference, updated by the likelihoods of Appendix~\ref{app:histories}; outside feedback applies report likelihoods even without implementation, and nondelivery is uninformative. With $\beta_\pi\ne0$ the belief also carries the hidden clock state, updated by each opportunity observation. Unknown types use Gauss--Jacobi nodes for the Beta prior. With uninformative opportunities the likelihoods are polynomials in $\pi$ of degree at most 85; an 81-node Gauss rule is exact up to degree 161, so in exact arithmetic 81 and 121 nodes give the same posteriors and comparing them measures rounding error. For the logistic likelihoods of informative opportunities, 81 nodes are compared with 161. Participation uses $P_j\le c_*$ without tolerance, with thresholds, mixed-acceptance posteriors and the prior mean represented exactly, so a forecast equal to the threshold is feasible, as at $A=1.70$, where $c_*=\bar\pi=0.4$.

\subsection{Node precision}\label{subsec:s5_precision}

All 136 node comparisons use the same random draws across node counts; none changes the participation classification, type-mean classification relative to $c_*$, forecast band, or any simulated choice, report or blocked day. Differences stay within $10^{-12}$ for posteriors and $10^{-10}$ for economic summaries scaled by $\max(1,|\text{81-node value}|)$, except for near-perfect experience at $R=0.7$ and $\gamma=0.95$, whose prior is singular at both ends. Its maximum posterior and scaled economic-summary differences are $2.39\times10^{-11}$ and $5.46\times10^{-9}$, far below its Monte Carlo interval widths. Of the 48 unknown-type simulations without a separate node comparison, 34 (ten used to set outside-feedback delivery probabilities) use the same 81-node rule; the other 14, no-report benchmarks with pooling proposals only, keep the prior belief.

\subsection{Belief calibration}\label{subsec:s5_calibration}

Choice-time forecasts and type means are compared with realized preferences and types in 20 belief bins for each environment over days 1--84, with standard errors clustered over 8,192 pairs; calibration is implied by the filter, so this checks its numerical implementation. All 2,399 qualifying bins, with at least 1,000 pair-days and 1,000 contributing pairs, lie within the critical absolute standardized residual of 5.061 (normal reference, Bonferroni family level 0.001), and the other 4,521 bins have insufficient support. The most informative bin is the main environment's $[0.50,0.55)$, just above the participation threshold, where blocked forecasts lie: the forecast's residual is $0.00044$ (standard error $0.0062$) and the type mean's $-0.00072$ ($0.0061$).

\subsection{Residuals from the payoff identity}\label{subsec:s5_accounting}

Proposition~\ref{prop:feedback_value} holds in expectation, not history by history. The residual is the simulated payoff change minus the right side of~\eqref{eq:feedback_value} on the same paired histories. Across 1,197 comparisons (133 feedback environments, each partner and the pair, three windows), its paired interval contains zero in 1,177; the 20 exceptions, fewer than the roughly 60 expected by chance at 95\% coverage, split 12, 7 and 1 across days 1--21, 1--30 and 1--84. In every window, unknown-type no-report simulations with uninformative opportunities and pooling proposals only recover the static no-information implementation rate and per-implementation responder utility within paired intervals.

\subsection{Monte Carlo intervals and ratios}\label{subsec:s5_intervals}

The estimate is the mean of the eight replication changes, each averaging 1,024 pairs over the window, with interval $\pm t_{7,0.975}$ times their standard deviation over $\sqrt8$: the Monte Carlo error of a mean conditional on the parameter values, not dispersion across pairs. Ratios are ratios of pooled means with first-order delta-method intervals from the paired influences $(\mathrm{num}_r-\widehat{\mathrm{ratio}}\,\mathrm{den}_r)/\overline{\mathrm{den}}$ of replications $r=1,\ldots,8$; when a ratio has a loss or cost in the denominator that is not detected positive, its numerator, its denominator and their difference are reported instead. Shares are not truncated to $[0,1]$.

\subsection{The complete comparisons and long-run losses}\label{subsec:s6_census}

The 133 feedback environments produce 399 payoff comparisons, each for one partner or the pair relative to the same environment without reports. Of these, 101 comparisons from 67 environments can still yield a gain under the analytical conditions (Table~\ref{tab:census}). Participation rules, opportunity values and forecast bounds (Corollaries~\ref{cor:who_gains} and~\ref{cor:known_type} and Proposition~\ref{cor:ceiling}) rule out gains in the other 298: 293 are detected negative and five are not detected, all five under the weak reference with pooling proposals only.

\begin{table}[htbp]
\centering\small
\caption{Comparisons in which analytical conditions do not rule out a gain, days 1--84}\label{tab:census}
\begin{tabularx}{\linewidth}{@{}>{\raggedright\arraybackslash}Xrrrr@{}}
\toprule
Persistence band & Comparisons & Negative & \shortstack{Not\\detected} & Positive\\
\midrule
Diary-compatible & 36 & 16 & 20 & 0\\
$\gamma=0.18$--$0.50$ & 17 & 11 & 3 & 3\\
$\gamma=0.70$--$0.90$ & 26 & 20 & 3 & 3\\
$\gamma=0.95$--$0.99$, including $R=0.1$ and $0.7$ at $\gamma=0.95$ & 22 & 9 & 2 & 11\\
\bottomrule
\end{tabularx}
\par\smallskip\begin{minipage}{\linewidth}\footnotesize\emph{Notes:} Entries count comparisons; the underlying payoff changes are per 100 pair-days relative to the same environment without reports. Negative, not detected and positive mean that the pointwise 95\% paired Monte Carlo interval lies below zero, contains zero or lies above zero, respectively. Intervals condition on fixed outside-feedback schedules. The first band is diary-compatible; all others lie outside the compatible range. Its 36 comparisons come from 24 environments.\end{minipage}
\end{table}

All 36 compatible comparisons use the baseline process; theory rules out gains in the other three compatible settings, which use only diary-shaped reports under the maintained mixture. Among the 20 undetected changes, the largest interval upper endpoint is 0.062 per 100 pair-days, for the responder under near-perfect outside feedback at $A=1.70$; all others are at most 0.013. The shorter windows detect no gain either: days 1--21 give 13 negative and 23 undetected changes, and days 1--30 give 16 negative and 20 undetected changes. None is detected positive in any of the three windows for any proposer weight $0\le\alpha_1<2$ on the responder's direct payoff (Section~\ref{sec:robustness}); $\alpha_1$ changes no simulated history and enters each payoff comparison linearly, so the whole range can be checked. Among the diary-compatible experience environments in Table~\ref{tab:spell_burden}, theory leaves open a responder gain in only four comparisons, strong and near-perfect reports at $A=2$ and near-perfect reports at $A=1.70$ and $1.75$, all of which are negative.

All 17 positive comparisons outside the compatible range use pooling proposals only: 15 concern the responder, two the pair and none the proposer. Eleven use outside feedback with a known type. Negative reports can release pairs excluded by the no-report benchmark from day 1; at $\gamma\ge0.90$, the stops this feedback itself causes also begin to protect the responder. The positive comparison closest to the compatible range is at $(R,\gamma)=(0.3,0.18)$, with a responder gain of $+0.0072$ $[0.0005,0.0140]$, less than 0.1\% of the benchmark; known-type outside feedback was not computed at the more persistent compatible settings. Known-type near-perfect experience at $\gamma=0.99$ also raises the responder's payoff, by $0.207$ $[0.003,0.410]$: its positive reports trigger additional stops, most of which begin above the payoff threshold and so protect the responder at first. The gain is detected in days 1--21 and 22--30 ($0.554$ and $0.625$), but not in days 31--84 ($0.002$): protection expires before these stops end (Corollary~\ref{cor:spell_path}), and most are right-censored on day 84. The remaining five positive comparisons have unknown types and $\gamma\ge0.95$: near-perfect outside feedback at 0.95 and 0.99, strong outside feedback at 0.99, near-perfect experience at 0.99, and near-perfect experience at 0.95 with $R=0.7$.

Proposition~\ref{cor:ceiling} applies to 40 experience environments. In 30, all diary-compatible, the ceiling lies below the payoff threshold; in the other ten it is not low enough to decide, including six compatible environments. No blocked opportunity lies above the payoff threshold in the complete simulated histories of the 38 unknown-type experience environments with uninformative opportunities and $\gamma\le0.30$.

The seven undetected outside-minus-experience comparisons in Section~\ref{sec:evidence} are unknown-type diary-shaped reports with pooling proposals only at $\gamma=0.05$, $0.50$ and $0.90$; the weak reference with pooling proposals only; known-type diary-shaped reports under the maintained mixture; and the two informative-opportunity environments under that mixture.

\paragraph{The long-run bounds in Section~\ref{subsec:type_knowledge}.}
Maintain the baseline process, pooling proposals only and diary-shaped reports. Let $\mathcal I_j^{-}$ be the public history just before observing whether day $j$ brings an opportunity; the superscript marks only this timing. Opportunities contain no type or preference information, so the type posterior and forecast $P_j$ are unchanged by observing the opportunity. Write $v_j=\operatorname{Var}(\pi\mid\mathcal I_j^{-})$. Given $\pi=z$, let $r_j(z)$ be the probability of preference $H$ that day. For $j\ge2$,
\[
r_j(z)=(1-\gamma)z+\gamma\widetilde q_{j-1}(z),
\]
where $\widetilde q_{j-1}(z)$ is the posterior probability of $H$ on the previous day, conditional on the type and public record through that day's end. Since $r_1(z)=z$, pooling public actions reveal no preference information and a binary report's posterior increases with its prior, induction makes $\widetilde q_{j-1}$ nondecreasing in $z$. The covariance of a random variable with a nondecreasing function of itself is nonnegative, hence
$\operatorname{Cov}(\pi,r_j(\pi)\mid\mathcal I_j^{-})\ge(1-\gamma)v_j$.

The law of total variance for a binary observation then implies that each report reduces expected posterior variance by at least $4(k_H-k_L)^2(1-\gamma)^2v_j^2$. Every state of the finite opportunity clock has a strictly positive opportunity probability and $\lambda\eta\zeta>0$. Conditional on $\mathcal I_j^{-}$, the probability of receiving a report on a feasible day therefore has a uniform lower bound $\varepsilon>0$. Setting $C_v=4\varepsilon(k_H-k_L)^2(1-\gamma)^2>0$ gives
\[
\mathbb E[v_{j+1}\mid\mathcal I_j^{-}]
\le v_j-C_v\mathbf1\{P_j\le c_*\}v_j^2,
\qquad
\mathbb E\sum_{j\ge1}\mathbf1\{P_j\le c_*\}v_j^2
\le\frac{v_1}{C_v}<\infty.
\]
Summing over days gives the second inequality. The nonnegative supermartingale $v_j$ converges to zero on the event of infinitely many feasible days; the type judgment then converges almost surely to the true type. If there are only finitely many feasible days, the last stop never ends on its own (Corollary~\ref{cor:spell_path}).

If feasible days are infinite and $\pi>c_*$, the type judgment eventually stays above the threshold by a positive margin. Independent participation selection gives any fixed run of $K$ days without participation conditional probability at least $(1-\lambda)^K>0$. A sufficiently long report-free run therefore occurs almost surely, allowing the forecast to approach the type judgment and cross $c_*$, after which the stop never ends on its own. Thus these pairs almost surely become permanently excluded under experience, and their opportunity weight retains the long-run time average $\bar G=0.1575$ after the stop. Nonnegative losses and Fatou's lemma imply that the lower limits of long-run average expected losses per 100 pair-days are at least
\[
100\,\bar G\Pr(\pi>c_*)\,(1.375,\;0.25)
=15.75\times0.35672775\ldots\times(1.375,\;0.25)
>(7.7,\;1.4),
\]
for the pair and responder, respectively. The weight bounds follow from $2-P$ and $1.5-2P$ in~\eqref{eq:weights} and the forecast ceiling $5189/8303<0.625$ in Proposition~\ref{cor:ceiling}. Under the same prior, the corresponding expected losses from type revelation are $7.26658\ldots$ and $0.48701\ldots$, both strictly below these bounds.
\FloatBarrier

\section{Behavior under fixed continuations}\label{sec:s7}

The comparisons behind Sections~\ref{sec:evidence} and~\ref{sec:robustness} ask whether a player gains by deviating today. They fix the player's information and compare today's pure plans, with prescribed behavior from the next day through day 84. Counts are numbers of evaluated states. Profitability requires a positive lower endpoint of the pointwise 95\% Monte Carlo gain interval after subtracting a bound on the payoffs that a continuation ending before day 84 leaves out (512 paired continuations, $t_{511}$ reference).

\subsection{Feasible states and refusal readings}\label{subsec:s7_feasible}

Responder counts here and in Section~\ref{sec:robustness} are out of 24 states, eight sampled from each period of Section~\ref{subsec:precision}. The proposer comparisons cover 2,112 information sets: $A\in\{2,1.75,1.70\}$, both feedback sources, continuations of up to 14 days at $\delta\in\{0,0.25,0.75,0.95\}$ and continuations through day 84 at $0.95$, for 8,976 comparisons in all (Section~\ref{sec:robustness}).

At the joint-acceptance states where one positive report would take tomorrow's forecast above $c_*$, refusal followed by $f$ is read as $L$, or neutrally as uninformative. With outside feedback, a neutrally interpreted refusal leaves subsequent information unchanged. Under experience at $\delta=0.95$, neutral readings remove every detected $H$ refusal gain. At $A=2$, 11 profitable $L$ refusals remain near $c_*$ ($0.480$--$0.500$): 6/24 under the maintained mixture and 5/24 with pooling proposals only; joining there gives $L$ almost nothing today and risks a false positive report. Under the maintained mixture, $A=1.70,1.75$ give 16/24 and 7/24, respectively; $\delta=0.75$ reduces but does not eliminate these refusals.

\subsection{Counter-proposals during exclusion}\label{subsec:s7_counter}

On an excluded encounter, a day with an opportunity on which participation is infeasible, a counter-proposal is the responder's only unilateral route to the activity (Section~\ref{sec:evidence}). The encounters are sampled from eight experience environments with pooling proposals only: diary-shaped reports at $\gamma=0.05,0.99$, near-perfect at $0.05,0.50,0.90,0.95,0.99$, and strong at $0.99$. Sampling takes at most eight points per environment, period (Section~\ref{subsec:precision}) and group: intermediate judgments $c_*\le m<\widehat p$, split by whether the forecast exceeds $\widehat p$, and low judgments $m<c_*$. The 425 excluded encounters before day 84 include 132 actual $L$, on which the responder's preference that day is $L$. All pure plans of a responder offered $f$ are compared through day 84 at counter-proposal readings $1,0.99,0.98,0.95,0.90$; at reading $d_H/\kappa=5/6$, where today's payoff to a responder who wants the activity is zero; and at the neutral reading $P_j$, under which a counter-proposal conveys no information. Opposite-preference cases reverse today's preference with private history fixed, so each preference state is evaluated at all 425 points.

A counter-proposal read as certain $H$ yields no detected gain. For intermediate judgments the continuation gain is exactly zero: a later report cannot move a posterior of one, so the stop continues either way. At tested readings $0.90$--$1$, profitable $L$ cases, including opposite-preference cases, require near-perfect reports with $\gamma\ge0.50$ or strong reports at $\gamma=0.99$; at $\gamma=0.50$ or with strong reports, gains are detected only at reading $0.90$. At reading $0.98$ and near-perfect $\gamma=\delta=0.95$, actual $L$ profits at 8/12 intermediate points, or 9/20 including low judgments. The two diary-shaped environments and near-perfect $\gamma=0.05$ yield no detected $L$ gain at readings $0.90$--$1$; at $\delta=0.95$, $L$ gains detectably at 5 of their 103 encounters under reading $5/6$ and at 57 under the neutral reading, including opposite-preference cases. At readings at least $0.90$, neither $L$ at $\delta=0.75$ nor $H$ at either discount factor gains detectably.

Under the two weaker readings, counter-proposals often pay. At $\delta=0.95$, actual $L$ gains at 66/132 points under $5/6$ and 111/132 neutrally (244/425 and 360/425 including opposite-preference cases); at $0.75$, neutral gains fall to 48/132 (134/425). Including opposite-preference cases, $H$ gains solely through continuation at 266/425 and 291/425 under $5/6$ ($\delta=0.95,0.75$, respectively); neutral readings give 353/425 and 364/425, usually with gains already today. In state $h$, the proposer's gain from accepting a counter-proposal has a lower endpoint of at least $0.430$ throughout. In state $l$, the proposer favors refusal, except at nine encounters under the neutral reading at $\delta=0.95$ (six acceptances, three undetermined).

\end{document}